\documentclass[pre,amsmath,amssymb,aps,preprint]{revtex4-1}
\usepackage{graphicx}
\usepackage{dcolumn}
\usepackage{bm}

\begin{document}
\preprint{APS/123-QED}

\title{Direct coupling information measure from non-uniform embedding}
\author{D. Kugiumtzis}
 \email{dkugiu@gen.auth.gr}
 \homepage{http://users.auth.gr/dkugiu}
 \affiliation{Department of Mathematical, Physical and Computational Sciences, Faculty of
Engineering, Aristotle University of Thessaloniki, Thessaloniki 54124, Greece }

\begin{abstract}
A measure to estimate the direct and directional coupling in
multivariate time series is proposed. The measure is an extension
of a recently published measure of conditional Mutual Information
from Mixed Embedding (MIME) for bivariate time series. In the
proposed measure of Partial MIME (PMIME), the embedding is on all
observed variables, and it is optimized in explaining the response
variable. It is shown that PMIME detects correctly direct
coupling, and outperforms the (linear) conditional Granger
causality and the partial transfer entropy. We demonstrate that
PMIME does not rely on significance test and embedding parameters,
and the number of observed variables has no effect on its
statistical accuracy, it may only slow the computations. The
importance of these points is shown in simulations and in an
application to epileptic multi-channel scalp EEG.
\end{abstract}


\maketitle

\section{Introduction}

In the recent years the study of causality in multivariate time
series, has gained much attention, also due to the advances of
complex networks from time series, and has contributed in the
understanding of complex systems \citep{Zanin12b}. Considering the
global system as a network, the interest in this work is in the
direct effect a driving sub-system, observed through a variable
$X$, may have on the evolution of a response sub-system, observed
through a variable $Y$. This is to be distinguished from an
indirect effect $X$ may have on $Y$ via other sub-systems, say
$Z$, where the observed variables in $Z$ are referred to as
confounding variables.

There are established linear measures of direct causality, such as
the conditional Granger Causality index (CGCI) \citep{Geweke84}.
Though many nonlinear directional coupling measures have been
proposed in the last decade \cite{Hlavackova07}, there are only
few extensions accounting for indirect effects, such as the
partial phase synchronization \citep{Schelter06c} and the partial
transfer entropy (PTE) \citep{Vakorin09,Papana12}. A possible
reason for this unbalanced production of measures might be the
increased data requirements when adding confounding variables in
the calculations. For example, for the same delay embedding with
embedding dimension $m$ (and delay $\tau$) for $X$ and $Y$, the
transfer entropy (TE) measuring the causal effect from $X$ to $Y$
requires the estimation of a joint probability distribution of
dimension $2m+1$ ($m$ for $X$, $m$ for $Y$ and 1 for the future of
$Y$). Extending TE to PTE when totally $K$ variables are observed,
the dimension becomes $Km+1$, and eventually PTE fails for a large
$m$ or $K$. This is indeed a common practical setting, e.g.
electroencephalograms (EEG), climatic records, and stock
portfolio, and there have been some suggestions on reducing the
dimension \citep{Shibuya11,Marinazzo12,Runge12}.

Dimensionality reduction is the first drawback we intend to
successfully address with the proposed measure. The next drawback
is related to the embedding parameters $m$ and $\tau$. In real
settings, one does not know aforehand the best choice of embedding
parameters, and recent works have shown that the measure
performance is very much dependent on them \cite{Papana11}. The
third drawback is the need for a statistical test of significance,
which for nonlinear measures is computationally intensive
requiring resampling test using surrogate data.

We recently proposed a non-uniform embedding scheme, bypassing the
problem of selecting the embedding parameters, and derived a
measure for bivariate directional coupling, the conditional Mutual
Information from Mixed Embedding (MIME) \cite{Vlachos10}. For this
we used information criteria and found that the $k$-nearest
neighbors (kNN) estimate of entropies, and consequently mutual
information (MI), is stable and efficient, as it adapts the local
neighborhood to the dimension of the state space \citep{Kraskov04}
(for a similar approach based on entropy and binning estimate see
\cite{Faes11}). Here, we extend the measure MIME to multivariate
time series, and form the partial MIME (PMIME) that can detect
direct coupling. The idea is first to reconstruct a point (vector)
in the subspace of the joint state space of lagged variables $X$,
$Y$ and $Z$, derived from the non-uniform embedding scheme with
the purpose of explaining best the evolution of $Y$. The derived
mixed embedding vector contains only the most relevant components
from all variables, avoiding thus large dimension that would
deteriorate the estimation. The presence of components of $X$ in
this vector indicates that $X$ has some effect on the evolution of
$Y$ and then the derived information measure PMIME is positive,
whereas the absence indicates no effect and then PMIME is exactly
zero.

We explain the measure in detail in Section ~\ref{sec:measure}. In
Section~\ref{sec:simulations}, we demonstrate the effectiveness of
PMIME, compared also to PTE and CGCI, on a number of simulated
systems and a multi-channel scalp EEG recording. We conclude in
Section~\ref{sec:conclusions}.

\section{The measure of partial mutual information from mixed embedding}
\label{sec:measure}

Let $\{x_t,y_t,z_{1,t},\ldots,z_{K-2,t}\}_{t = 1}^{n}$ be a multivariate time series of $K$ variables $X,Y,Z_1,\ldots,Z_{K-2}$, and we want to estimate the effect of $X$ on $Y$ conditioning on $Z\! = \!\{Z_1,\ldots,Z_{K-2}\}$.
The future of $Y$ at each time step $t$ is generally represented by a vector of $T$ feature values, $\mathbf{y}_t^T\! = \![y_{t+1},\ldots,y_{t+T}]$. This is an extension of the one step ahead, $\mathbf{y}_t^1\! = \!y_{t+1}$, and can be more appropriate in some settings, e.g. a relatively dense sampling for continuous-timed systems. The lags of $X$, $Y$ and $Z$ are searched within a range given by a maximum lag for each variable, e.g. $L_x$ for $X$ and $L_y$ for $Y$. When all variables are of the same type, e.g. EEG signals, it is natural to assume the same maximum lag $L$ for all variables. Let us denote the set of all lagged variables at time $t$ as $W_t$, containing the components $x_t,x_{t-1},\ldots,x_{t-L_x}$ of $X$ and the same for the other variables.

We use an iterative scheme to form the mixed embedding vector $\mathbf{w}_t \in W_t$ starting with an empty embedding vector, $\mathbf{w}_t^0 \! = \! \emptyset$ \cite{Vlachos10}. In the first iteration, termed first embedding cycle, we find the component in $W_t$ being most correlated to $\mathbf{y}_t^T$ given by the kNN estimate of MI, $w_t^1 \! = \! \mbox{argmax}_{w \in W_t} I(\mathbf{y}_t^T;w)$, and we have $\mathbf{w}_t^1\! = \![w_t^1]$. In the second embedding cycle, the mixed embedding vector is augmented by the component $w_t^2$ of $W_t$, giving most information about $\mathbf{y}_t^T$ additionally to the information already contained in $w_t^1$, i.e. $w_t^2 \! = \! \mbox{argmax}_{w \in W_t} I(\mathbf{y}_t^T;w|w_t^1)$, where the conditional mutual information (CMI) is again estimated by kNN, and the mixed embedding vector is $\mathbf{w}_t^2\! = \![w_t^1,w_t^2]$. The progressive vector building stops at the embedding cycle $j$ and we have $\mathbf{w}_t\! = \!\mathbf{w}_t^{j-1}$, if the additional information of $w_t^j$ selected at the embedding cycle $j$ is not large enough. In \cite{Vlachos10}, we quantified this with the termination criterion
\begin{equation}
I\big(\mathbf{y}_t^T;\mathbf{w}_t^{j-1}\big)/I\big(\mathbf{y}_t^T;\mathbf{w}_t^{j}\big)> A
\label{eq:termination}
\end{equation}
for a threshold $A < 1$.

The obtained mixed embedding vector $\mathbf{w}_t$ may contain any of the lagged variables $X,Y,Z_1,\ldots,Z_{K-2}$, and the interest in terms of the causality $X\! \rightarrow \! Y$ is whether there are any components of $X$ in $\mathbf{w}_t$. Let us denote the components of $X$ in $\mathbf{w}_t$ as $\mathbf{w}_t^x$, for $Y$ as $\mathbf{w}_t^y$ and for the other variables in $Z$ as $\mathbf{w}_t^z$. To quantify the causal effect of $X$ on $Y$ conditioned on the other variables in $Z$, we define PMIME as
\begin{equation}
R_{X \! \rightarrow \! Y | Z}\! = \!
{I(\mathbf{y}_t^T; \mathbf{w}_t^x \mid \mathbf{w}^y, \mathbf{w}^z) \over I(\mathbf{y}_t^T; \mathbf{w}_t)}.
\label{eq:PMIME}
\end{equation}
The numerator is the CMI of the future response vector and the part of the mixed embedding vector formed by lags of the driving variable, accounting for the rest part of the vector.
The form of CMI is similar to PTE, but in PTE the uniform delay embedding vectors of $X$, $Y$ and $Z$ are used and the delay parameters have to be set. The normalization in eq.(\ref{eq:PMIME}) with the MI of the future response vector and the whole mixed embedding vector restricts $R_{X \! \rightarrow \! Y | Z}$ in [0,1], and it is zero if there are no driving components in the mixed embedding vector ($\mathbf{w}_t^x\! = \!\emptyset$), meaning there is no direct causal effect from $X$ on $Y$, and it is one if the mixed embedding vector is totally dominated by the driving variable ($\mathbf{w}_t^y\! = \!\mathbf{w}_t^z\! = \!\emptyset$). The latter is rather unlikely to be met in practice and in general we expect $R_{X \! \rightarrow \! Y | Z}$ to be closer to zero than to one.

The free parameters in PMIME are the maximum time lags for each variable, e.g. $L_X$, the time horizon $T$ in the future response vector $\mathbf{y}_t^T$ and the threshold $A$ in the termination criterion. The selection of maximum lags is not critical and can be arbitrarily large at the cost of excessive computations. A rule of thump is to have a small number of lags for maps (discontinuous series of observations), and a larger number of lags for flows (smoothly changing observations), which for oscillating time series should cover one or more oscillation periods \cite{Kugiumtzis96}. The time horizon $T$ is also dependent on the underlying dynamics. Nevertheless $T\! = \!1$ is widely used in works on linear and nonlinear causality measures, but we have argued that $T>1$ may be more appropriate in cases of densely sampled time series \cite{Vlachos10}.

The threshold $A$ is the only inherent parameter of PMIME. For
MIME, it was found after a simulation study that $A\! = \!0.95$ is
an appropriate choice to avoid false positives, i.e. components of
$X$ entering $\mathbf{w}_t$ in the absence of coupling. We extend
this study here and compare the fixed threshold $A$ to an adjusted
threshold for the significance of
$I(\mathbf{y}_t^T;w_t^j|\mathbf{w}_t^{j-1})$, the CMI for the
selected component $w_t^j$ at the embedding cycle $j$. As the null
distribution for the null hypothesis H$_0$:
$I(\mathbf{y}_t^T;w_t^j|\mathbf{w}_t^{j-1})\! = \!0$, is not
known, we form it empirically by shuffling randomly the components
of the vector $w_t^j$ and the rows of the matrix
$\mathbf{w}_t^{j-1}$. This random shuffling scheme aims at
obtaining the most independent joint distribution that gives
largest bias in the estimation of CMI, setting higher significance
threshold and thus making the termination criterion more
stringent. Then if the original
$I(\mathbf{y}_t^T;w_t^j|\mathbf{w}_t^{j-1})$ is larger than the
$(1-\alpha)\%$ percentile of the ensemble of the randomized
$I(\mathbf{y}_t^T;w_t^j|\mathbf{w}_t^{j-1})$, we accept $w_t^j$ as
significant and proceed to the next embedding cycle, otherwise the
mixed embedding scheme terminates and $\mathbf{w}_t\! =
\!\mathbf{w}_t^{j-1}$.

We found that the adjusted threshold criterion is more adaptive
than the fixed threshold to system complexity, time series length
and noise level. For illustration, we consider the system of $K$
coupled H\'{e}non maps, defined as
\begin{equation}
\hspace{-5mm}
 \begin{array}{ll}
x_{i,t} = 1.4-x_{i,t-1}^2+0.3x_{i,t-2}, & \mbox{for} \,\, i=1,K  \\
x_{i,t} =
1.4-0.5C(x_{i-1,t-1}+x_{i+1,t-1})+(1-C)x_{i,t-1}^2+0.3x_{i,t-2}, &
\mbox{for} \,\, j=2,\ldots,K-1
 \end{array}
\label{eq:Henon}
\end{equation}
where $C$ is the coupling strength. For the example of $K\! =
\!3$, it is shown in Table~\ref{tab:threshold} that for weak
coupling ($C\! = \!0.1$), $A\! = \!0.95$ is too conservative and a
larger $A$, such as 0.97 or even better 0.99, is needed to include
components of the driving variable in the mixed embedding vector
for the two true direct couplings.
\begin{table}[h!]
\setlength{\tabcolsep}{3mm}
\centerline{\begin{tabular}{lccc} \hline
 & $X_1\! \rightarrow \! X_2$ & $X_3\! \rightarrow \! X_2$ & $X_2\! \rightarrow \! X_1$ \\ \hline
\multicolumn{4}{c}{noise-free} \\ \hline
$A\! = \!0.95$ & 1 & 2 & 0\\
$A\! = \!0.97$ & 41 & 25 & 0\\
$A\! = \!0.99$ & 72 & 70 & 3\\
$\alpha\! = \!0.01$ & 21 & 11 & 0\\
$\alpha\! = \!0.05$ & 51 & 35 & 0\\
$\alpha\! = \!0.1$ & 59 & 47 & 0\\ \hline
\multicolumn{4}{c}{20$\%$ noise} \\ \hline
$A\! = \!0.95$ & 22 & 11 & 0\\
$A\! = \!0.97$ & 59 & 46 & 5\\
$A\! = \!0.99$ & 79 & 92 & 36\\
$\alpha\! = \!0.01$ & 20 & 4 & 0\\
$\alpha\! = \!0.05$ & 48 & 37 & 3\\
$\alpha\! = \!0.1$ & 63 & 57 & 8 \\ \hline
\end{tabular}}
\caption{
Number of times PMIME is positive for 100 realizations of three coupled H\'{e}non maps ($C\! = \!0.1$) with true coupling $X_1\! \rightarrow \! X_2$ and $X_3\! \rightarrow \! X_2$, and false coupling $X_2\! \rightarrow \! X_1$. The parameters are $n\! = \!512$, $L\! = \!5$, $T\! = \!1$, and the stopping criterion is set by a fixed threshold $A$ and an adjusted threshold determined by $\alpha$.}
\label{tab:threshold}
\end{table}
However, in the presence of noise (observational Gaussian white
noise with standard deviation (SD) 20\% of the data SD), a larger
$A$ allows for components of non-driving variables entering the
mixed embedding vector, giving small false direct couplings. The
choice of $A$ should balance these two effects and it seems that
in practice a fixed threshold cannot be optimized. On the other
hand, the adjusted threshold seems to work well for both
noise-free and noisy time series, and the choice of $\alpha\! =
\!0.05$ balances well sensitivity, i.e. probability of having
positive PMIME for true direct couplings, and specificity, i.e.
probability of having zero PMIME when there is no direct coupling.

\section{Simulation Study}
\label{sec:simulations}

Next we compare PMIME (with the adjusted threshold at $\alpha\! =
\!0.05$) to the conditional Granger causality index (CGCI)
\citep{Geweke84}, and the partial transfer entropy (PTE)
\citep{Vakorin09,Papana12}, respectively. We report the best
obtained results for CGCI and PTE optimizing the parameter $m$ for
the model order in CGCI and the embedding dimension in PTE.
To assess statistically the sensitivity and specificity of the
measures we compute the measures on 100 realizations from each
system. PMIME is considered significant if it is positive, whereas
the significance of CGCI and PTE is determined by the surrogate
data test (for the null hypothesis of no coupling) using
time-shifted surrogates at a significance level $\alpha=0.05$
\cite{QuianQuiroga02b}.

\begin{figure}[htb]
\centerline{\hbox{\includegraphics[width=5cm]{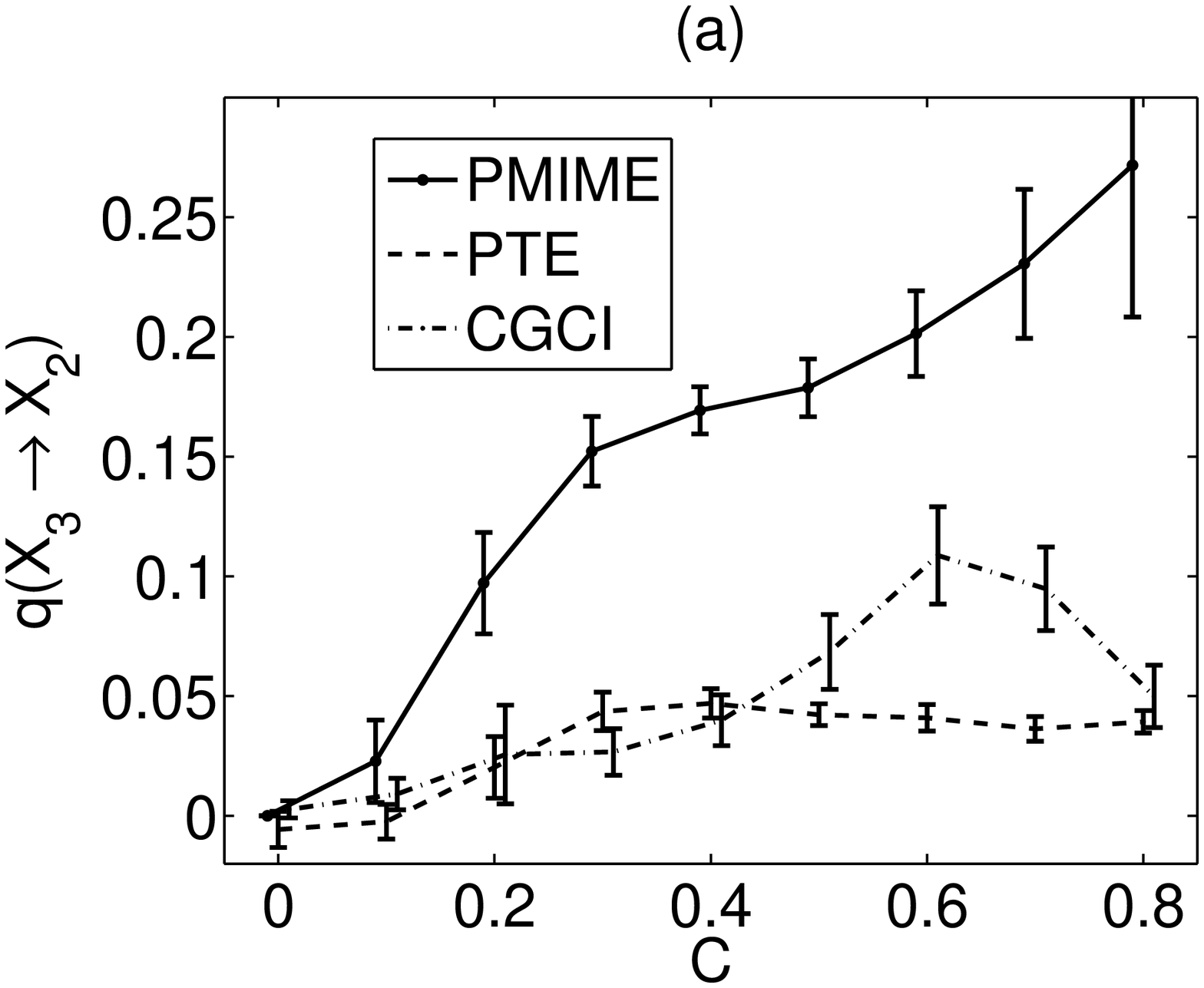}
\includegraphics[width=5cm]{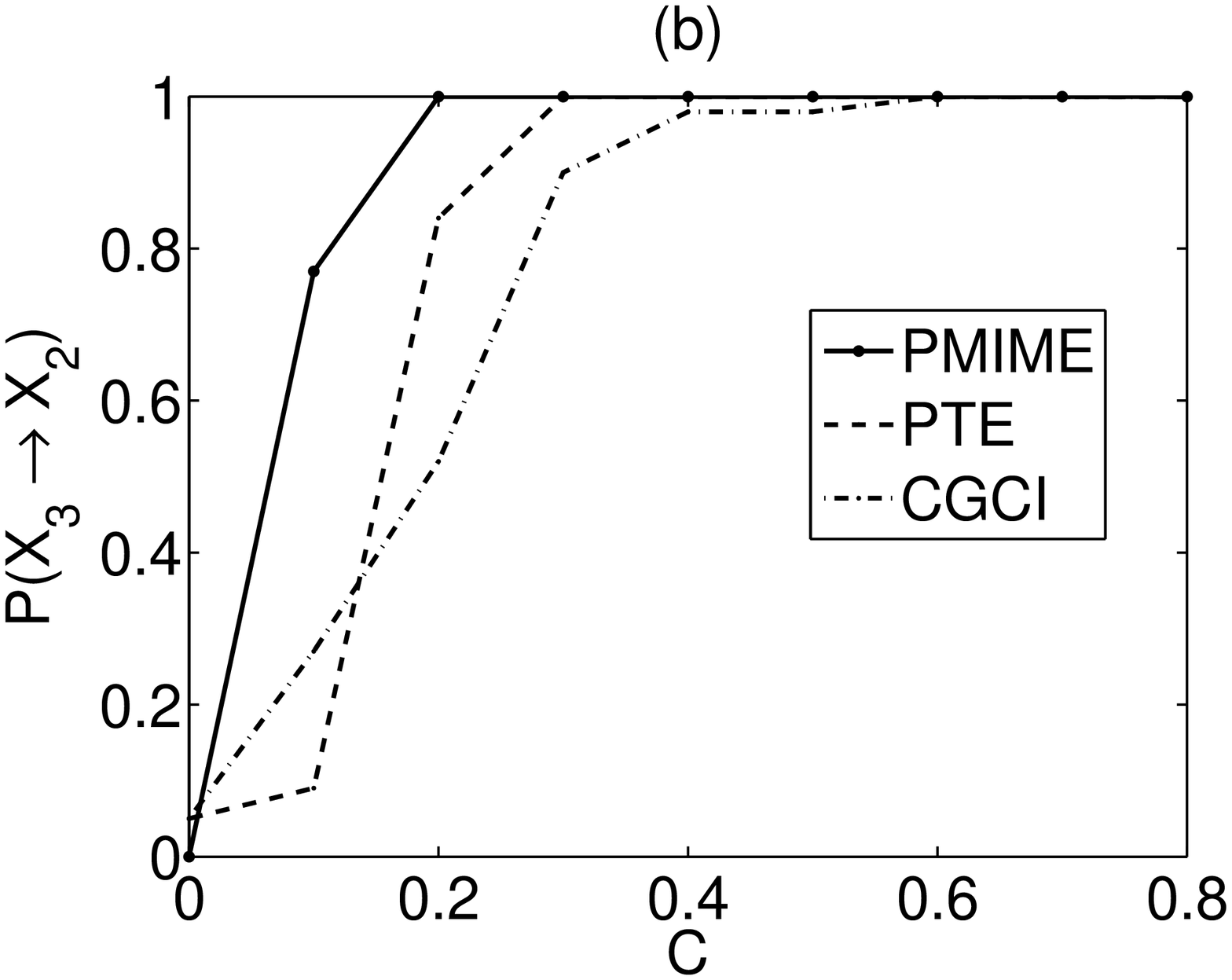}}}
\centerline{\hbox{\includegraphics[width=5cm]{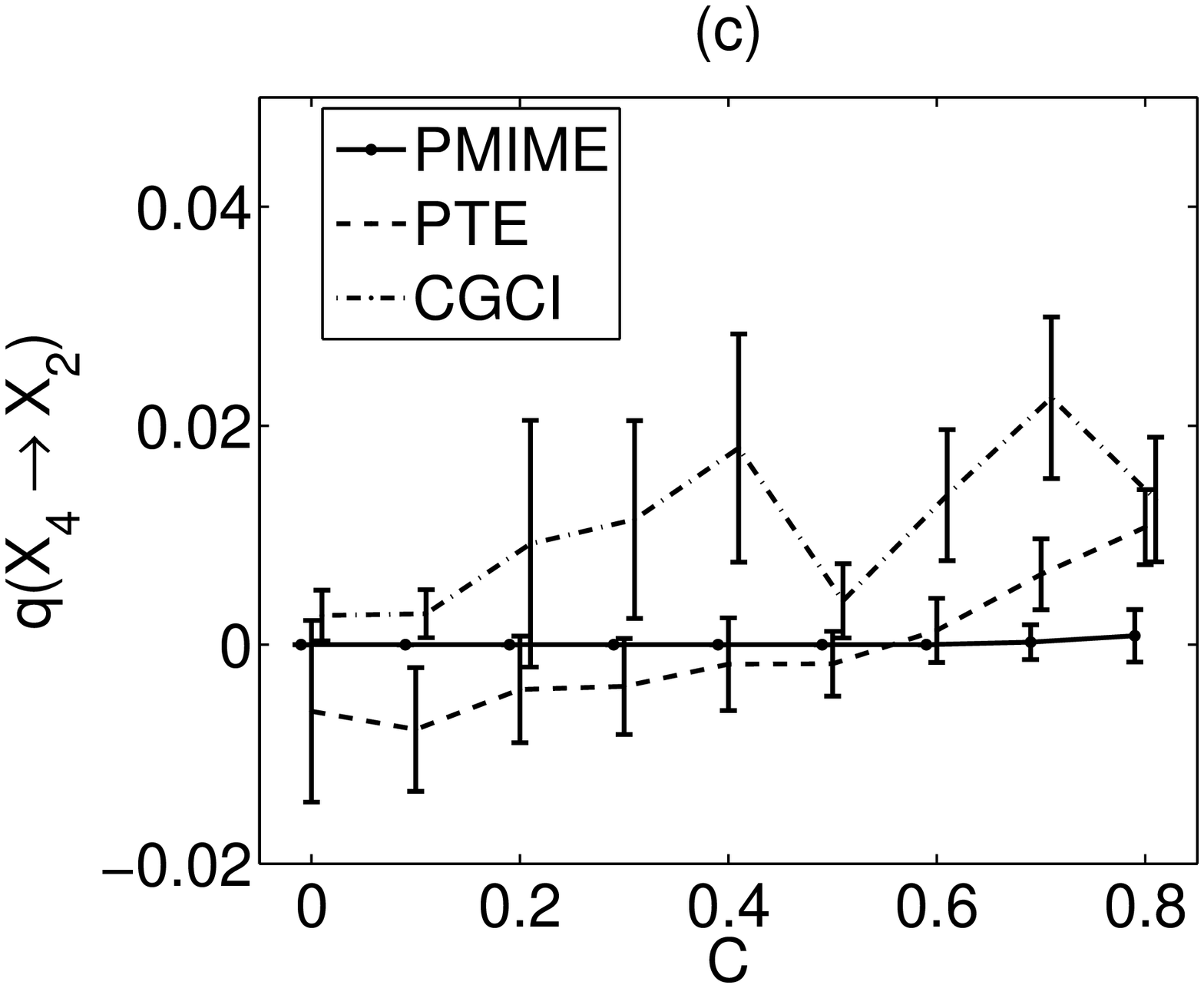}
\includegraphics[width=5cm]{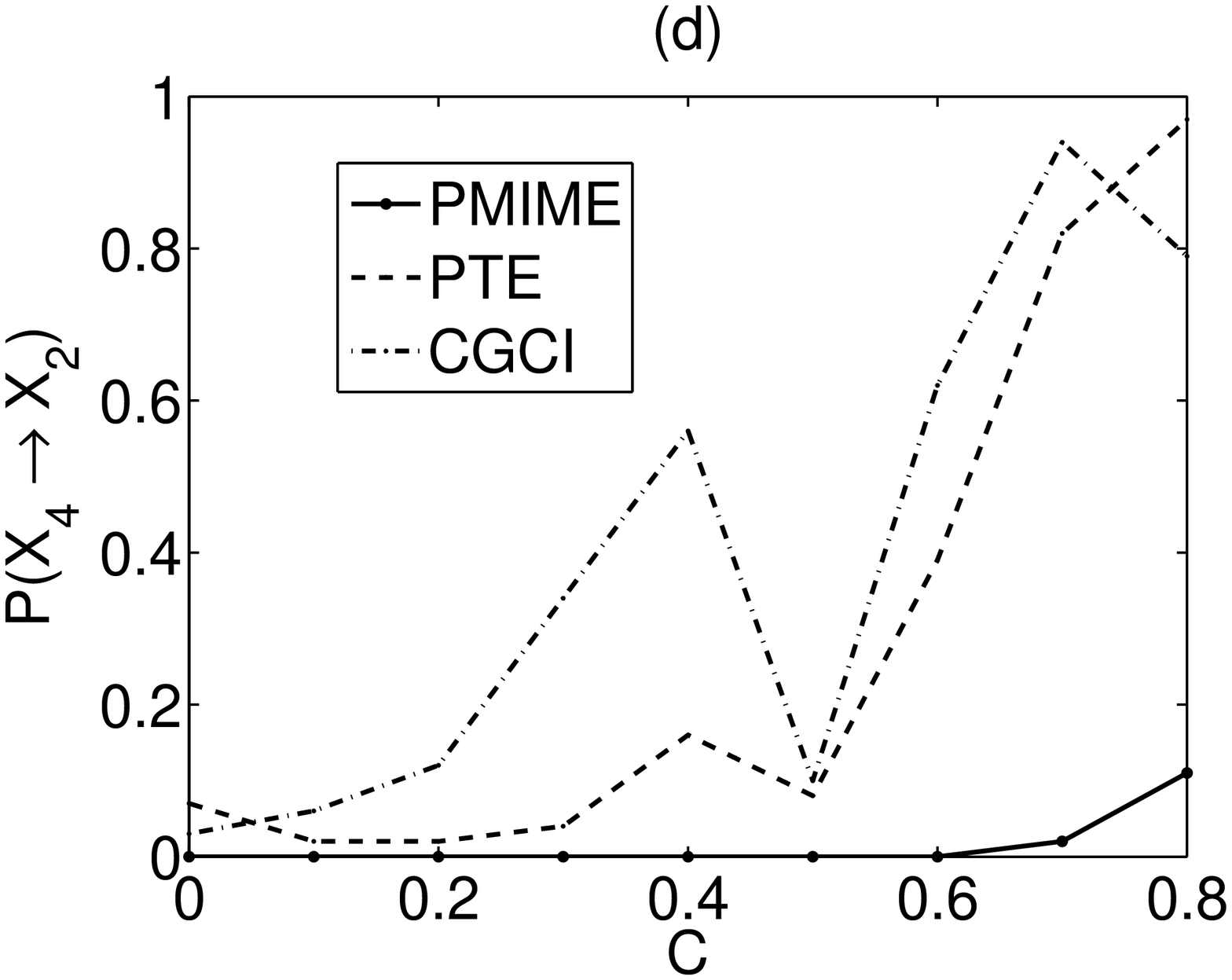}}}
\caption{Estimated coupling $q$, where $q$ is any of PMIME, PTE and CGCI given in the legend, as a function of coupling strength $C$ for the system of $K\! = \!5$ coupled H\'{e}non maps: (a) $X_3 \! \rightarrow \! X_2$, and (c) $X_4 \! \rightarrow \! X_2$. The parameters are $n\! = \!1024$, $T\! = \!1$ for all measures, $m\! = \!2$ for PTE and CGCI. Each error bar denotes the mean and SD over 100 realizations, and the error bars are slightly displaced horizontally for better visualization. The panels in (b) and (d) show the estimated probability $P$ of detecting the coupling in (a) and (c), respectively, which is the relative frequency PTE and CGCI found significant at the significance level $\alpha\! = \!0.05$ using 100 time-shifted surrogates, or the relative frequency of PMIME being positive.}
\label{fig:HenonVaryC}
\end{figure}

Before we show detailed results on a number of linear stochastic,
nonlinear stochastic and chaotic systems, we demonstrate the
superiority of PMIME in terms of sensitivity and specificity on
the system of $K$ coupled H\'{e}non maps. As shown for $K\! = \!5$
in Fig.~\ref{fig:HenonVaryC}a, for the true direct coupling $X_3\!
\rightarrow \! X_2$ PMIME increases more than the other measures
with the coupling strength $C$ and up to $C\! = \!0.8$. The larger
increase of PMIME with $C$, particularly for small $C$, is
justified by the statistical significance of the measures
(Fig.~\ref{fig:HenonVaryC}b). On the other hand, for the indirect
coupling $X_4\! \rightarrow \! X_2$, PMIME is zero for all $C$ (a
slight deviation is observed only for very large $C$), whereas PTE
increases slowly with $C$ and CGCI fluctuates at some positive
level (Fig.~\ref{fig:HenonVaryC}c), both tending to be more
significant with the increase of $C$ (Fig.~\ref{fig:HenonVaryC}d).

\begin{table}[htb]
\setlength{\tabcolsep}{2mm}
\centerline{\begin{tabular}{p{25mm}p{25mm}p{25mm}p{25mm}} \hline
 & $K\! = \!5$ & $K\! = \!15$ & $K\! = \!25$\\ \hline
\multicolumn{4}{c}{$X_{(K-1)/2} \! \rightarrow \! X_{(K-1)/2+1}$} \\ \hline
PMIME & 0.105(1.00) & 0.063(0.92) & 0.061(0.79)\\
PTE & 0.021(0.89) & 0.001(0.15) & 0.000(0.10)\\
CGCI & 0.040(0.85) & 0.188(0.68) & 0.230(0.67)\\
\hline
\multicolumn{4}{c}{$X_{(K-1)/2} \! \rightarrow \! X_{(K-1)/2+2}$} \\ \hline
PMIME & 0.000(0.00) & 0.000(0.01) & 0.001(0.02)\\
PTE & -0.005(0.04) & 0.001(0.02) & 0.000(0.07)\\
CGCI & 0.009(0.15) & 0.064(0.44) & 0.136(0.45)\\
\hline
\end{tabular}}
\caption{Mean of coupling measure and relative frequency for its statistical significance in parentheses from 100 realizations of coupled H\'{e}non maps ($C\! = \!0.2$) with varying number of variables $K$. Results are shown for the true direct coupling (first block) and the indirect coupling (second block) of the variables in the center part of the chain of the $K$ variables. The parameter setup is as for Fig.~\ref{fig:HenonVaryC}.}
\label{tab:HenonVaryK}
\end{table}
A challenging situation is when the number of variables $K$ increases.
We observed that even for the optimal $m$, PTE looses significance in detecting the true direct coupling, and CGCI tends to falsely detect direct coupling, whereas PMIME attains both high sensitivity and specificity, decreasing rather slowly with the increase of $K$. These features get more pronounced for the most interacting variables and as $K$ gets large, as shown in Table~\ref{tab:HenonVaryK} for the variables in the middle of the chain of the coupled H\'{e}non maps. We note that regardless of $K$ the mixed embedding vector for PMIME contains always few components, one (more seldom two) of which are from the driving variable in the presence of causal effect.

In the following, further results for the performance of PMIME and
comparison to PTE and CGCI are presented for multivariate time
series from different discrete and continuous systems and for
different time series lengths and levels of noise added to the
time series.

\subsection{Linear multivariate stochastic process - 1}

The first system is a linear vector autoregressive process of order 5 in 4 variables, VAR$_4(5)$ (model 1 in \cite{Winterhalder05})
\begin{equation*}
    \begin{array}{l}
    x_{1,t} = 0.8x_{1,t-1} + 0.65x_{2,t-4} + e_{1,t} \\
    x_{2,t} = 0.6x_{2,t-1} + 0.6x_{4,t-5} +e_{2,t} \\
    x_{3,t} = 0.5x_{3,t-3} - 0.6x_{1,t-1} + 0.4x_{2,t-4} + e_{3,t} \\
    x_{4,t} = 1.2x_{4,t-1} - 0.7x_{4,t-2} +e_{4,t}
    \end{array}
\end{equation*}
where $e_{i,t}$, $i=1,\ldots,4$, are white noise components having zero mean and unit covariance matrix. The true direct causality connections are $X_1 \rightarrow X_3$, $X_2 \rightarrow X_1$, $X_2 \rightarrow X_3$, and $X_4 \rightarrow X_2$.

For all discrete systems we use $T=1$, and for PMIME $L=5$, which here matches the larger lag in the process. For PTE and CGCI we vary the embedding dimension and model order, respectively, $m=2,\ldots,5$, in order to investigate for the best $m$ and show also the dependence of their performance on the parameter $m$. The results from 100 Monte Carlo realizations of the system VAR$_4(5)$ are shown in Table~\ref{tab:VAR_4(5)n512}.
\begin{table}[htb]
{\small
\setlength{\tabcolsep}{1mm}
\centerline{\begin{tabular}{lccccccc} \hline
 & PMIME & PTE($m=2$) & PTE($m=3$) & PTE($m=4$) & PTE($m=5$) & CGCI($m=2$) & CGCI($m=5$) \\ \hline
\fbox{$X_2 \rightarrow X_1$}  & 0.348(1.00)  & 0.034(0.53)  & 0.073(1.00)  & 0.116(1.00)  & 0.115(1.00)  & 0.175(1.00)  & 0.750(1.00)  \\
\fbox{$X_1 \rightarrow X_3$}  & 0.610(1.00)  & 0.160(1.00)  & 0.158(1.00)  & 0.134(1.00)  & 0.093(1.00)  & 0.608(1.00)  & 0.587(1.00)  \\
\fbox{$X_2 \rightarrow X_3$} & 0.073(0.99)  & 0.007(0.07)  & 0.013(0.25)  & 0.015(0.18)  & 0.015(0.09)  & 0.046(1.00)  & 0.361(1.00)  \\
\fbox{$X_4 \rightarrow X_2$}  & 0.487(1.00)  & 0.048(0.78)  & 0.049(0.80)  & 0.081(0.99)  & 0.126(1.00)  & 0.089(1.00)  & 0.622(1.00)  \\
$X_1 \rightarrow X_2$  & 0.002(0.09)  & 0.010(0.03)  & 0.013(0.11)  & 0.013(0.11)  & 0.013(0.14)  & 0.007(0.13)  & 0.010(0.08)  \\
$X_3 \rightarrow X_1$  & 0.000(0.02)  & 0.015(0.06)  & 0.012(0.07)  & 0.010(0.07)  & 0.011(0.05)  & 0.011(0.28)  & 0.010(0.07)  \\
$X_1 \rightarrow X_4$  & 0.003(0.11)  & 0.007(0.01)  & 0.012(0.05)  & 0.015(0.11)  & 0.017(0.20)  & 0.004(0.05)  & 0.010(0.03)  \\
$X_4 \rightarrow X_1$  & 0.000(0.03)  & 0.012(0.03)  & 0.013(0.02)  & 0.012(0.02)  & 0.013(0.03)  & 0.008(0.23)  & 0.010(0.07)  \\
$X_3 \rightarrow X_2$  & 0.001(0.03)  & 0.009(0.05)  & 0.013(0.10)  & 0.013(0.09)  & 0.014(0.06)  & 0.006(0.07)  & 0.010(0.07)  \\
$X_2 \rightarrow X_4$  & 0.002(0.04)  & 0.010(0.04)  & 0.013(0.05)  & 0.018(0.06)  & 0.021(0.09)  & 0.004(0.03)  & 0.010(0.06)  \\
$X_3 \rightarrow X_4$  & 0.001(0.04)  & 0.005(0.04)  & 0.011(0.08)  & 0.015(0.10)  & 0.019(0.10)  & 0.004(0.03)  & 0.010(0.03)  \\
$X_4 \rightarrow X_3$  & 0.000(0.00)  & 0.001(0.07)  & 0.004(0.01)  & 0.008(0.07)  & 0.011(0.06)  & 0.003(0.10)  & 0.010(0.03)  \\
 \hline
\end{tabular}}
\caption{Mean of coupling measure and relative frequency for its statistical significance in parentheses from 100 realizations of the system VAR$_4(5)$ and $n=512$. For PTE the results are shown for $m=2,\ldots,5$ and for CGCI for $m=2,5$. The true direct couplings are shown in a frame box.}
\label{tab:VAR_4(5)n512}
}
\end{table}
PMIME is high and always positive for the four direct couplings and essentially zero for the other couplings. The largest frequency of false positive PMIME is for $X_1 \rightarrow X_4$ (11 in 100 realizations), but still the PMIME values are very small (the mean is 0.003). Regarding the true direct couplings, the weakest causal effect is estimated by PMIME for $X_2 \rightarrow X_3$ (mean 0.073), but still PMIME is positive almost always (99 in 100 realizations). This true direct coupling cannot be estimated by PTE for any $m$, and the best rejection rate of H$_0$ of no causal effect is for $m=3$ (25 rejections in 100 significance randomization tests using time-shifted surrogates). However, the selection $m=3$ is not appropriate for $X_4 \rightarrow X_2$, as it gives only 80 rejections, which is much less than the highest rejection rate of 100\% obtained by PMIME and CGCI, and also by PTE for $m=5$. This example demonstrates how PMIME resolves the ambiguity in the selection of the appropriate embedding for PTE. The selection of a suitable order $m$ may be an issue also for the linear measure CGCI, as a small $m$ does not give good specificity (for two non-existing direct couplings the rejection rate is 23\% and 28\%) and sensitivity (though the power of the test is 1.0 for all four true direct couplings, the mean CGCI is much smaller for $m=2$ than for $m=5$ in three of the four couplings).

\subsection{Linear multivariate stochastic process - 2}

The second linear VAR process is of order 4 in 5 variables, VAR$_5(4)$ (model 1 in \cite{Schelter06e})
\begin{equation*}
    \begin{array}{l}
    x_{1,t} = 0.4x_{1,t-1} - 0.5x_{1,t-2} +0.4x_{5,t-1}+e_{1,t} \\
    x_{2,t} = 0.4x_{2,t-1} - 0.3x_{1,t-4} + 0.4x_{5,t-2}+e_{2,t} \\
    x_{3,t} = 0.5x_{3,t-1} - 0.7x_{3,t-2} - 0.3x_{5,t-3} + e_{3,t} \\
    x_{4,t} = 0.8x_{4,t-3} + 0.4x_{1,t-2} + 0.3x_{2,t-2} +e_{4,t} \\
    x_{5,t} = 0.7x_{5,t-1} - 0.5x_{5,t-2} - 0.4x_{4,t-1} + e_{5,t}
    \end{array}
\end{equation*}
The simulation setup is the same as for the first linear system, and the results are shown in Table~\ref{tab:VAR_5(4)n512}.
\begin{table}[htb]
{\small
\setlength{\tabcolsep}{1mm}
\centerline{\begin{tabular}{lccccccc} \hline
 & PMIME & PTE($m=2$) & PTE($m=3$) & PTE($m=4$) & PTE($m=5$) & CGCI($m=2$) & CGCI($m=5$) \\ \hline
\fbox{$X_1 \rightarrow X_2$}  & 0.224(1.00)  & 0.012(0.26)  & 0.010(0.22)  & 0.019(0.69)  & 0.017(0.61)  & 0.023(0.62)  & 0.108(1.00)  \\
\fbox{$X_1 \rightarrow X_4$}  & 0.196(1.00)  & 0.022(0.50)  & 0.016(0.56)  & 0.012(0.49)  & 0.010(0.27)  & 0.101(1.00)  & 0.208(1.00)  \\
\fbox{$X_5 \rightarrow X_1$}  & 0.411(1.00)  & 0.061(0.99)  & 0.044(1.00)  & 0.037(0.99)  & 0.032(0.96)  & 0.243(1.00)  & 0.225(1.00)  \\
\fbox{$X_2 \rightarrow X_4$}  & 0.110(0.95)  & 0.008(0.08)  & 0.009(0.24)  & 0.005(0.16)  & 0.005(0.11)  & 0.033(0.87)  & 0.119(1.00)  \\
\fbox{$X_5 \rightarrow X_2$}  & 0.400(1.00)  & 0.052(1.00)  & 0.036(0.95)  & 0.030(0.95)  & 0.026(0.85)  & 0.185(1.00)  & 0.194(1.00)  \\
\fbox{$X_5 \rightarrow X_3$}  & 0.171(1.00)  & 0.002(0.00)  & 0.011(0.39)  & 0.011(0.30)  & 0.009(0.26)  & 0.031(0.86)  & 0.092(1.00)  \\
\fbox{$X_4 \rightarrow X_5$}  & 0.494(1.00)  & 0.070(1.00)  & 0.048(1.00)  & 0.039(0.97)  & 0.032(0.93)  & 0.405(1.00)  & 0.320(1.00)  \\
$X_2 \rightarrow X_1$  & 0.017(0.22)  & 0.001(0.05)  & -0.001(0.05)  & -0.000(0.02)  & 0.000(0.04)  & 0.004(0.04)  & 0.010(0.07)  \\
$X_1 \rightarrow X_3$  & 0.007(0.13)  & 0.002(0.05)  & -0.001(0.04)  & -0.000(0.02)  & 0.000(0.03)  & 0.012(0.24)  & 0.010(0.04)  \\
$X_3 \rightarrow X_1$  & 0.013(0.13)  & -0.001(0.05)  & 0.000(0.03)  & -0.001(0.07)  & -0.000(0.05)  & 0.004(0.03)  & 0.008(0.04)  \\
$X_4 \rightarrow X_1$  & 0.009(0.12)  & -0.002(0.05)  & -0.002(0.04)  & -0.001(0.05)  & -0.000(0.03)  & 0.004(0.02)  & 0.010(0.08)  \\
$X_1 \rightarrow X_5$  & 0.014(0.16)  & -0.000(0.03)  & -0.001(0.06)  & 0.001(0.03)  & 0.002(0.07)  & 0.004(0.08)  & 0.010(0.04)  \\
$X_2 \rightarrow X_3$  & 0.004(0.10)  & 0.003(0.08)  & -0.001(0.05)  & 0.001(0.02)  & 0.000(0.05)  & 0.016(0.42)  & 0.010(0.03)  \\
$X_3 \rightarrow X_2$  & 0.009(0.13)  & -0.001(0.06)  & 0.001(0.08)  & 0.000(0.01)  & 0.000(0.05)  & 0.007(0.14)  & 0.010(0.04)  \\
$X_4 \rightarrow X_2$  & 0.009(0.15)  & -0.001(0.06)  & 0.000(0.02)  & 0.000(0.05)  & 0.001(0.03)  & 0.006(0.11)  & 0.010(0.02)  \\
$X_2 \rightarrow X_5$  & 0.009(0.12)  & 0.001(0.06)  & 0.000(0.05)  & 0.003(0.05)  & 0.002(0.01)  & 0.004(0.04)  & 0.011(0.06)  \\
$X_3 \rightarrow X_4$  & 0.004(0.12)  & 0.000(0.04)  & -0.002(0.05)  & -0.002(0.04)  & -0.000(0.03)  & 0.008(0.37)  & 0.010(0.03)  \\
$X_4 \rightarrow X_3$  & 0.005(0.07)  & 0.001(0.05)  & -0.001(0.05)  & -0.001(0.03)  & -0.001(0.04)  & 0.020(0.50)  & 0.009(0.06)  \\
$X_3 \rightarrow X_5$  & 0.013(0.17)  & -0.001(0.04)  & -0.001(0.05)  & 0.002(0.05)  & 0.001(0.02)  & 0.004(0.06)  & 0.011(0.01)  \\
$X_5 \rightarrow X_4$  & 0.007(0.15)  & 0.033(0.81)  & 0.004(0.10)  & 0.002(0.07)  & 0.004(0.11)  & 0.131(1.00)  & 0.011(0.05)  \\
 \hline
\end{tabular}}
\caption{Mean of coupling measure and relative frequency for its statistical significance in parentheses from 100 realizations of the system VAR$_5(4)$ and $n=512$. For PTE the results are shown for $m=2,\ldots,5$ and for CGCI for $m=2,5$. The true direct couplings are shown in a frame box.}
\label{tab:VAR_5(4)n512}
}
\end{table}
The results for VAR$_5(4)$ are similar to these for the
VAR$_4(5)$. This example was included to show that for small time
series from stochastic systems PMIME may be falsely positive at a
rate higher than the nominal rate of $\alpha=0.05$. Here, the
highest false positive rate was 22\% for $X_2 \rightarrow X_1$,
but still PMIME was very small (in average one tenth of the
weakest true direct coupling). On the other hand, PTE does not
have overall high sensitivity and moreover there is no optimal
$m$, e.g. for $X_2 \rightarrow X_4$ the rejection rate is very
small with highest rate being 24\% for $m=3$, while for $X_1
\rightarrow X_2$ the highest rejection rate is 69\% for $m=4$.
CGCI is again smaller for $m=2$ and the significance test has
large size (higher false rejection rate than the nominal type I
error of $\alpha=0.05$) and smaller power, which all improve with
the increase of $m$.

Note that for the two linear stochastic processes, CGCI is the most appropriate measure of causality, but PMIME is comparable to CGCI.

\subsection{Nonlinear multivariate stochastic process}

Next we consider the nonlinear VAR process of order 1 in 3 variables, NLVAR$_3(1)$ (model 7 in \cite{Gourevitch06})
\begin{equation*}
    \begin{array}{l}
    x_{1,t} = 3.4x_{1,t-1}(1-x_{1,t-1}^2)\mbox{e}^{-x_{1,t-1}^2} +0.4e_{1,t} \\
    x_{2,t} = 3.4x_{2,t-1}(1-x_{2,t-1}^2)\mbox{e}^{-x_{2,t-1}^2} +0.5x_{1,t-1}x_{2,t-1} +0.4e_{2,t} \\
    x_{3,t} = 3.4x_{3,t-1}(1-x_{3,t-1}^2)\mbox{e}^{-x_{3,t-1}^2} +0.3x_{2,t-1} +0.5x_{1,t-1}^2 + 0.4e_{3,t}
    \end{array}
\end{equation*}
The simulation setup is the same as for the previous systems, and the results are shown in Table~\ref{tab:NLVAR_3(1)n512}.
\begin{table}[htb]
{\small
\setlength{\tabcolsep}{1mm}
\centerline{\begin{tabular}{lccccccc} \hline
 & PMIME & PTE($m=2$) & PTE($m=3$) & PTE($m=4$) & PTE($m=5$) & CGCI($m=2$) & CGCI($m=5$) \\ \hline
\fbox{$X_1 \rightarrow X_2$}  & 0.272(1.00)  & 0.056(0.98)  & 0.027(0.53)  & 0.020(0.39)  & 0.016(0.19)  & 0.006(0.08)  & 0.013(0.09)  \\
\fbox{$X_1 \rightarrow X_3$}  & 0.226(1.00)  & 0.045(0.86)  & 0.022(0.45)  & 0.016(0.26)  & 0.011(0.13)  & 0.005(0.07)  & 0.010(0.03)  \\
\fbox{$X_2 \rightarrow X_3$}  & 0.171(0.97)  & 0.033(0.71)  & 0.018(0.34)  & 0.015(0.27)  & 0.014(0.21)  & 0.090(1.00)  & 0.096(1.00)  \\
$X_2 \rightarrow X_1$  & 0.006(0.08)  & -0.006(0.04)  & -0.005(0.04)  & -0.004(0.04)  & -0.002(0.04)  & 0.004(0.04)  & 0.010(0.08)  \\
$X_3 \rightarrow X_1$  & 0.009(0.15)  & -0.008(0.01)  & -0.005(0.05)  & -0.003(0.04)  & -0.002(0.09)  & 0.004(0.00)  & 0.009(0.02)  \\
$X_3 \rightarrow X_2$  & 0.004(0.07)  & -0.009(0.04)  & -0.007(0.06)  & -0.002(0.05)  & 0.000(0.07)  & 0.004(0.07)  & 0.010(0.06)  \\
 \hline
\end{tabular}}
\caption{Mean of coupling measure and relative frequency for its statistical significance in parentheses from 100 realizations of the system NLVAR$_3(1)$ and $n=512$. For PTE the results are shown for $m=2,\ldots,5$ and for CGCI for $m=2,5$. The true direct couplings are shown in a frame box.}
\label{tab:NLVAR_3(1)n512}
}
\end{table}
Here again PMIME attains the best possible sensitivity (only for
one true direct coupling there are three zero PMIME values) and
good specificity (only for one false coupling the rate of positive
PMIME values is well above the level of 5\%, being 15\%, and again
the mean PMIME is almost two orders of magnitude smaller than for
the true direct couplings). The largest lag in the process is one
and therefore PTE looses sensitivity with the increase of $m$. For
example, for $X_2 \rightarrow X_3$ the rejection rate is 71\%
(already not high enough) for $m=2$, and drops with $m$ down to
21\% for $m=5$. As expected, CGCI has very small sensitivity
regardless of $m$, and can only identify one true direct coupling,
the linear one $X_2 \rightarrow X_3$.

\subsection{Coupled H\'{e}non maps}

The system of $K$ coupled chaotic H\'{e}non maps was previously
defined in eq.(\ref{eq:Henon}). For $K \geq 3$, complete
synchronization is not observed for any pair of variables as $C$
increases, but the time series of the driven variables explode for
$C>1.0$ dependent on $K$, so $C$ is studied in the range
$[0,0.8]$. Again 100 realizations for each system scenario are
generated, for PTE and CGCI the free parameter is $m=2$ (embedding
dimension and model order, respectively), for PMIME the standard
parameter for maps $L=5$ is used, and for all measures the time
ahead is $T=1$. Note that for this system, the choice $L=5$ is not
suitable, as only lags up to 2 are present in the difference
equations, whereas for PTE the parameter $m$ is optimal.

\begin{figure}
\centerline{\hbox{\includegraphics[width=5cm]{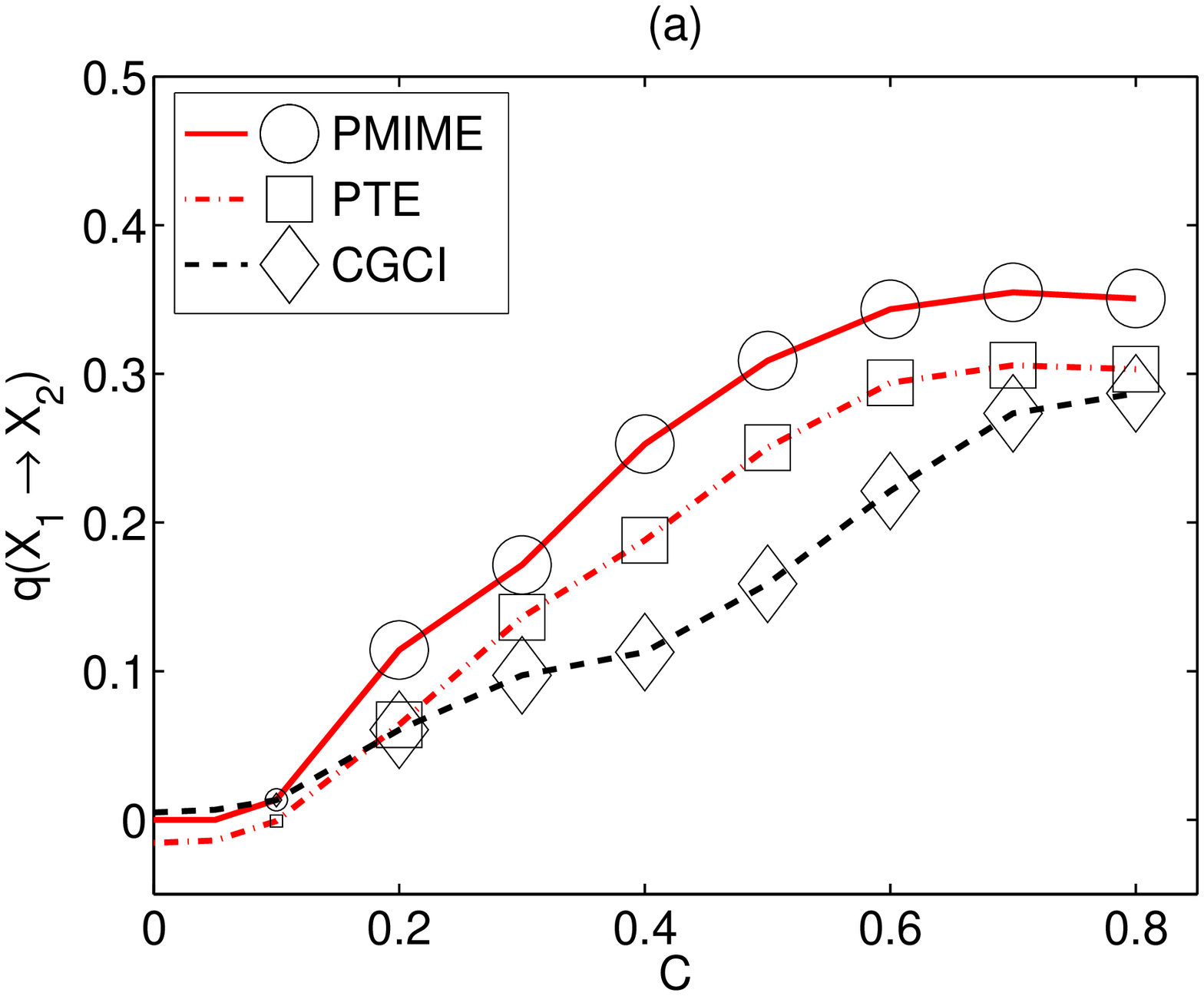}
\includegraphics[width=5cm]{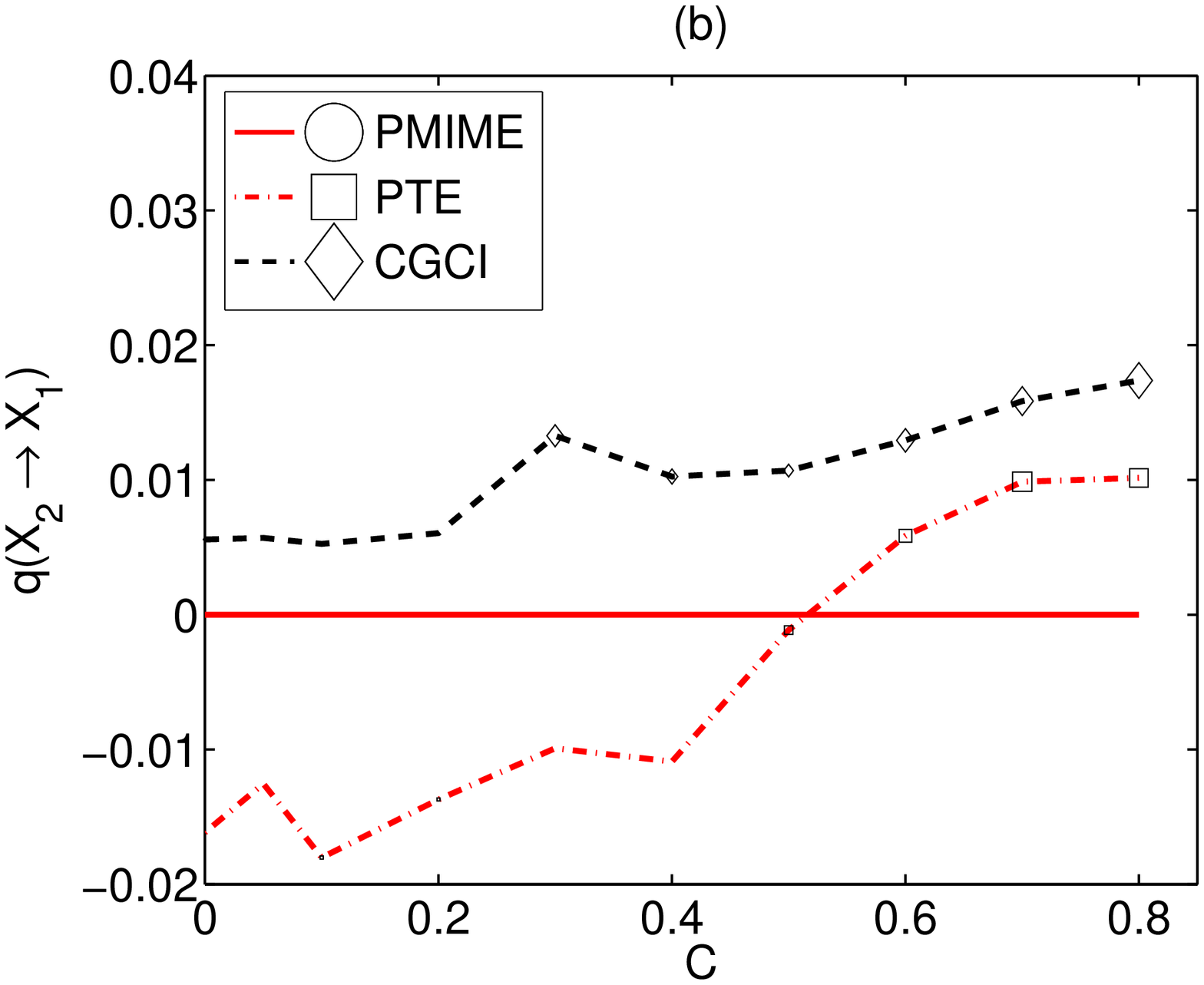}
\includegraphics[width=5cm]{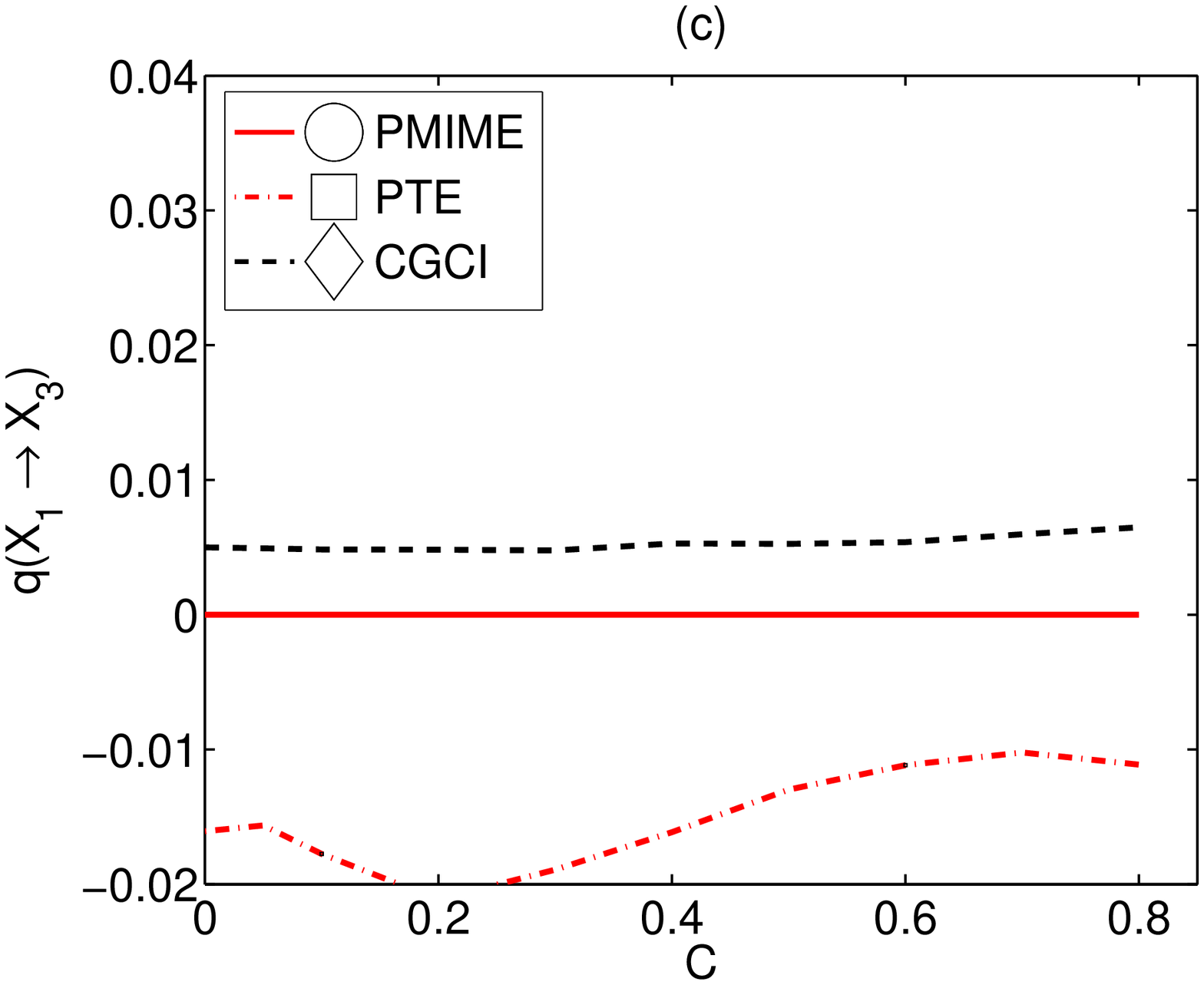}}}
\caption{PMIME, PTE and CGCI measures (denoted collectively $q$) are given as functions of the coupling strength $C$ for the true direct causality $X_1 \rightarrow X_2$ in (a), the non-existing couplings $X_2 \rightarrow X_1$ in (b) and $X_1 \rightarrow X_3$ in (c) for the coupled H\'{e}non maps of $K=3$ and for time series length $n=512$. The number of rejections in 100 realizations of the randomization test determines the size of a symbol displayed for each measure and $C$, where in the legend the size of the symbols regards 100 rejections.}
\label{fig:HenonK3diffCnoisefree}       
\end{figure}

\begin{figure}
\centerline{\hbox{\includegraphics[width=5cm]{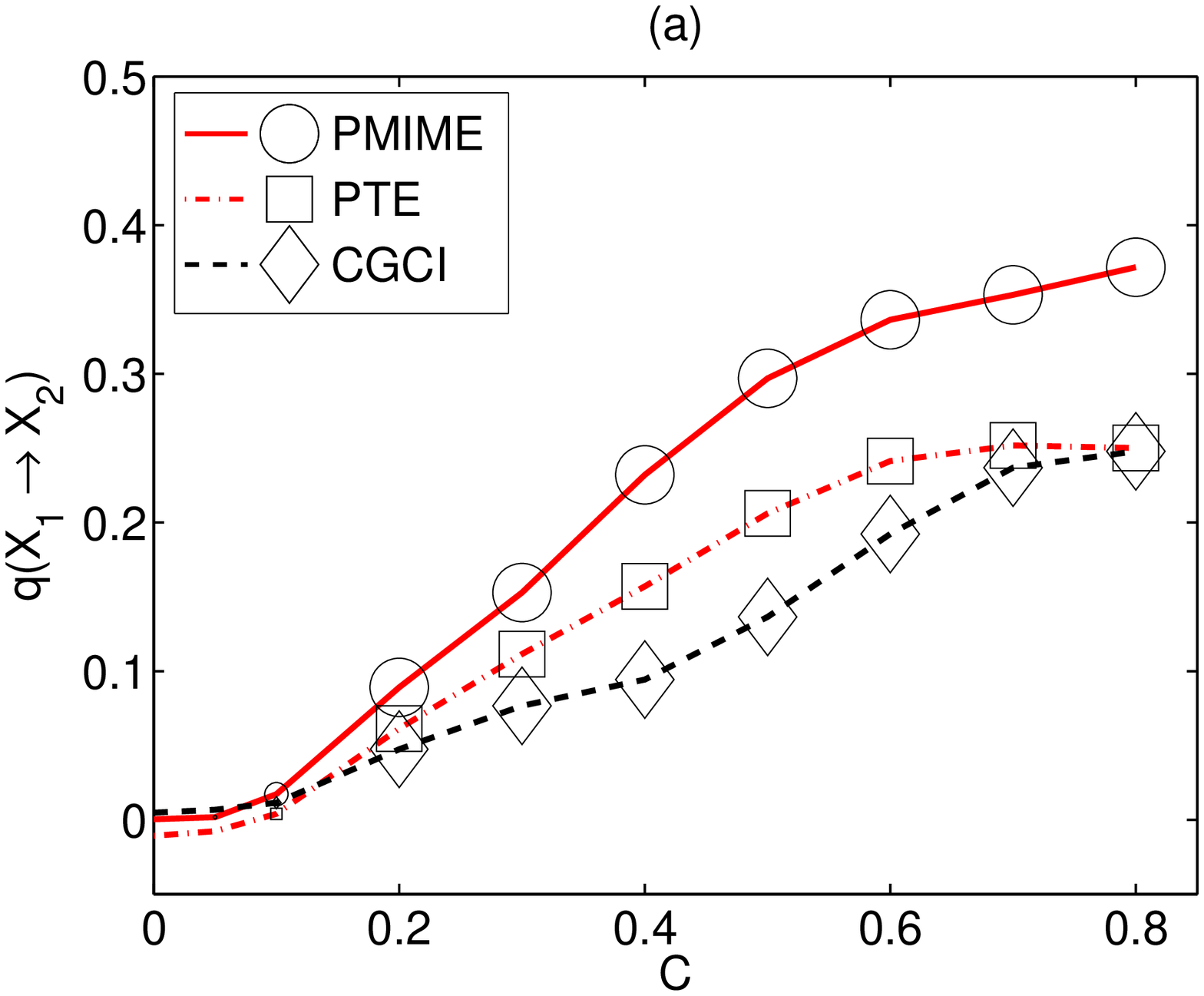}
\includegraphics[width=5cm]{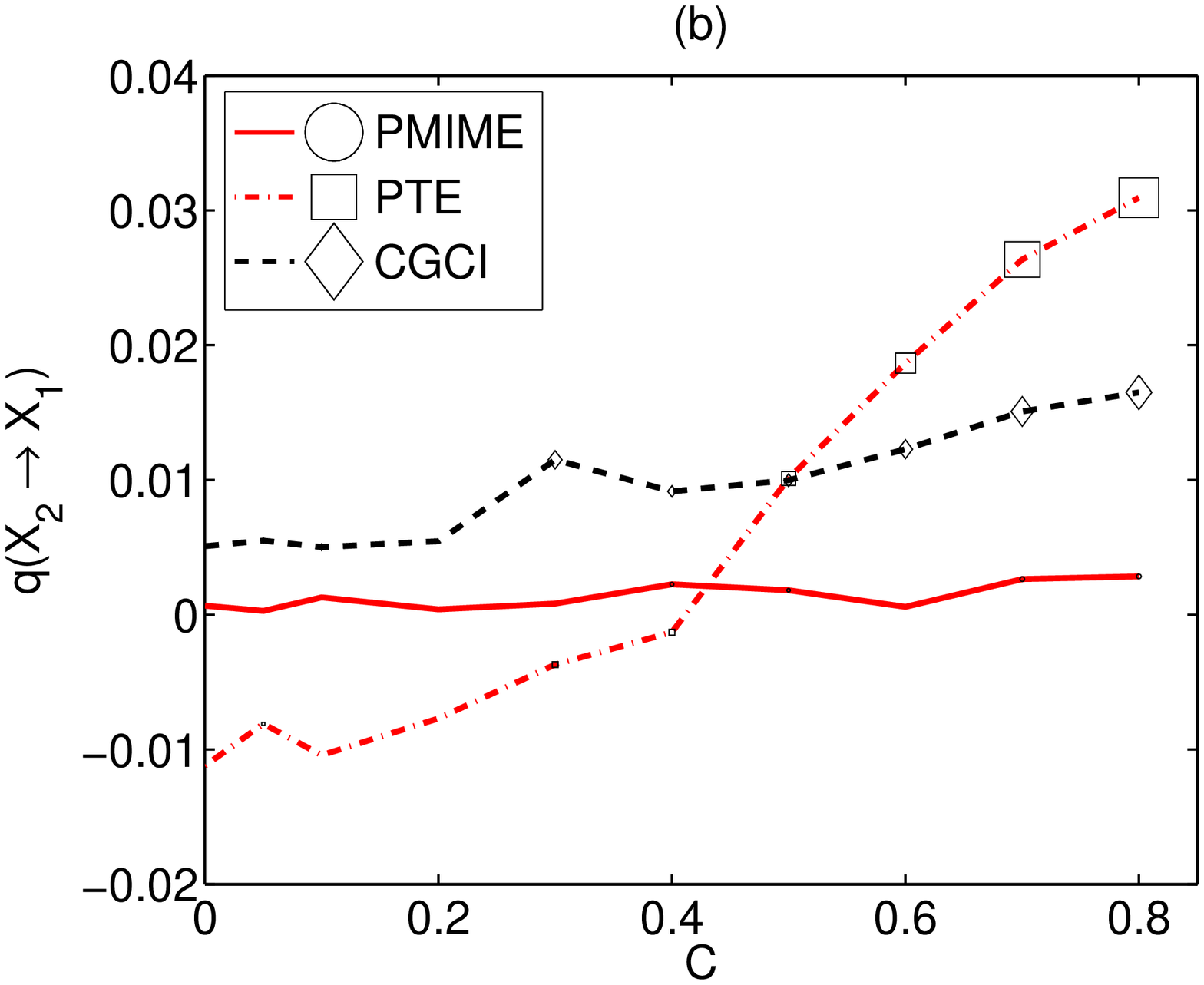}
\includegraphics[width=5cm]{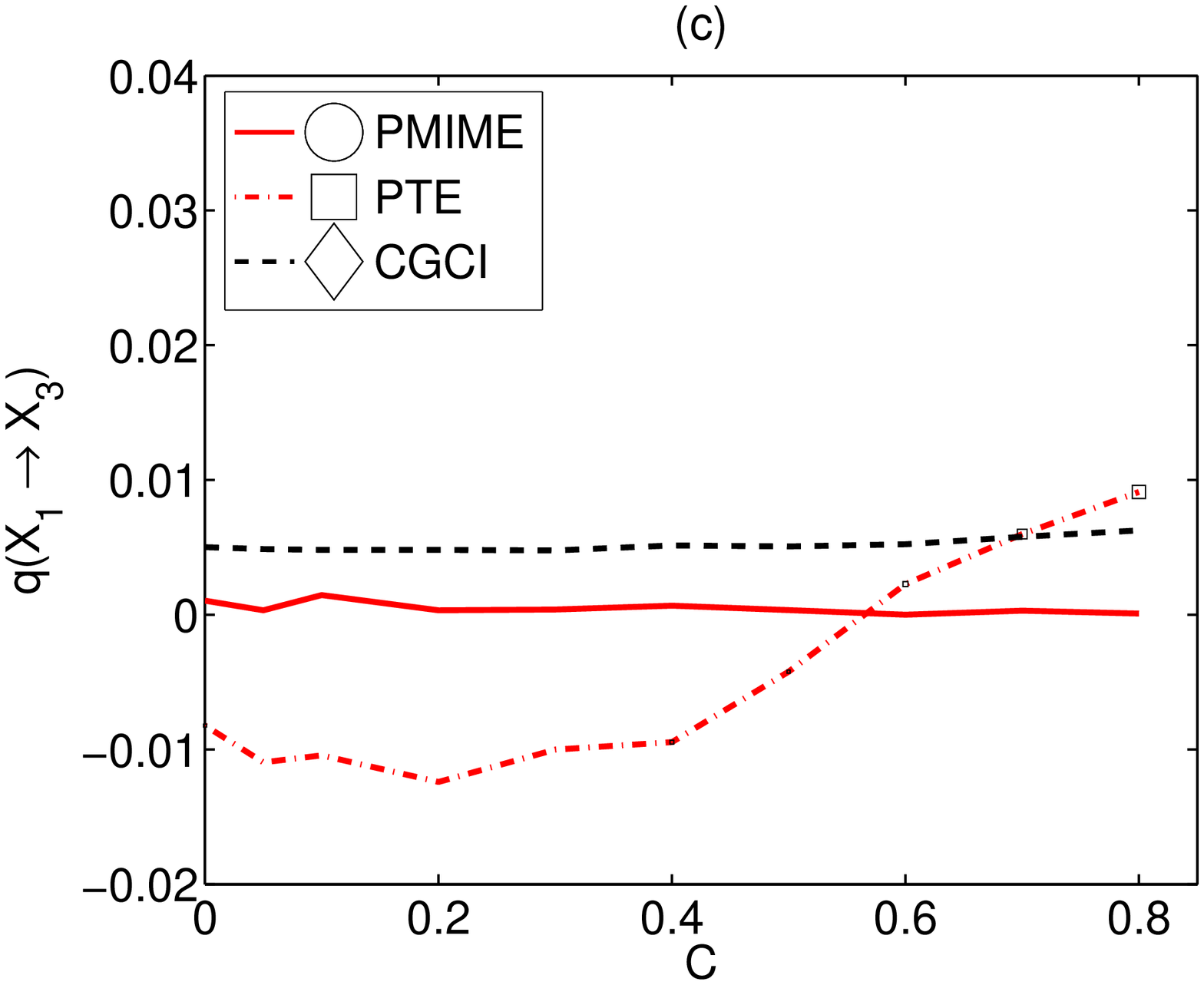}}}
\caption{As Figure~\ref{fig:HenonK3diffCnoisefree}, but when white
noise with SD 20\% of the data SD is added.}
\label{fig:HenonK3diffCnoise20}       
\end{figure}

Some additional results to these shown earlier are shown below.
First, the case of $K=3$ is presented for small time series of
length $n=512$ and for the whole range of $C$. The true direct
couplings are $X_1 \rightarrow X_2$ and $X_3 \rightarrow X_2$ and
they are equivalent in strength. There is a symmetry in the
coupling structure and therefore only three couplings are shown in
Figure~\ref{fig:HenonK3diffCnoisefree}.
All measures confidently detect the true direct coupling for $C
\geq 0.1$, but have very small power to detect it for $C=0.05$ and
their average from 100 realizations is only slightly above the
zero level (see Figure~\ref{fig:HenonK3diffCnoisefree}a). Note
that the zero level for PTE is negative. This is better seen for
the non-existing couplings in
Figure~\ref{fig:HenonK3diffCnoisefree}b and c, while PMIME is
always exactly zero. Moreover, for $X_2 \rightarrow X_1$, PTE
increases with $C$ and for larger $C$ it is found significant more
often than the nominal level ($\alpha=0.05$), and the same holds
for CGCI. The  biased detection of false couplings with PTE (and
CGCI) is more evident in the presence of noise, as shown in
Figure~\ref{fig:HenonK3diffCnoise20}, where white noise with SD
being 20\% of the data SD is added for each variable.
Note that PMIME is not affected by noise and achieves the same power in detecting the true direct couplings, while it remains at the zero level when there is no direct coupling.

\begin{figure}
\centerline{\includegraphics[width=11cm]{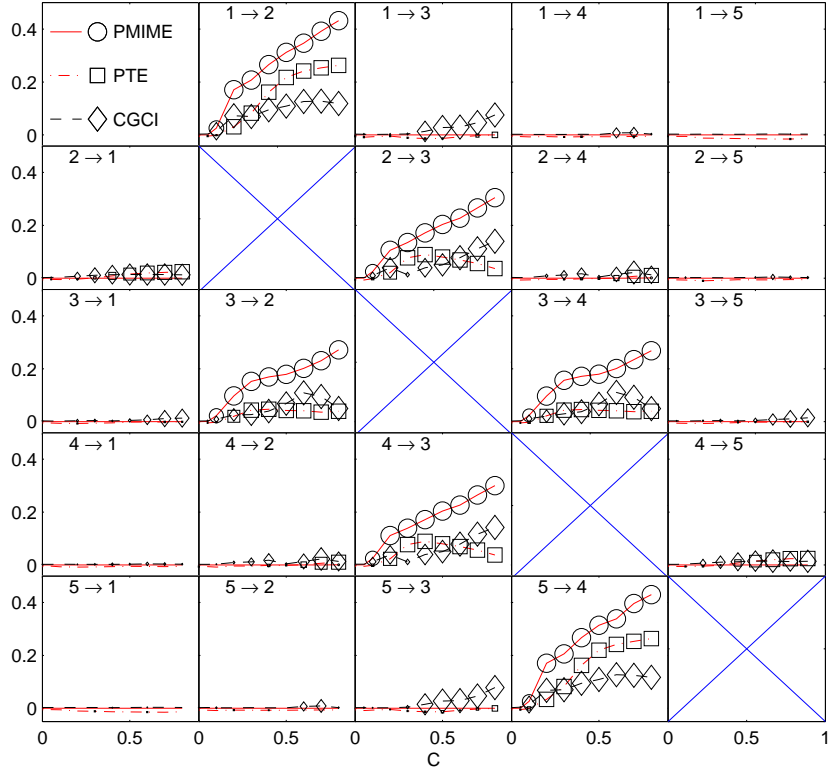}}
\caption{Matrix plot of all possible couplings of $K=5$ variables of the coupled H\'{e}non maps ($n=1024$). Each panel regards a coupling of the variables with indices as shown at the top of the panel, and the organization of each panel is as for Figure~\ref{fig:HenonK3diffCnoisefree}.}
\label{fig:HenonK5diffCnoisefree}       
\end{figure}

For $K=5$, the efficiency of PMIME as opposed to PTE and CGCI persists, as shown in the matrix plot of Figure~\ref{fig:HenonK5diffCnoisefree} for all possible pairs.
The off-diagonal panels correspond to true direct couplings, the panels in column 1 and 5 correspond to non-existing coupling and all the other panels correspond to indirect couplings. PMIME estimates with high confidence the correct direct couplings for all $C \geq 1$. PTE is high and monotonically increasing only for the direct couplings $X_1 \rightarrow X_2$ and $X_5 \rightarrow X_4$, while for the other direct couplings it tends to decrease for $C > 0.3$. It is pointed in \cite{Chicharro12} that PTE may not be monotonic to $C$ due to changes in the inter-dependence structure, but here PMIME does not seem to be affected. The interpretation of the PTE results as to the identification of the true direct couplings is difficult because small and significant PTE is observed both for true direct couplings and spurious couplings ($X_2 \rightarrow X_1$ and $X_4 \rightarrow X_2$). The same holds for CGCI, which is significant also for indirect couplings ($X_1 \rightarrow X_3$ and $X_5 \rightarrow X_3$).

\begin{figure}
\centerline{\includegraphics[width=11cm]{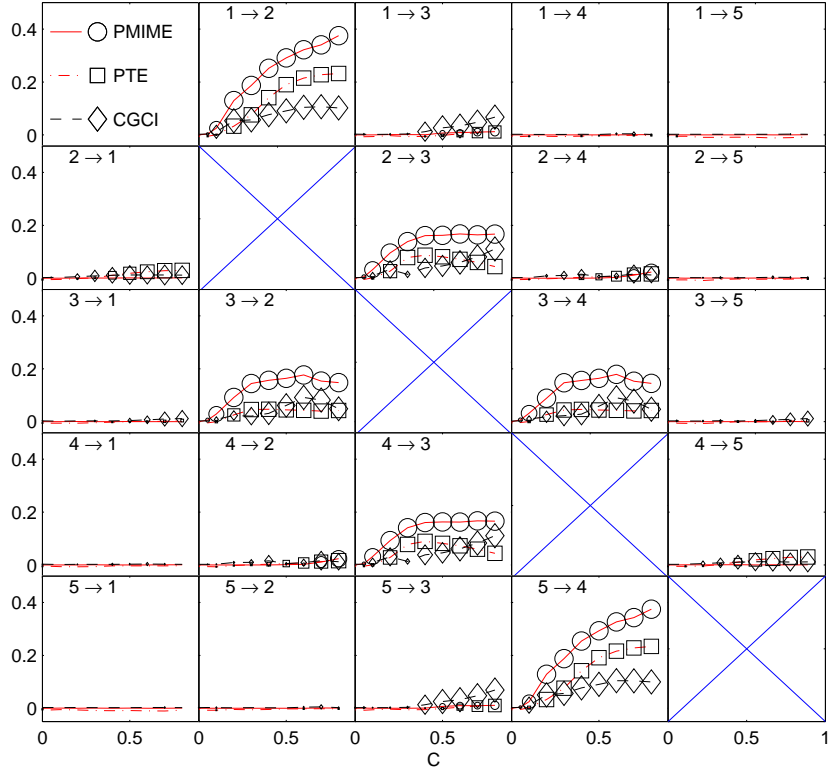}}
\caption{As Figure~\ref{fig:HenonK5diffCnoisefree} but for 20\% additive white noise.}
\label{fig:HenonK5diffCnoise20}       
\end{figure}

About the same results are obtained when 20\% white noise is added to the data, as shown in Figure~\ref{fig:HenonK5diffCnoise20}.
PMIME is somehow smaller in magnitude but still can distinguish well the true direct couplings even for small $C$. On the other hand, PTE tends to be significant for more non-existing direct couplings than for the noise-free case ($X_1 \rightarrow X_3$ and $X_5 \rightarrow X_3$) following well CGCI to spurious detection of couplings.

\begin{figure}
\centerline{\hbox{\includegraphics[width=5cm]{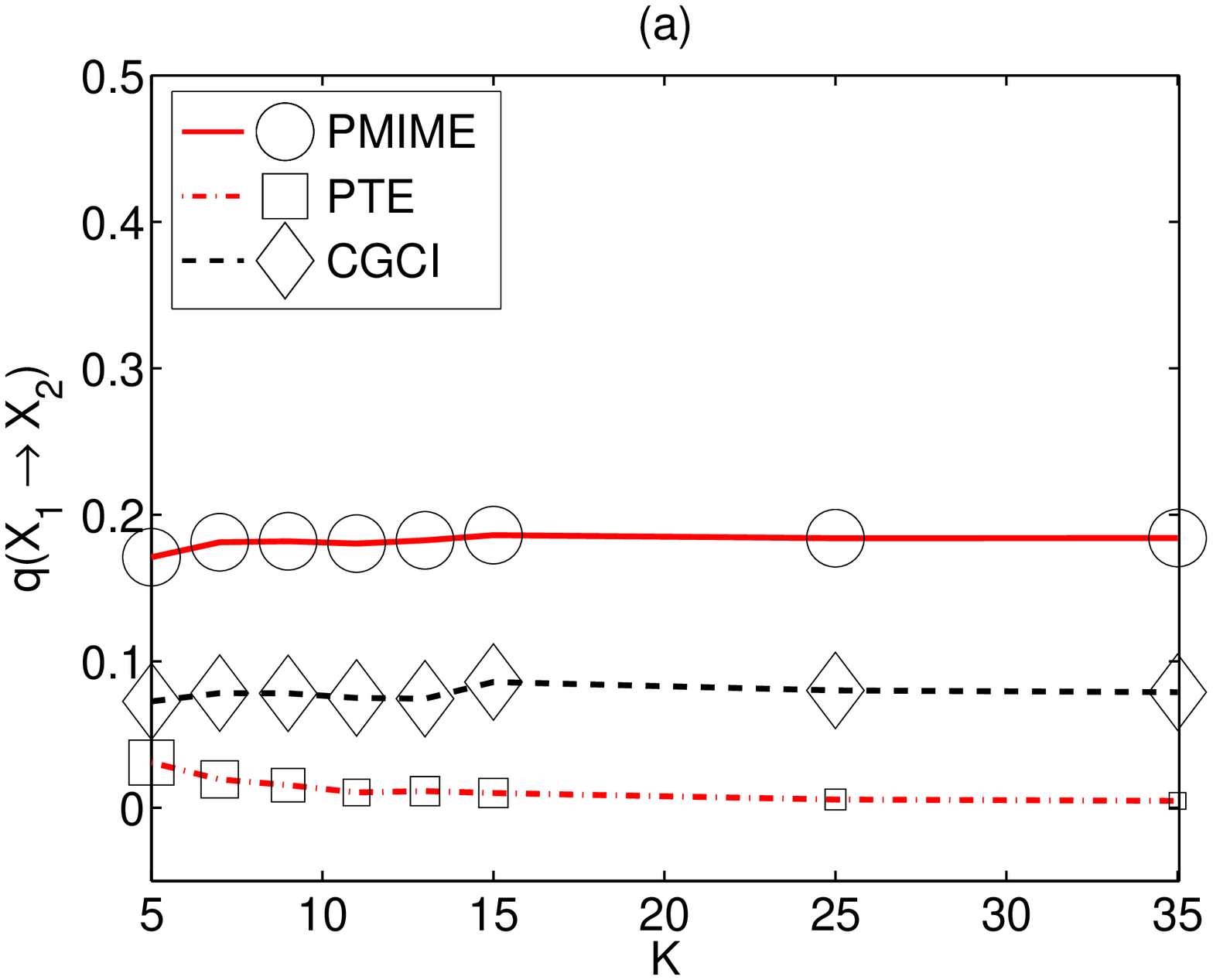}
\includegraphics[width=5cm]{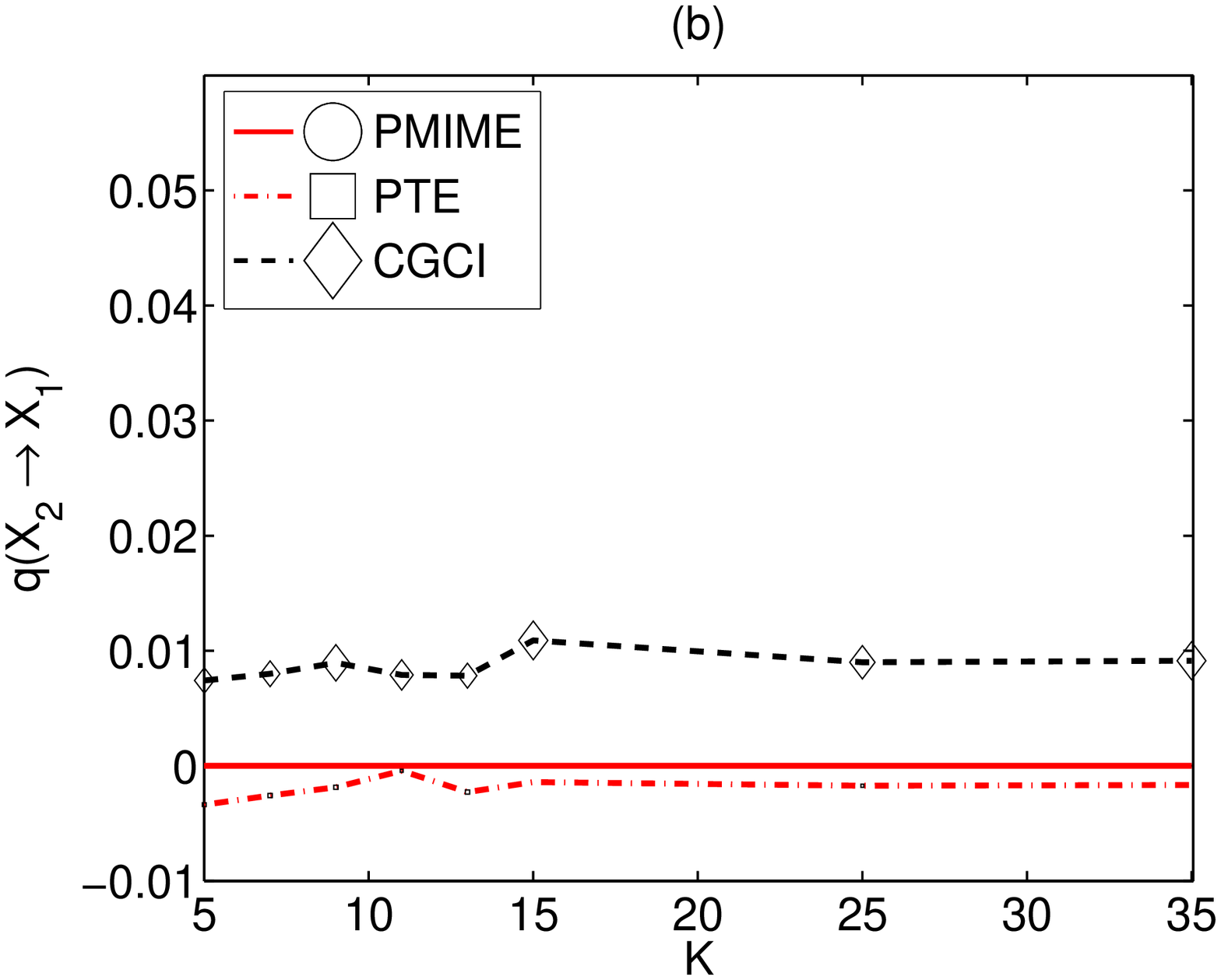}}}
\centerline{\hbox{\includegraphics[width=5cm]{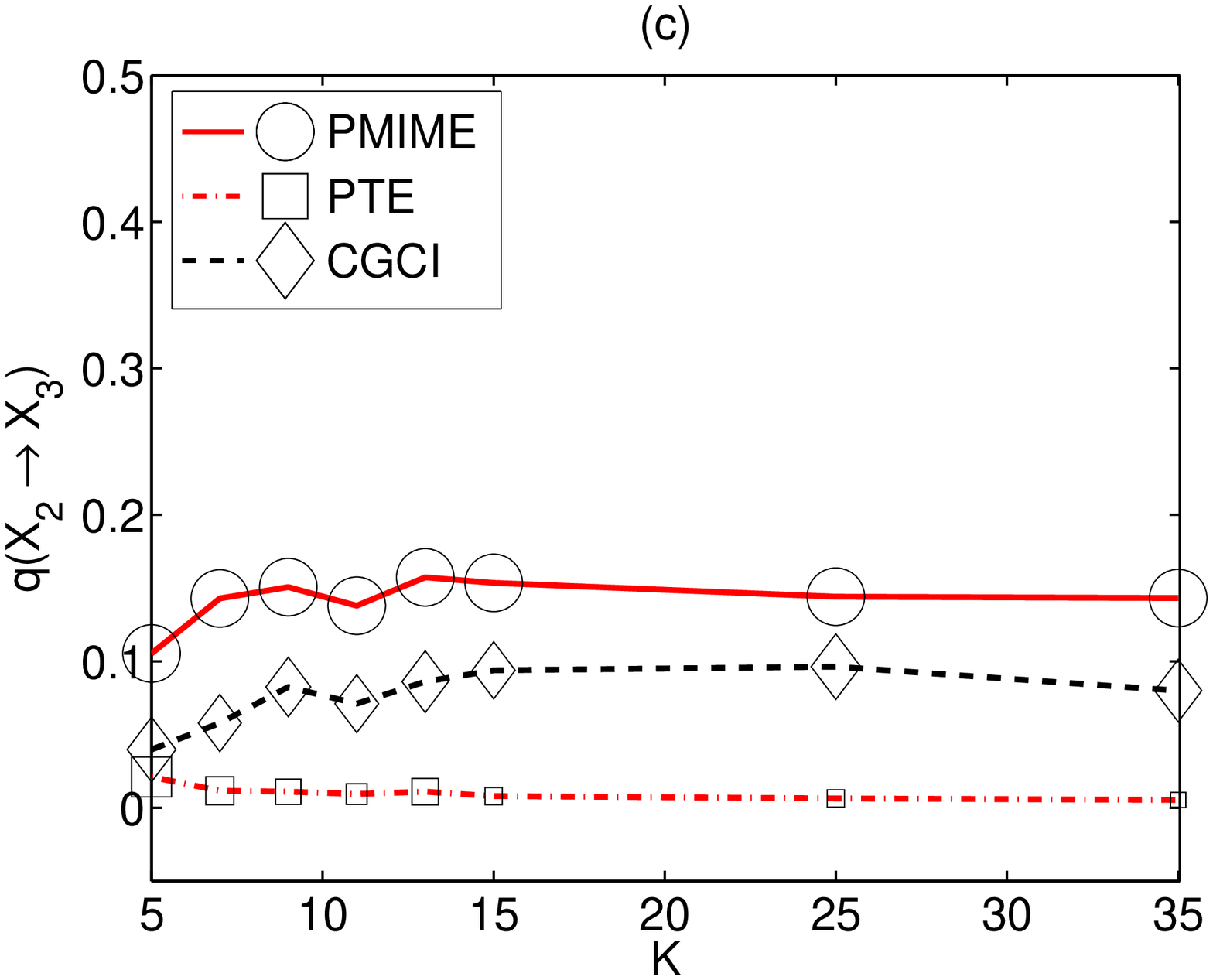}
\includegraphics[width=5cm]{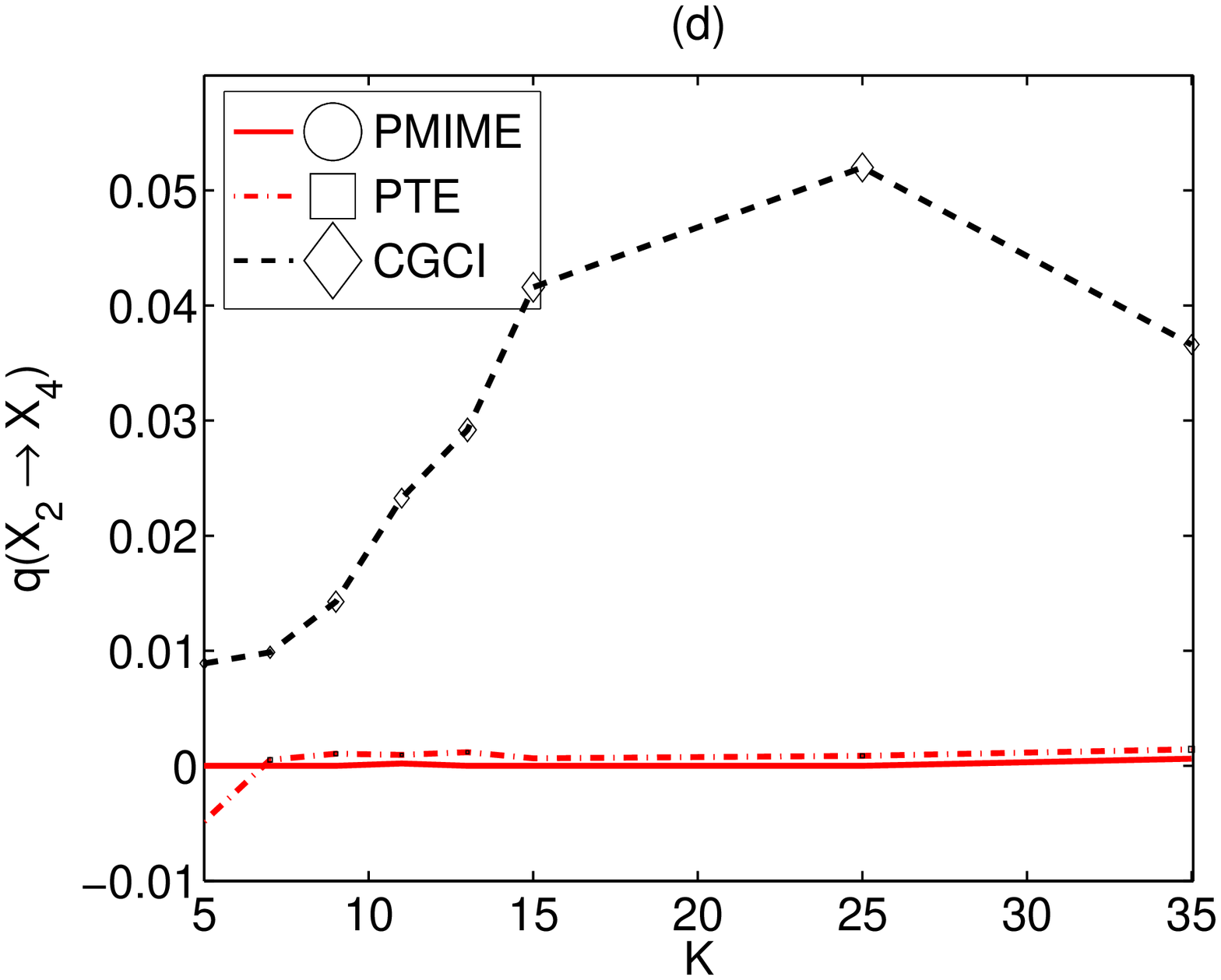}}}
\caption{PMIME, PTE and CGCI measures (denoted collectively $q$) are given as functions of the number of variables $K$ of the coupled H\'{e}non maps for time series length $n=1024$. (a) True direct causality $X_1 \rightarrow X_2$. (b) Non-existing coupling $X_2 \rightarrow X_1$. (c) True direct causality $X_2 \rightarrow X_3$. (d) True indirect causality $X_2 \rightarrow X_4$. The number of rejections in 100 realizations of the randomization test determines the size of a symbol displayed for each measure and $C$, where in the legend the size of the symbols regards 100 rejections.}
\label{fig:HenonC02diffKnoisefree}       
\end{figure}

In Table II, the performance of the measures PMIME, PTE and CGCI
was shown for $K=5,15,25$, coupled H\'{e}non maps. In
Figure~\ref{fig:HenonC02diffKnoisefree} more detailed results are
shown.
For the two true direct couplings $X_1 \rightarrow X_2$ and $X_2 \rightarrow X_3$ for different $K$ in Figure~\ref{fig:HenonC02diffKnoisefree}a and c, respectively, PMIME is at the same significantly positive magnitude for as large $K$ as 35, an impressive result for a relatively small time series of length $n=1024$. The same holds for CGCI, which for the second coupling is smaller for smaller $K$, possibly due to the additional causal effect of other neighboring variables. On the other hand, PTE decreases both in magnitude and in statistical significance with $K$. In the case of the non-existing coupling $X_2 \rightarrow X_1$ in Figure~\ref{fig:HenonC02diffKnoisefree}b, PMIME is always zero for any $K$, PTE is also statistically insignificant for any $K$, while CGCI is positive and statistically significant at about half of the 100 realizations. For the indirect causal connection $X_2 \rightarrow X_4$ in Figure~\ref{fig:HenonC02diffKnoisefree}d, PMIME and PTE are again at the zero level and CGCI gets larger but less statistically significant. Thus CGCI is biased towards giving spurious direct causality, PTE cannot identify the direct causal effects, while PMIME attains optimal sensitivity and specificity.

\begin{figure}
\centerline{\hbox{\includegraphics[width=5cm]{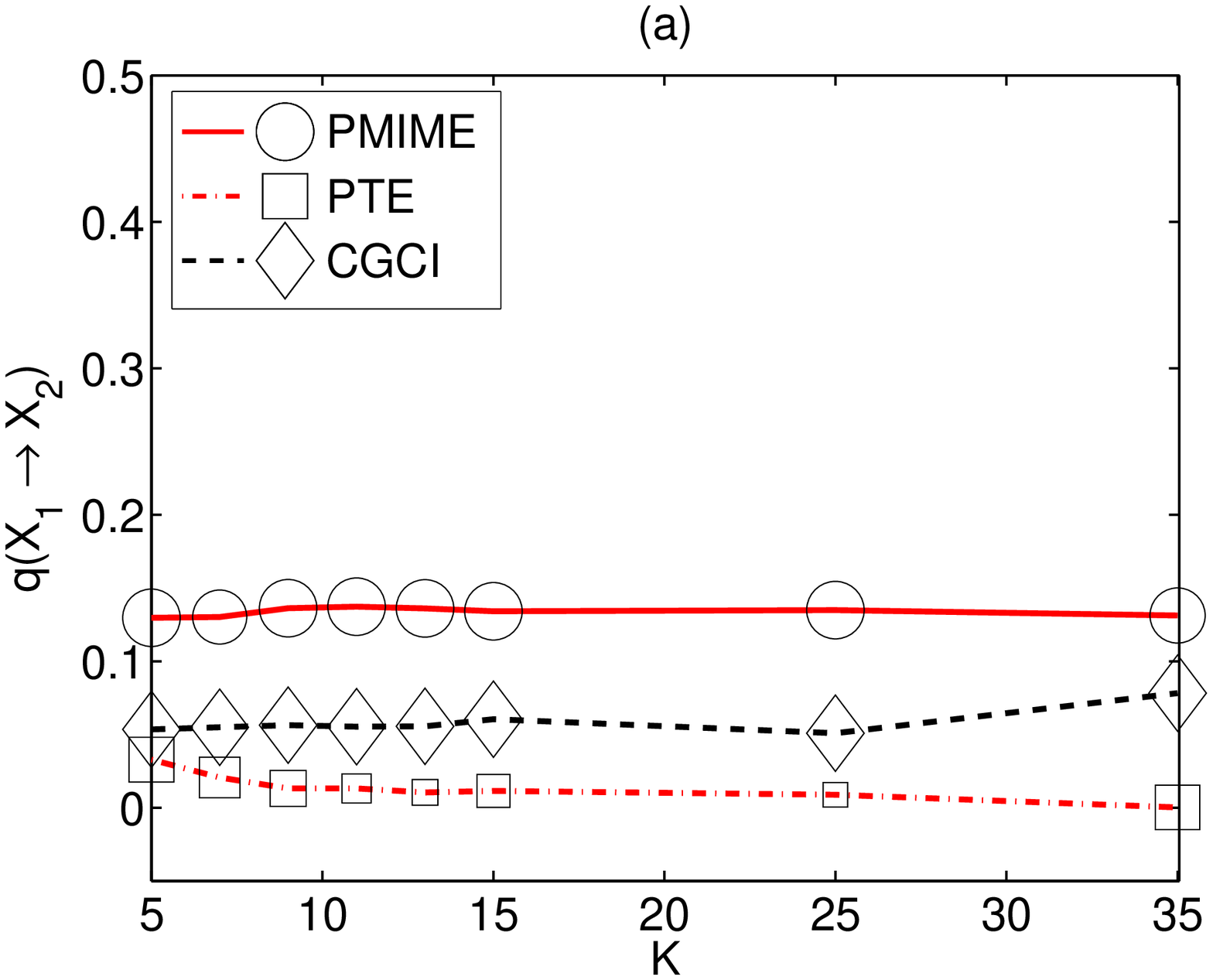}
\includegraphics[width=5cm]{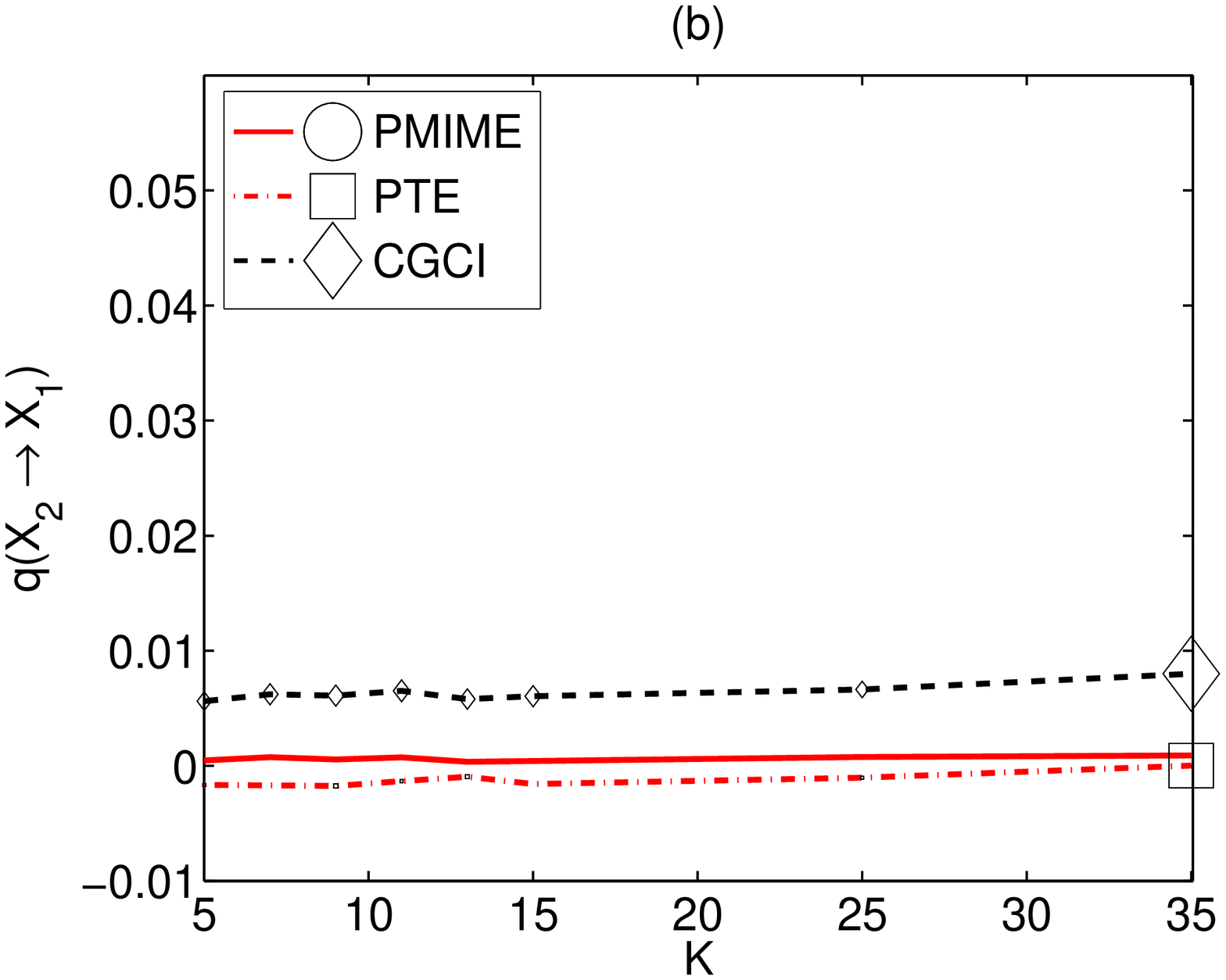}}}
\centerline{\hbox{\includegraphics[width=5cm]{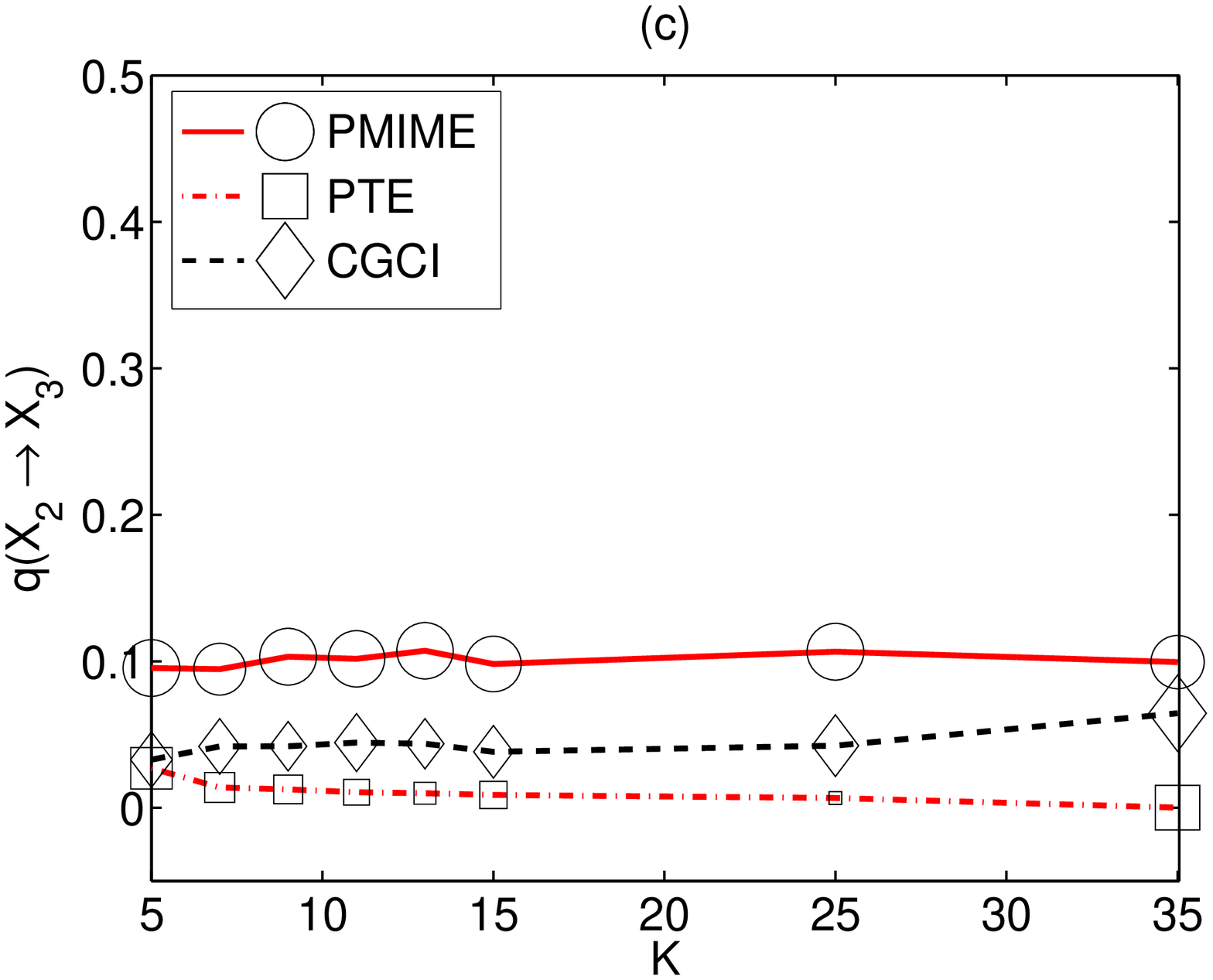}
\includegraphics[width=5cm]{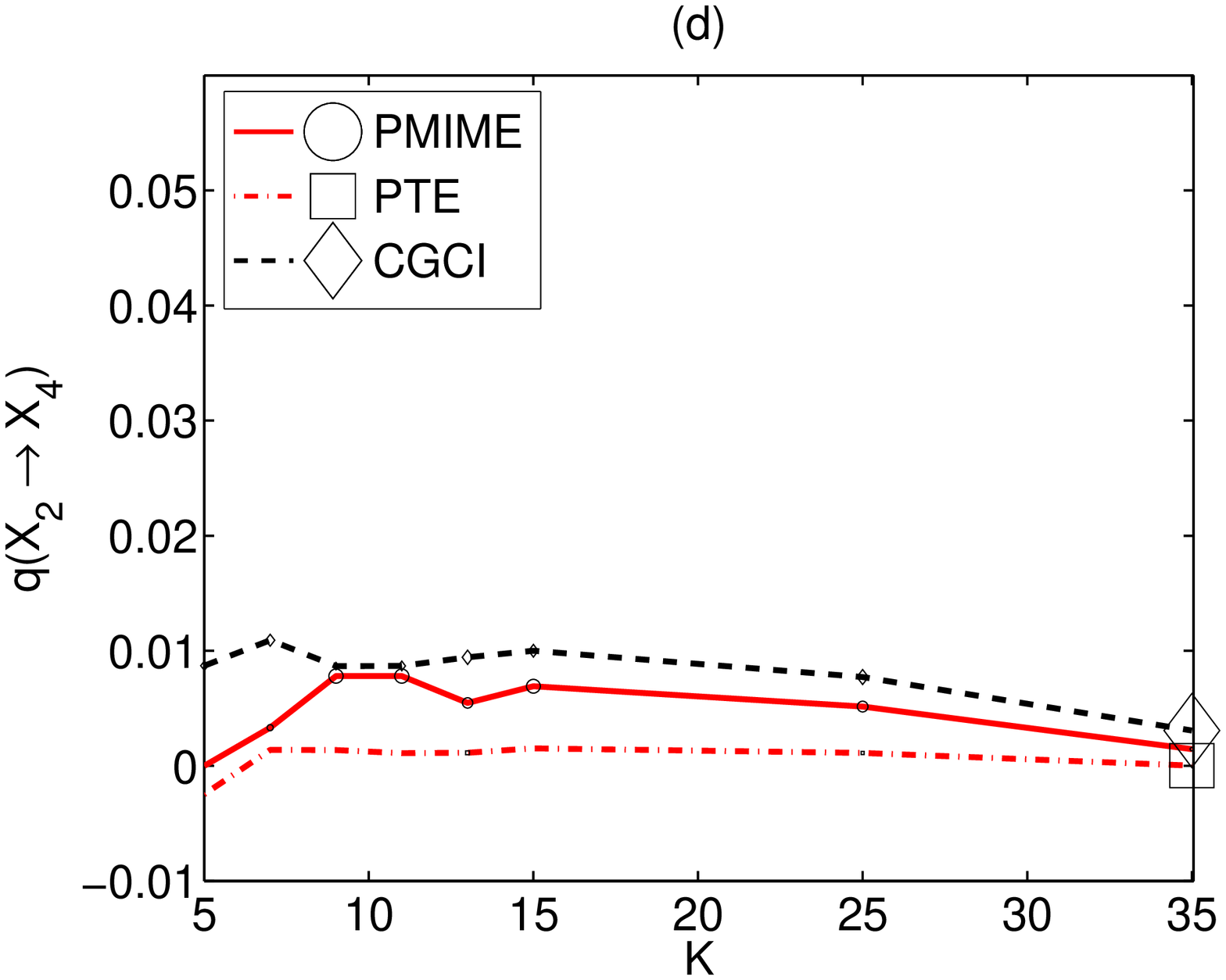}}}
\caption{As Figure~\ref{fig:HenonC02diffKnoisefree} but for 20\% additive white noise.}
\label{fig:HenonC02diffKnoise20}       
\end{figure}

The proper performance of PMIME persists also when noise is added to the time series, as shown for the same examples in Figure~\ref{fig:HenonC02diffKnoise20} in the presence of additional 20\% white noise.
CGCI and PTE exhibit the same shortcomings, and CGCI actually improves its specificity as it decreases in the case of indirect coupling. For the latter case, PMIME gets positive in a small percentage of the 100 realizations which is at the level of significance of the surrogate data test for the termination criterion.

\subsection{Coupled Lorenz system}

Next we study the system of three coupled identical Lorenz subsystems defined as
\begin{equation*}
\begin{array}{l}
\dot{x}_{1} = -10x_{1} + 10y_{1} \\
\dot{y}_{1} = -x_{1}z_{1} + 28x_{1} - y_{1} \\
\dot{z}_{1} = x_{1}y_{1} -8/3z_{1}
\end{array}
\quad\quad
\begin{array}{l}
\dot{x}_{i} = -10x_{i} + 10y_{i} + C(x_{i-1}-x_{i}) \\
\dot{y}_{i} = -x_{i}z_{i} + 28x_{i} - y_{i} \\
\dot{z}_{i} = x_{i}y_{i} -8/3z_{i}
\end{array}
\quad i=2,3
\end{equation*}
The system of differential equations is solved using the explicit Runge-Kutta (4,5) method in Matlab and the time series are generated at a sampling time of 0.01 time units. The first variable of each subsystem is observed, denoted respectively as $X_1$, $X_2$ and $X_3$, and the direct couplings are $X_1 \rightarrow X_2$ and $X_2 \rightarrow X_3$. The same coupling strength $C$ is used for both couplings and for this setting it was assessed by observing the generated trajectories and characteristics of the observed time series (delay mutual information, correlation dimension and cross correlation) that complete synchronization is approached for $C>8$, so the measures were computed for $C=0,0.5,1,\ldots,8$. For each $C$, 100 realizations are generated, and we set $m=3$ for PTE and CGCI, $L=15$ for PMIME, and for both PMIME and PTE the future vector is formed for the time horizon $T=3$, i.e. $\mathbf{y}_t^3=[y_{t+1},y_{t+2},y_{t+3}]$ for the response variable $Y$, where $Y$ is any of the variables $X_1$, $X_2$ and $X_3$. The three steps ahead are chosen to represent better the time evolution of the continuous system, as suggested also in \cite{Vlachos10}. For CGCI, the option of a larger step ahead $T>1$ is not considered and it is computed for $T=1$.

\begin{figure}
\centerline{\hbox{\includegraphics[width=7.5cm]{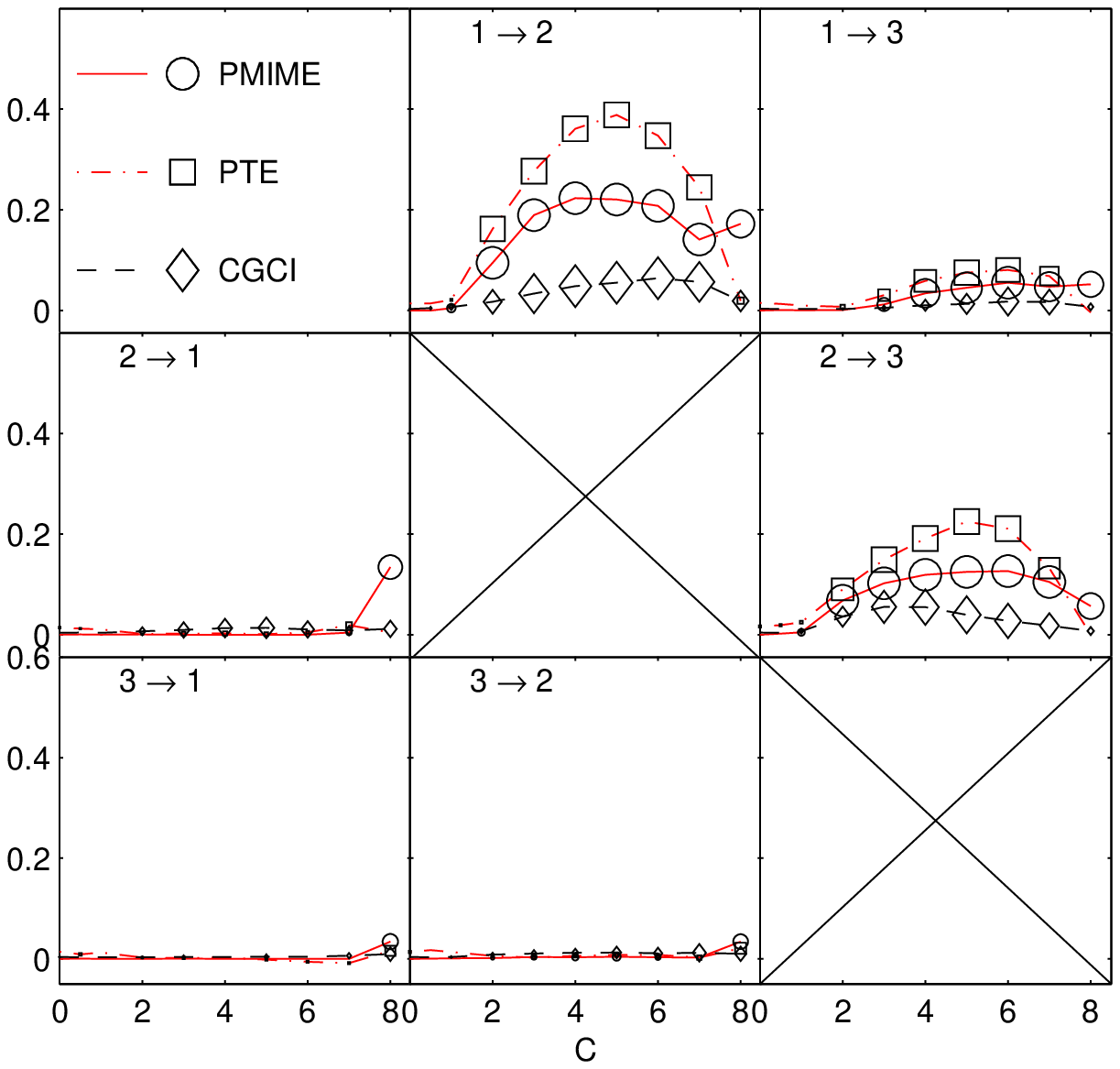}
\includegraphics[width=7.5cm]{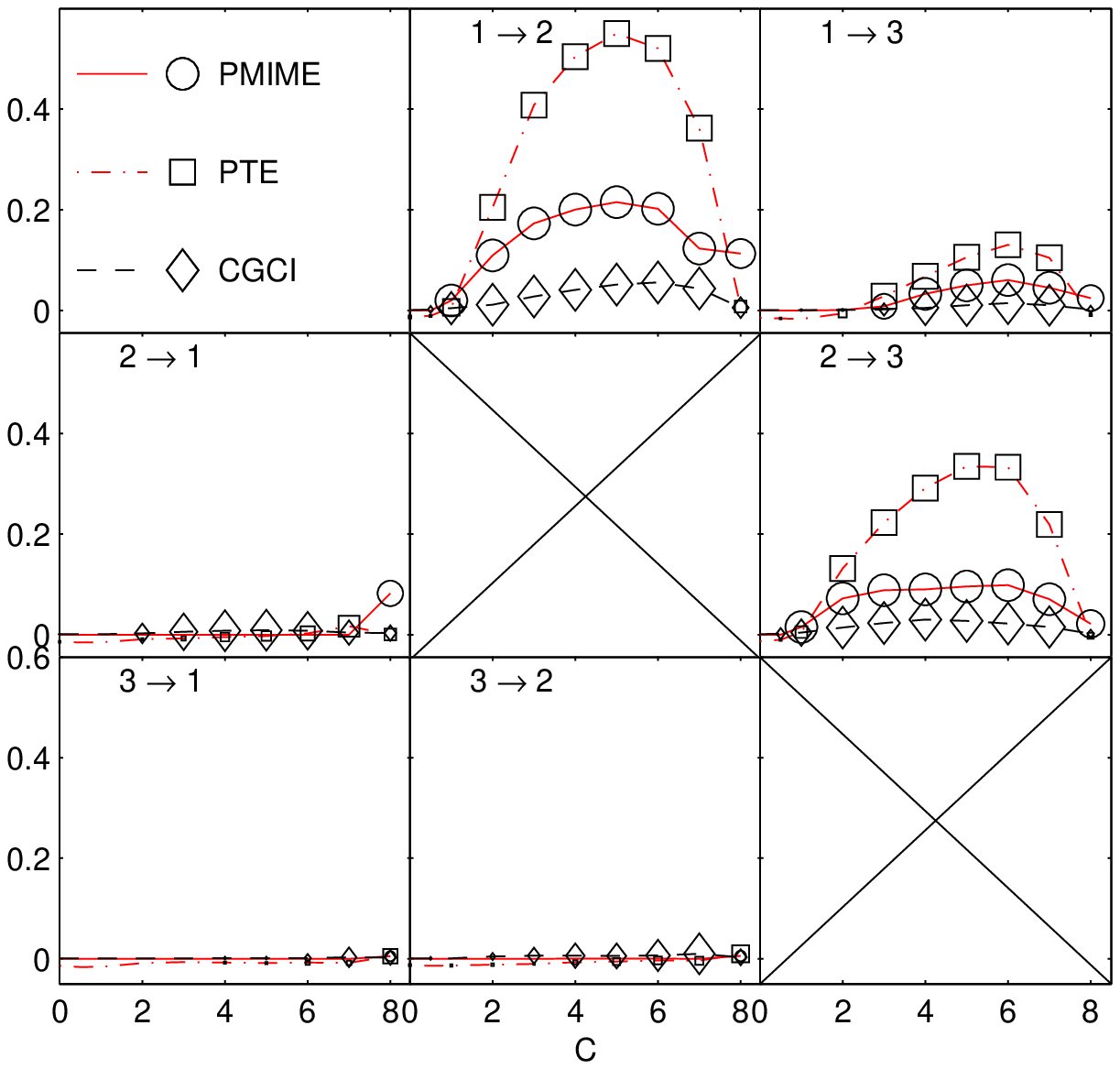}}}
\caption{Matrix plot of all possible couplings of the Lorenz systems of $K=3$ subsystems and noise-free time series of length $n=1024$ on the left panel and $n=4096$ on the right panel. The organization of the panels is as for Figure~\ref{fig:HenonK5diffCnoisefree}.}
\label{fig:LorenzdiffCnoisefree}       
\end{figure}

\begin{figure}
\centerline{\hbox{\includegraphics[width=7.5cm]{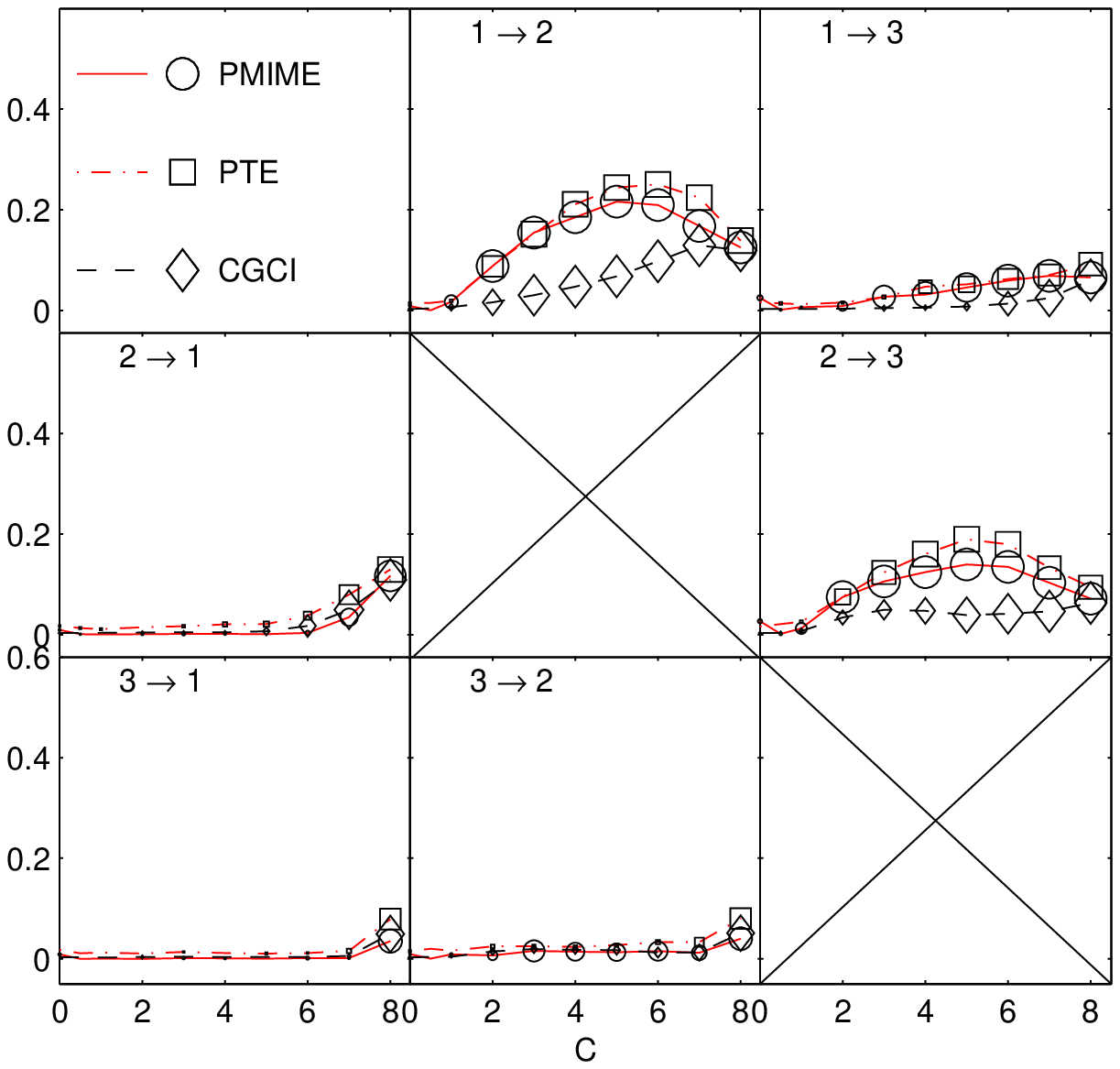}
\includegraphics[width=7.5cm]{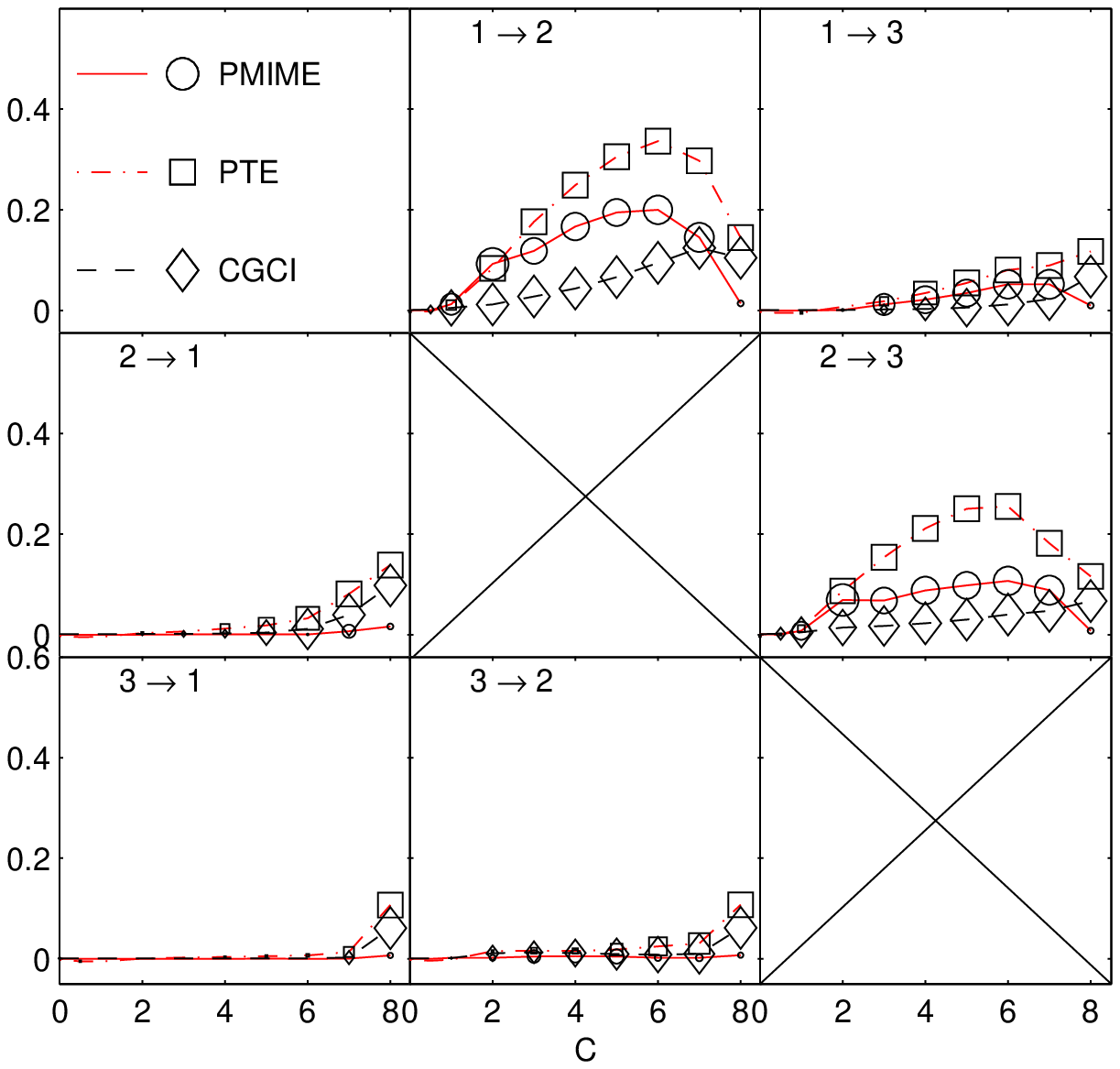}}}
\caption{The same as Figures~\ref{fig:LorenzdiffCnoisefree} but with 20\% white noise added to the time series.}
\label{fig:LorenzdiffCnoise20}       
\end{figure}

The results of the simulations on noise-free time series of length $n=1024$ and $n=4096$ are shown in Figure~\ref{fig:LorenzdiffCnoisefree}, and when 20\% white noise is added in Figure~\ref{fig:LorenzdiffCnoise20}.
First, it is noted that the scale is not the same for the three
measures and their magnitude is not a safe criterion of
comparison. All three measures capture well the two true direct
couplings for $C \geq 2$ with a rejection rate of the null
hypothesis of no coupling at about 100\%. For example, for
noise-free data, $C=2$ and $n=1024$, the rejection rates for $X_2
\rightarrow X_3$ are 99\% for PMIME, 87\% for PTE and 49\% for
CGCI, and change to 100\% for all measures when $n=4096$. However,
for weaker coupling with $C=1$ only PMIME detects confidently the
true direct courplings, and for $X_2 \rightarrow X_3$ it is 100\%
for $n=4096$ dropping to 23\% for $n=1024$, whereas for PTE the
respective rejection rates are 47\% and 11\%, while for CGCI are
64\% and 13\%. Thus though for stronger coupling all measures can
detect well the direct true couplings, for smaller $C$ PMIME shows
significantly better sensitivity. Regarding specificity, PMIME is
also scoring best. For example, for the indirect coupling $X_1
\rightarrow X_3$ and $C=2$, PMIME as well as CGCI give small
rejection rate at the nominal significance level 5\%, while PTE
gives rejection rate 16\% for $n=1024$ getting double for
$n=4096$. For larger $C$, the rejection rate gets larger for all
measures, indicating that for stronger coupling the indirect
causal effects cannot be distinguished. For the cases of no
coupling, all measures are at about the zero level, but only PMIME
is statistically insignificant. For example for the non-existing
coupling $X_2 \rightarrow X_1$, the rejection rate of PTE is 16\%
for $n=1204$ and becomes double for $n=4096$, while for CGCI the
respective rejection rates are much higher (87\% and 97\%), but
PMIME gives no positive value for both $n$. PMIME has high
rejection rates for $C=8$ because then the three variables are
almost completely synchronized and the lagged variables may
exhibit similar causal effects and thus the algorithm for mixed
embedding does not systematically pick up a particular set of
lagged variables.

The performance of the measures turns out to be persistent to the
presence of noise (see Figure~\ref{fig:LorenzdiffCnoise20}). PMIME
tends to be biased towards detecting false direct couplings for
small time series lengths ($n=1024$), but improves for larger time
series lengths ($n=4096$). However, PTE and CGCI seem to suffer
from lack of specificity for increasing $C$ also when $n$
increases.

\begin{figure}
\centerline{\hbox{\includegraphics[width=5cm]{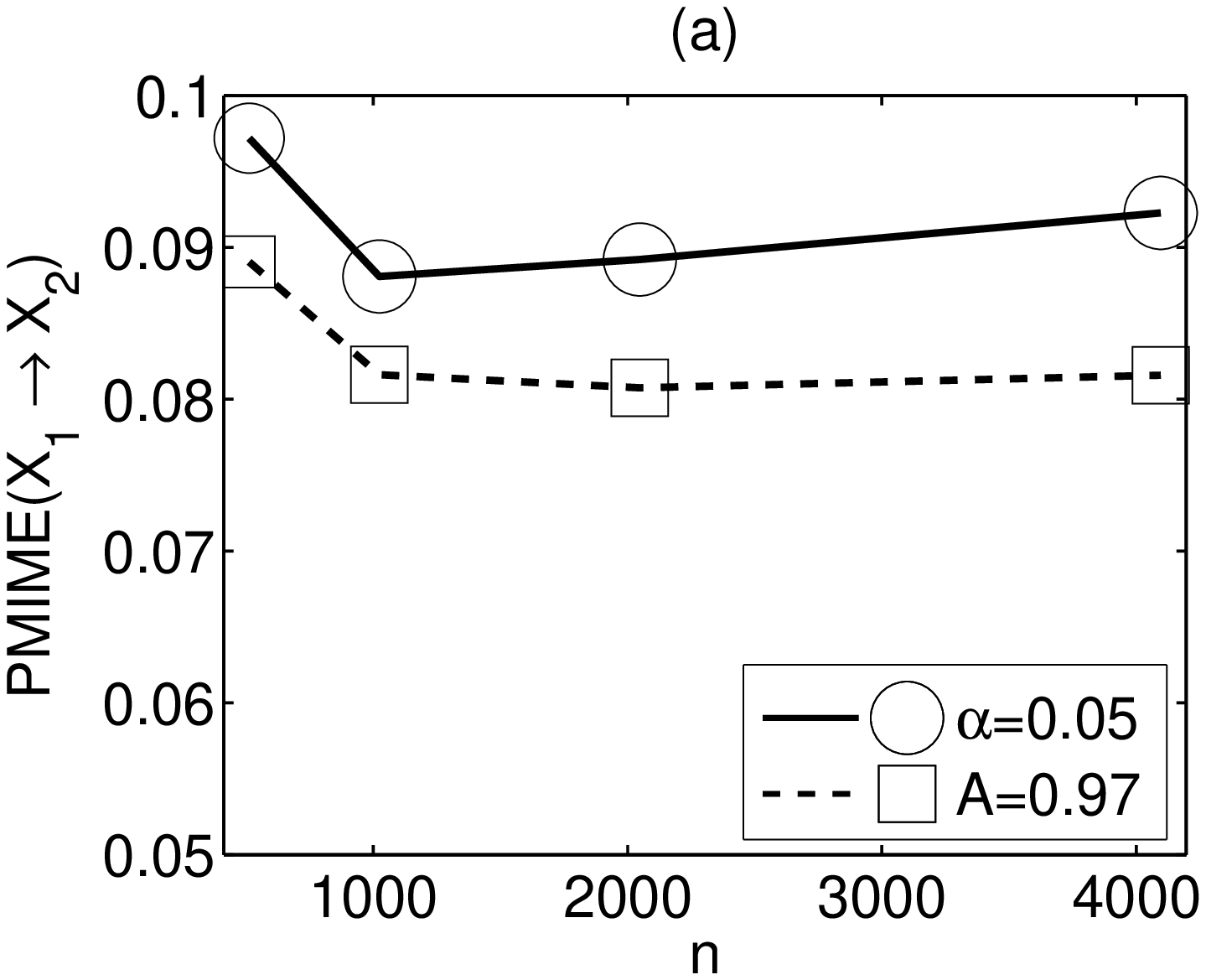}
\includegraphics[width=5cm]{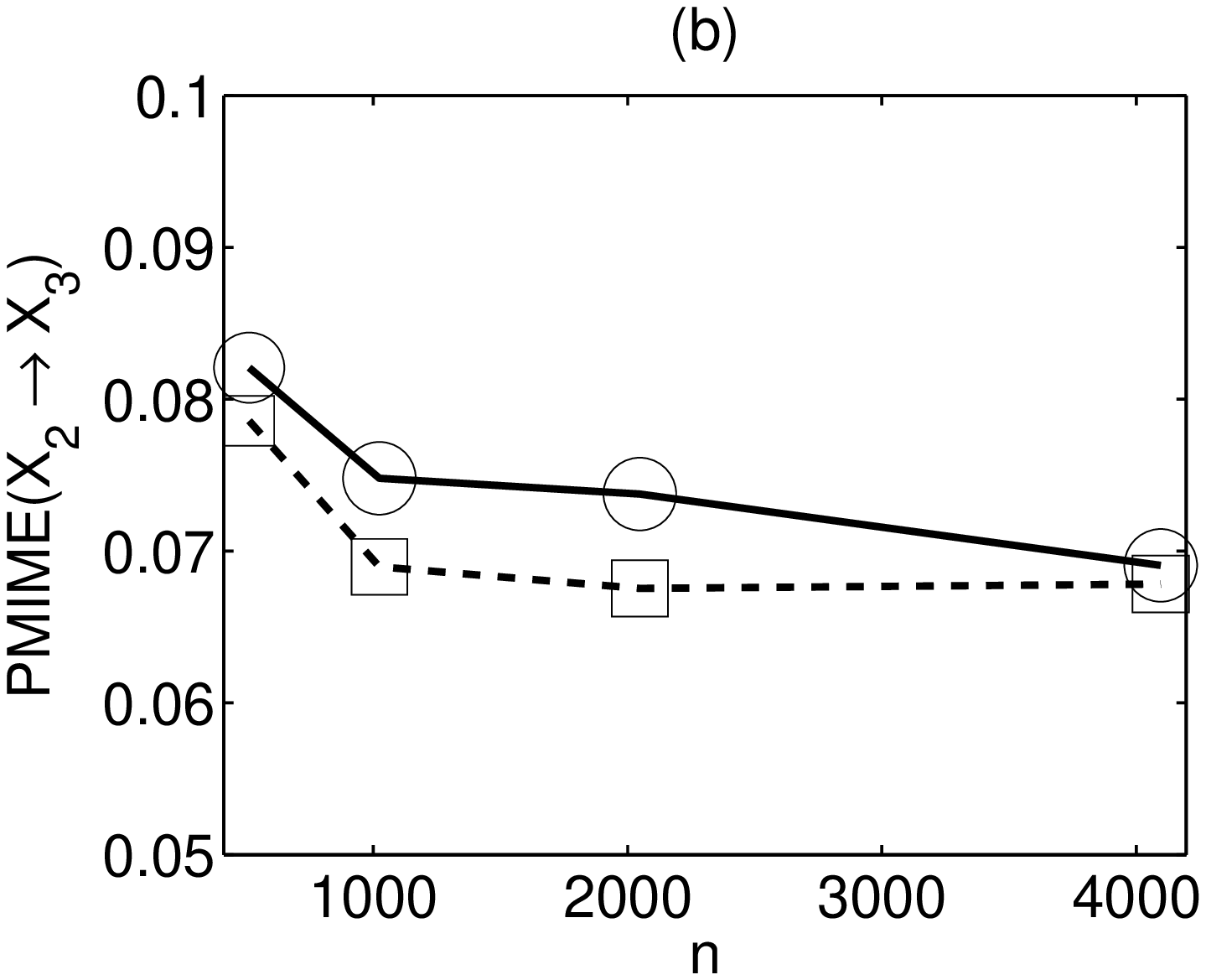}
\includegraphics[width=5cm]{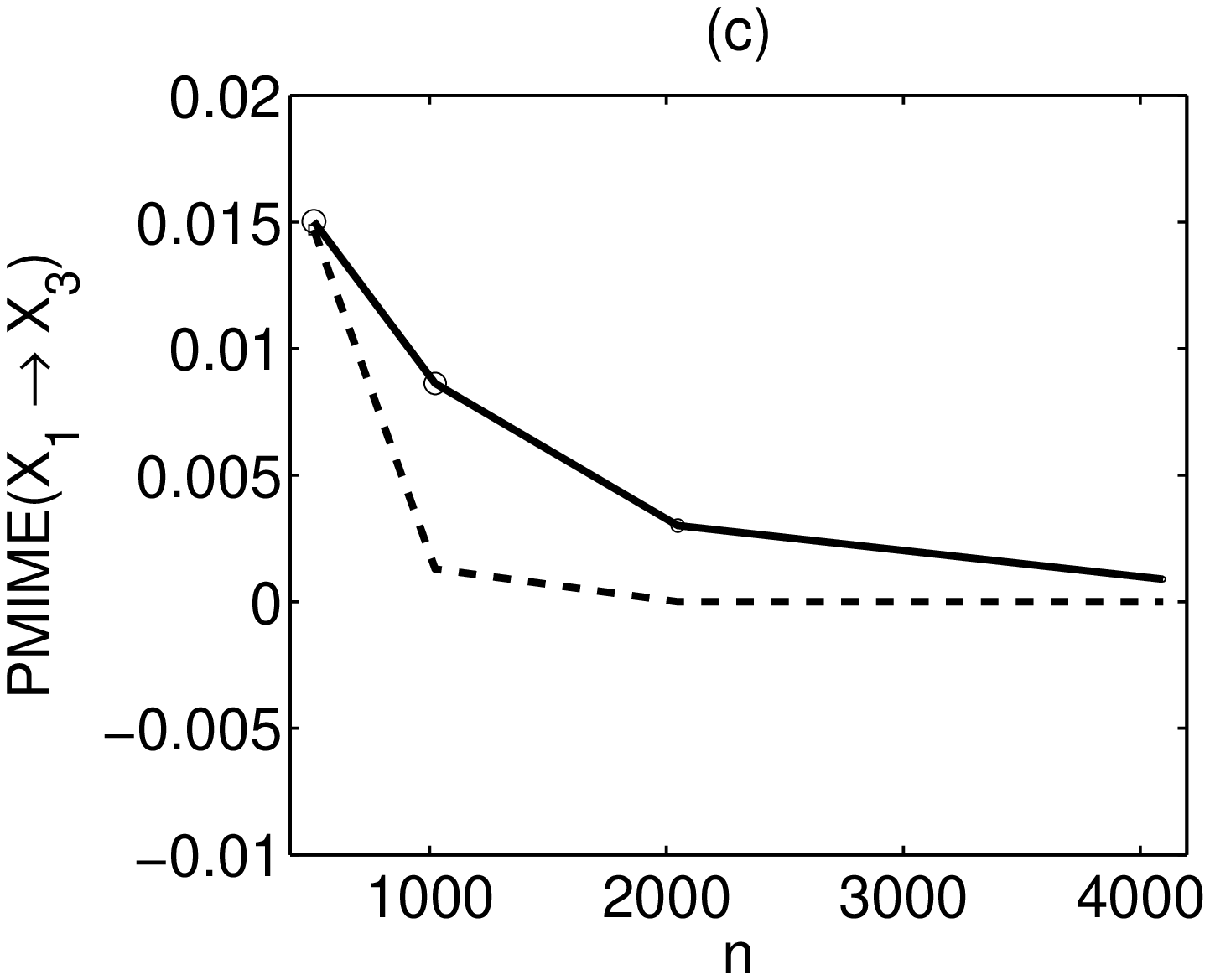}}}
\centerline{\hbox{\includegraphics[width=5cm]{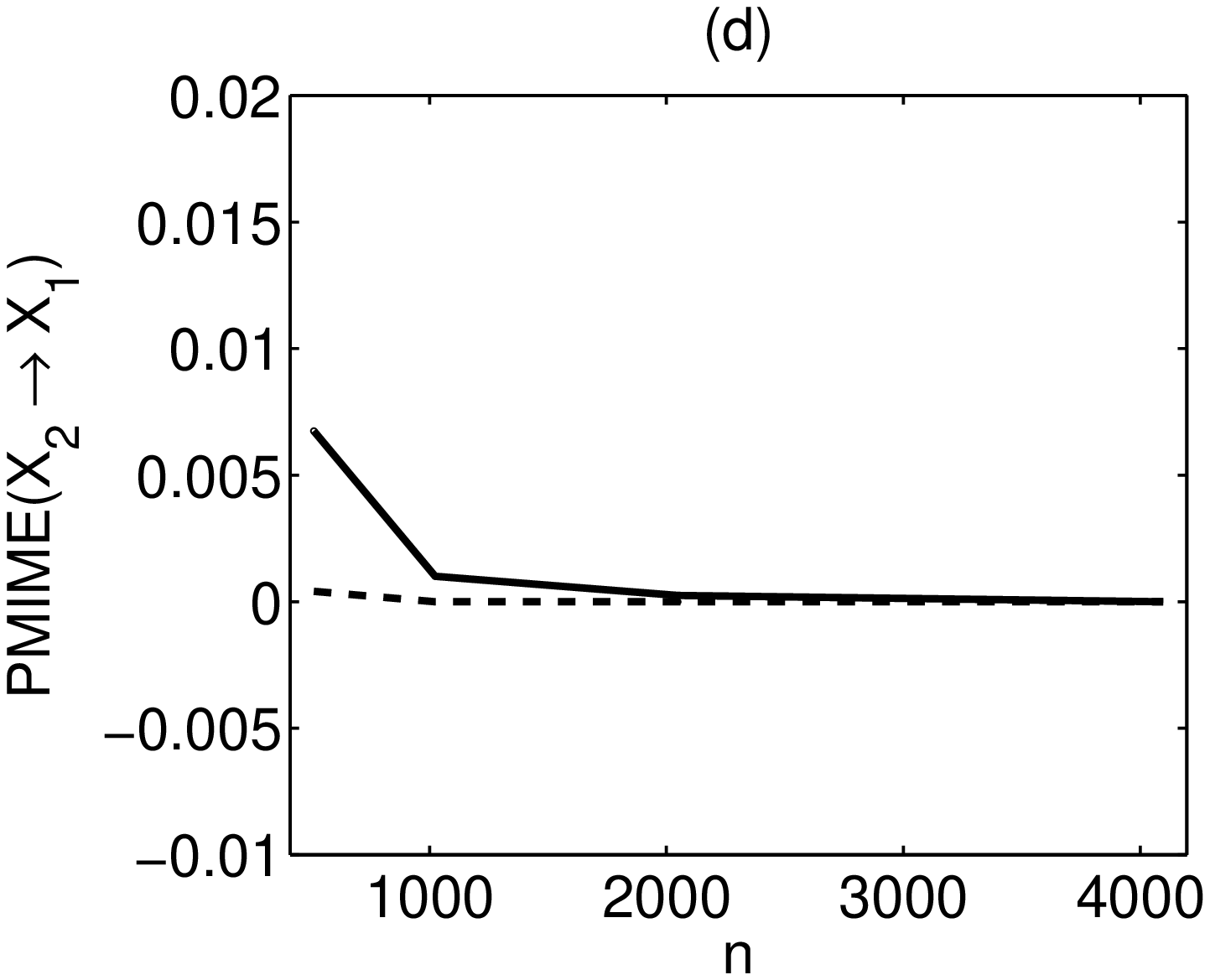}
\includegraphics[width=5cm]{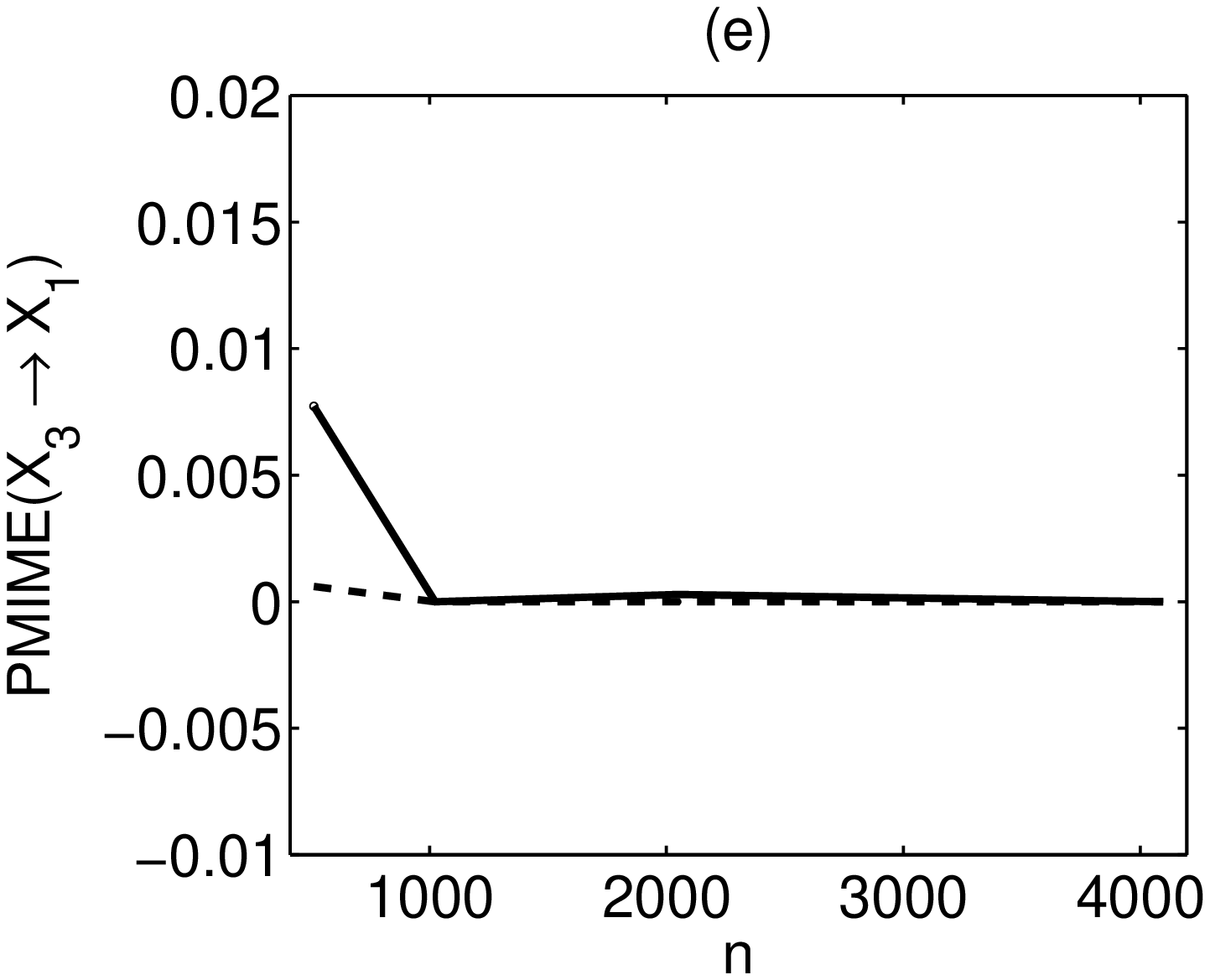}
\includegraphics[width=5cm]{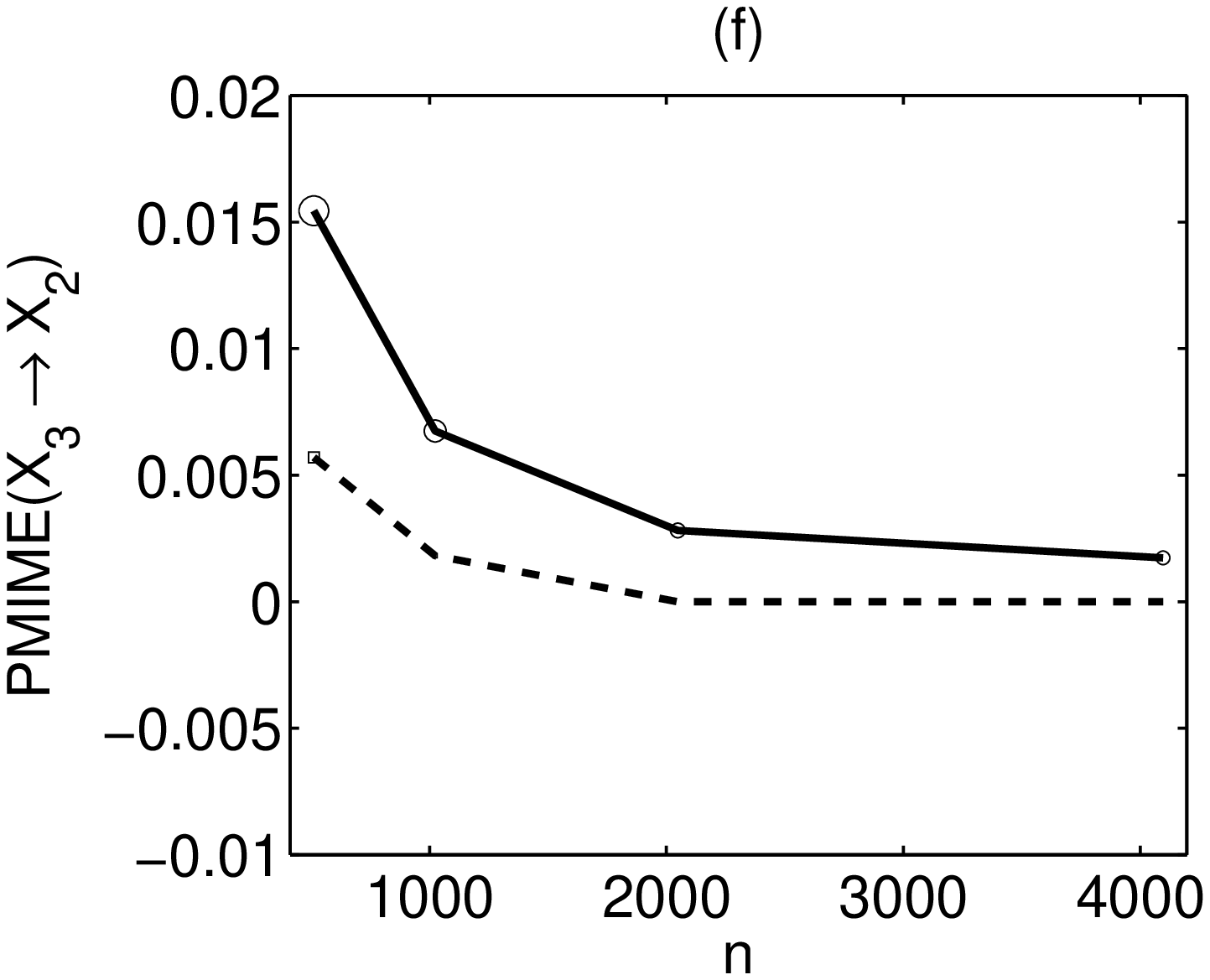}}}
\caption{PMIME measure with adapted threshold ($\alpha=0.05$) and fixed threshold ($A=0.97$) are given as functions of the time series length $n$ for the true direct causality $X_1 \rightarrow X_2$ in (a) and $X_2 \rightarrow X_3$ in (b), the indirect causality $X_1 \rightarrow X_3$ in (c), and the non-existing couplings $X_2 \rightarrow X_1$ in (d), $X_3 \rightarrow X_1$ in (e), and $X_3 \rightarrow X_2$ in (f), for 3 Lorenz subsystems coupled with coupling strength $C=2$. The number of times in 100 realizations PMIME is positive determines the size of a symbol displayed for each threshold type and $n$, where in the legend of panel (a) the size of the symbols regards the maximum of 100 positive PMIME values.}
\label{fig:Lorenzdiffnnoise20}       
\end{figure}

Further investigation of the low specificity of PMIME for small $n$ indicated that this is merely due to the use of the adapted threshold for the termination criterion, i.e. the significance test for the conditional mutual information regarding the selected candidate lagged variable at a significance level $\alpha=0.05$. It seems that for this particular case (coupled Lorenz system, 20\% noise), a fixed threshold of $A=0.97$ is more suitable, as shown in Figure~\ref{fig:Lorenzdiffnnoise20} for $C=2$.
For the two true direct couplings
(Figure~\ref{fig:Lorenzdiffnnoise20}a and b) both threshold types
give positive PMIME for all 100 realizations, but the adapted
threshold gives larger PMIME values, which indicates that the
termination criterion is less stringent and allows more components
of the driving variable in the mixed embedding vector. This seems
to have a negative consequence for the cases of indirect coupling
$X_1 \rightarrow X_3$ (Figure~\ref{fig:Lorenzdiffnnoise20}c) and
the non-existing coupling $X_3 \rightarrow X_2$
(Figure~\ref{fig:Lorenzdiffnnoise20}f), as the adapted threshold
allows other than the lagged response variable components to enter
in the mixed embedding vector, which results in positive PMIME
more often than chance when $n$ is small. Nevertheless, this
effect decreases with $n$. The use of the fixed threshold $A=0.97$
is more appropriate here as it does not produce this effect.
However, as shown in Table~\ref{tab:threshold} and suggested by
other simulations not shown here, the fixed threshold does not
adapt to different inter-dependence structures and data
conditions. For example when 20\% white noise is added to the
system in Figure~\ref{fig:Lorenzdiffnnoise20}, the fixed threshold
of $A=0.97$ gives still the highest specificity but has much lower
sensitivity than the adapted threshold, e.g. for the weak coupling
with $C=1$ and $n=4096$ the adapted threshold with $\alpha=0.05$
detects the true direct couplings (68 positive PMIME for $X_1
\rightarrow X_2$ and 58 $X_2 \rightarrow X_3$) and gives zero
PMIME otherwise, while the fixed threshold $A=0.97$ gives zero
PMIME also for the direct couplings.

\subsection{Coupled Mackey-Glass system}

The last simulated system is a continuous system of coupled
identical Mackey-Glass delayed differential equations defined as
\begin{equation*}
\dot x_i(t) = -0.1 x_i(t) + \sum_{j=1}^K \frac{C_{ij} x_j(t-\Delta_j)}{1+x_j(t-\Delta_j)^{10}}  \quad
\mbox{for} \quad i=1,\ldots,K
\label{eq:coupledMG}
\end{equation*}
For $K=2$ the system was first used in \cite{Senthilkumar08} and then in \cite{Vlachos10}.
The system is solved using the solver {\tt dde23} for delayed differential equations in Matlab and the time series are generated at a sampling time of 4 time units. For $K$ delayed differential equations $K$ time series are generated and the corresponding variables are denoted $X_i$, $i=1,\ldots,K$. When $\Delta_j=\Delta$, $j=1,\ldots,K$, the $K$ coupled subsystems are identical and we consider this case here. We also set $C_{ii}=0.2$ and let $C_{i,j}$ for $i\neq j$ determine the coupling structure.

For the future vector we set $\mathbf{y}_t^3=[y_{t+1},y_{t+\tau_1},y_{t+\tau_2}]$, where $\tau_1$ and $\tau_2$ are respectively the first minimum and maximum of the delayed mutual information for the response variable $Y$ (where $Y$ stands for any of $X_i$, $i=1,\ldots,K$). This choice is found in \cite{Vlachos10} to better represent the short-term time evolution of the response system as the Mackey-Glass system exhibits irregular oscillations. This future vector is used in PMIME and PTE, while for CGCI the option of a larger step ahead $T>1$ is not considered and it is computed for $T=1$.

\begin{figure}
\centerline{\hbox{\includegraphics[width=7.5cm]{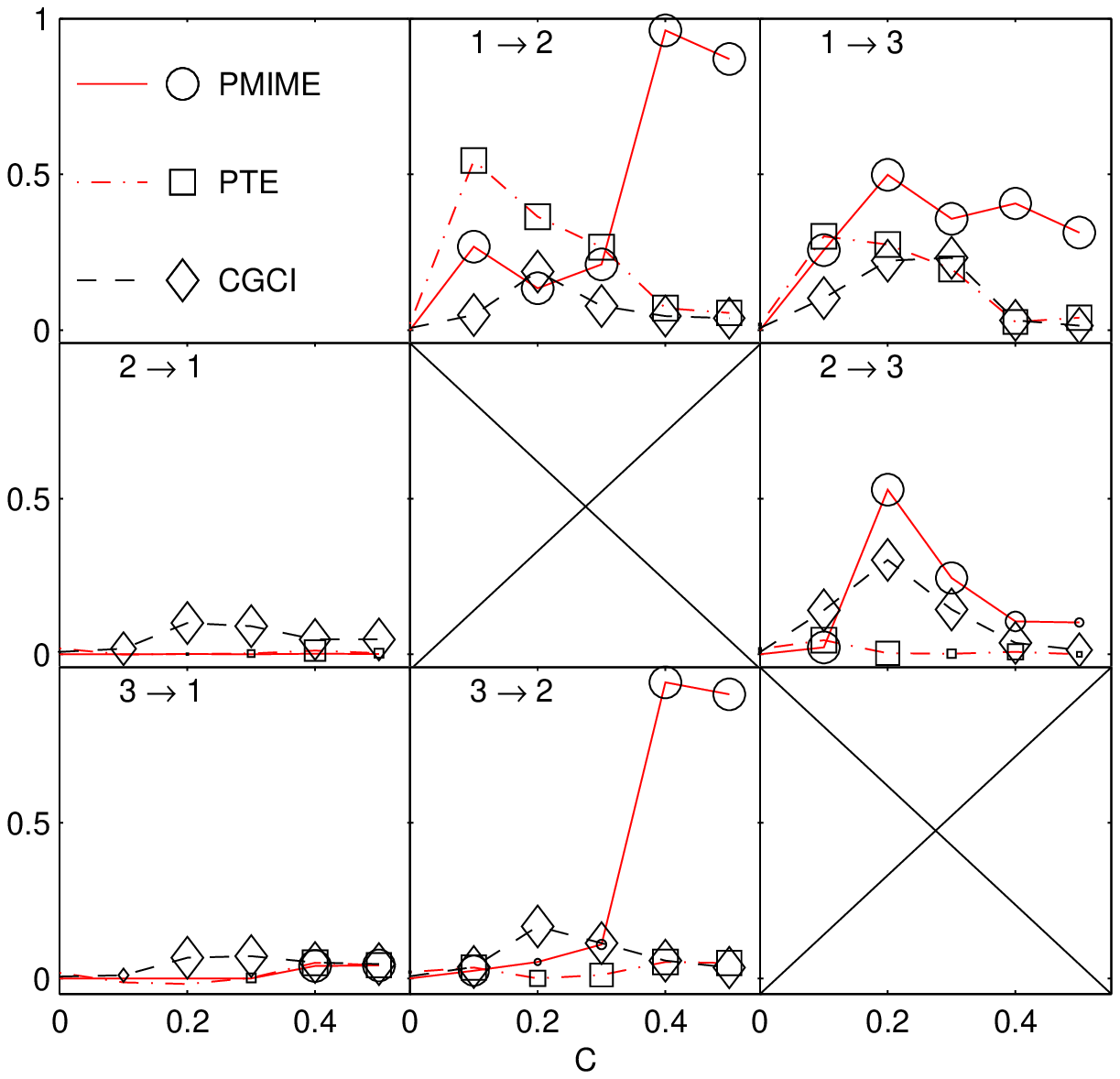}
\includegraphics[width=7.5cm]{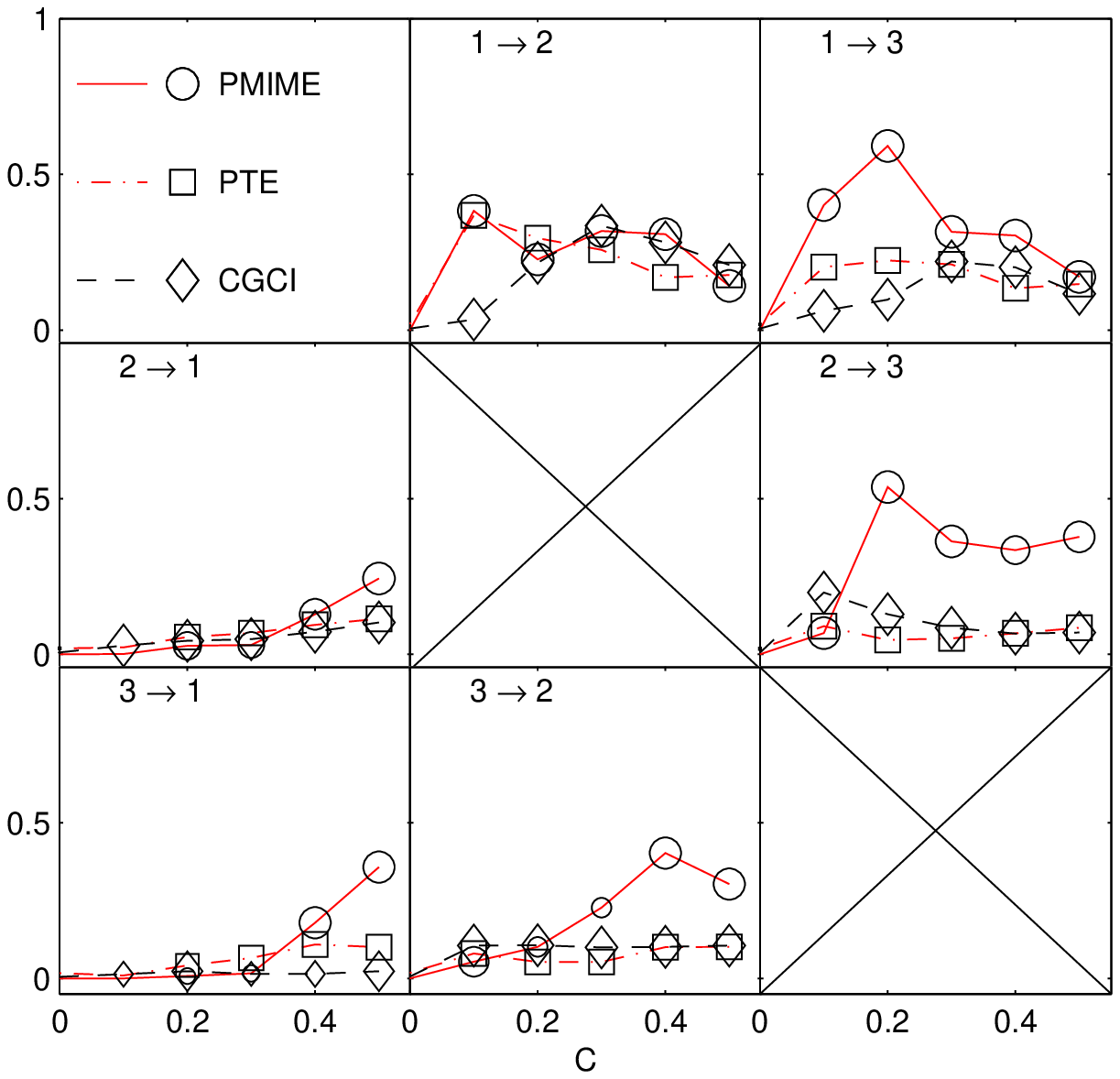}}}
\caption{Matrix plot of all possible couplings of the $K=3$ coupled Mackey-Glass subsystems and noise-free time series of length $n=2048$ on the left panel, and with 20\% additive white noise on the right panel. The organization of the panels is as for Figure~\ref{fig:HenonK5diffCnoisefree}.}
\label{fig:MGdiffCn2048}       
\end{figure}

First we consider the case of $K=3$, $\Delta_1=\Delta_2=\Delta_3=20$, $C_{1,2}=C_{1,3}=C_{2,3}=C$ and $C_{2,1}=C_{3,1}=C_{3,2}=0$. The results of the three measures on 100 realizations of length $n=2048$ of this system for varying coupling strength $C$ are shown in Figure~\ref{fig:MGdiffCn2048} for noise-free and noisy time series.
For the noise-free data, PMIME detects best the three direct
couplings (upper triangular panel components) and only for strong
coupling $C \geq 0.4$ the number of positive values decreases for
the coupling $X_2 \rightarrow X_3$. The latter holds also for PTE,
which is less sensitive for smaller coupling strengths, e.g. for
$C=0.3$ and $X_2 \rightarrow X_3$ PMIME gives 99 positive values
and PTE only 34 statistically significant values. CGCI gives the
highest sensitivity of 100\% rejection rate of the null hypothesis
of no-coupling, but this is of little benefit as it is followed by
a very low specificity giving about the same highest rejection
rate when there is no causal effect. PTE has also low specificity
for all $C$, but PMIME only for $C \geq 0.4$. For example for the
non-existing connection $X_3 \rightarrow X_2$ and $C=0.3$, PTE
gives 91 statistically significant values and PMIME 31 positive
values, and these are 61 and 20, respectively for $C=0.2$, whereas
CGCI gives constantly 100 statistically significant values.

For the noisy data, the sensitivity of the measures remain about the same, but the specificity gets lower for PTE and PMIME, with PMIME still performing better than PTE. For the setting $X_3 \rightarrow X_2$ and $C=0.3$ the positive PMIME values are 63 and the statistically significant PTE values are 100, and the same holds for $C=0.2$ and the other two non-existing couplings.

\begin{figure}
\centerline{\hbox{\includegraphics[width=4.5cm]{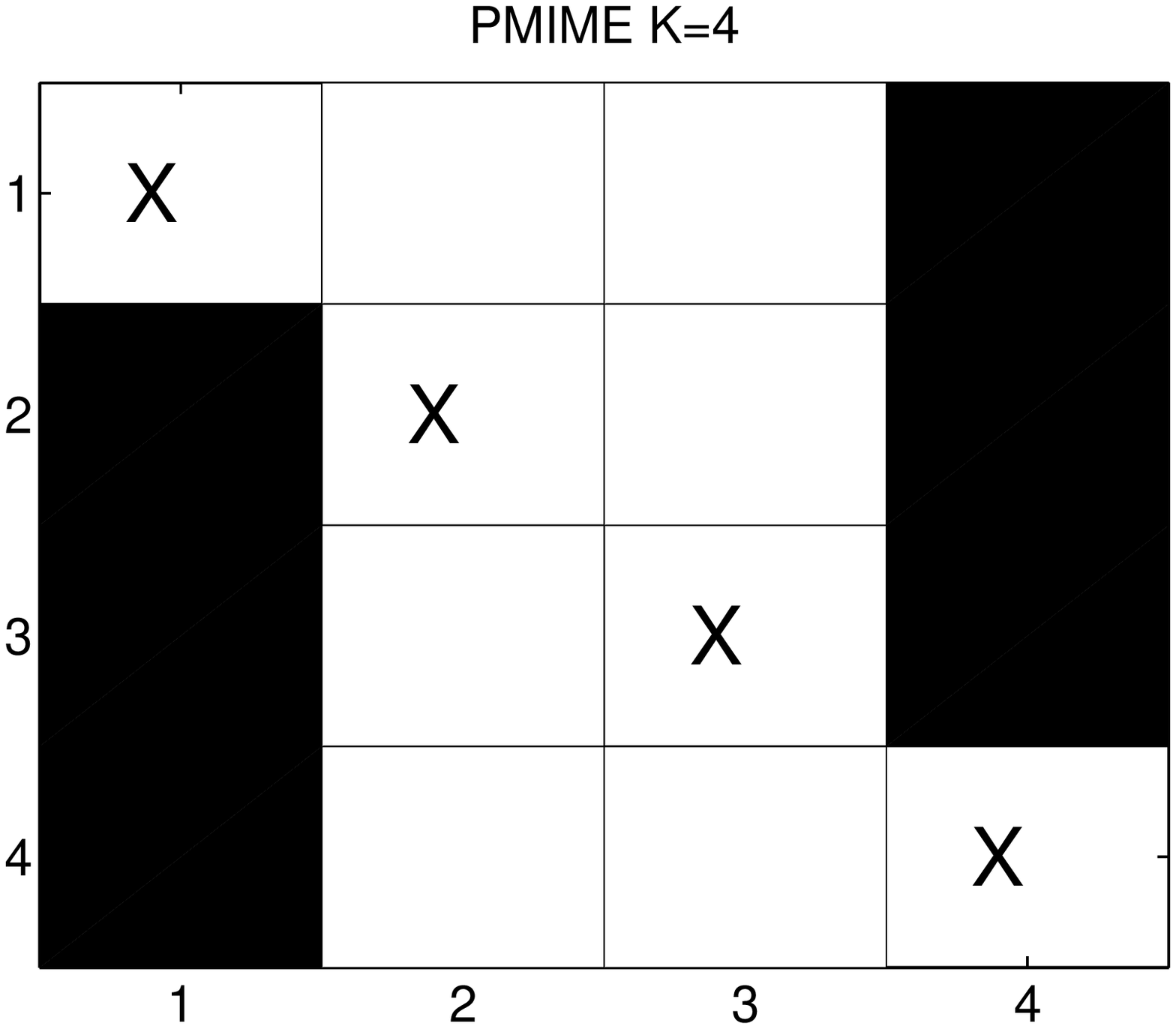}
\includegraphics[width=4.5cm]{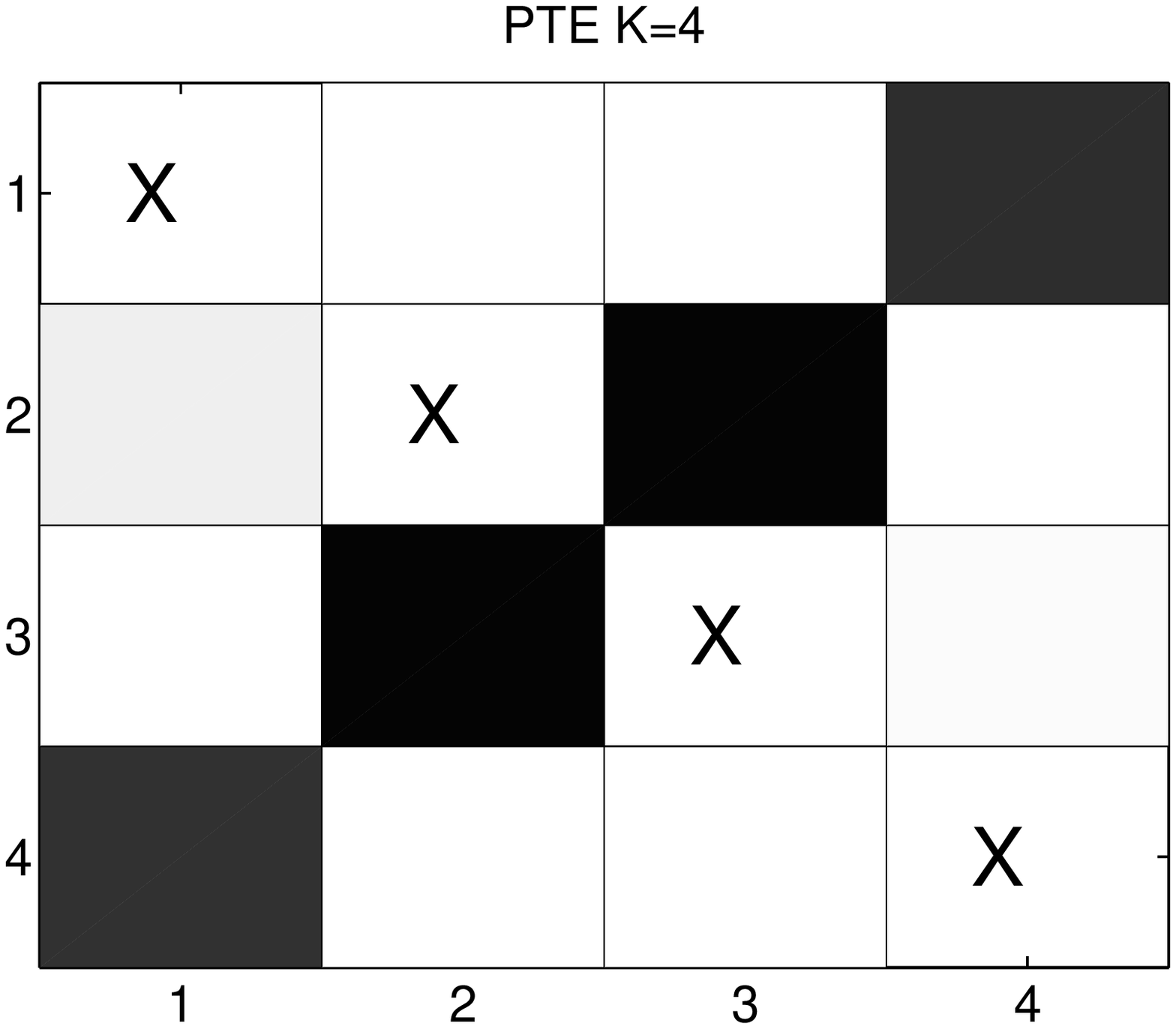}
\includegraphics[width=4.5cm]{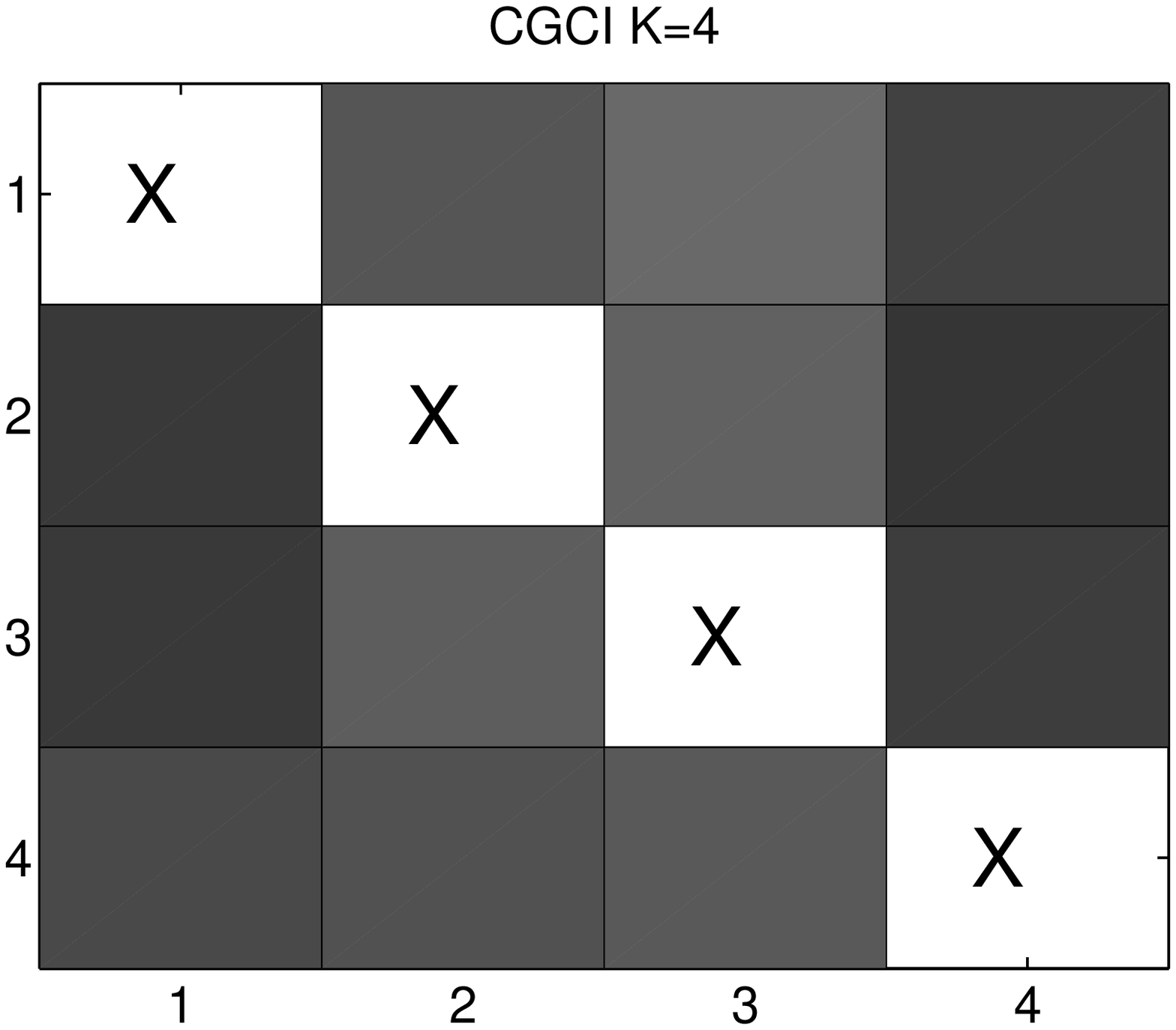}}}
\centerline{\hbox{\includegraphics[width=4.5cm]{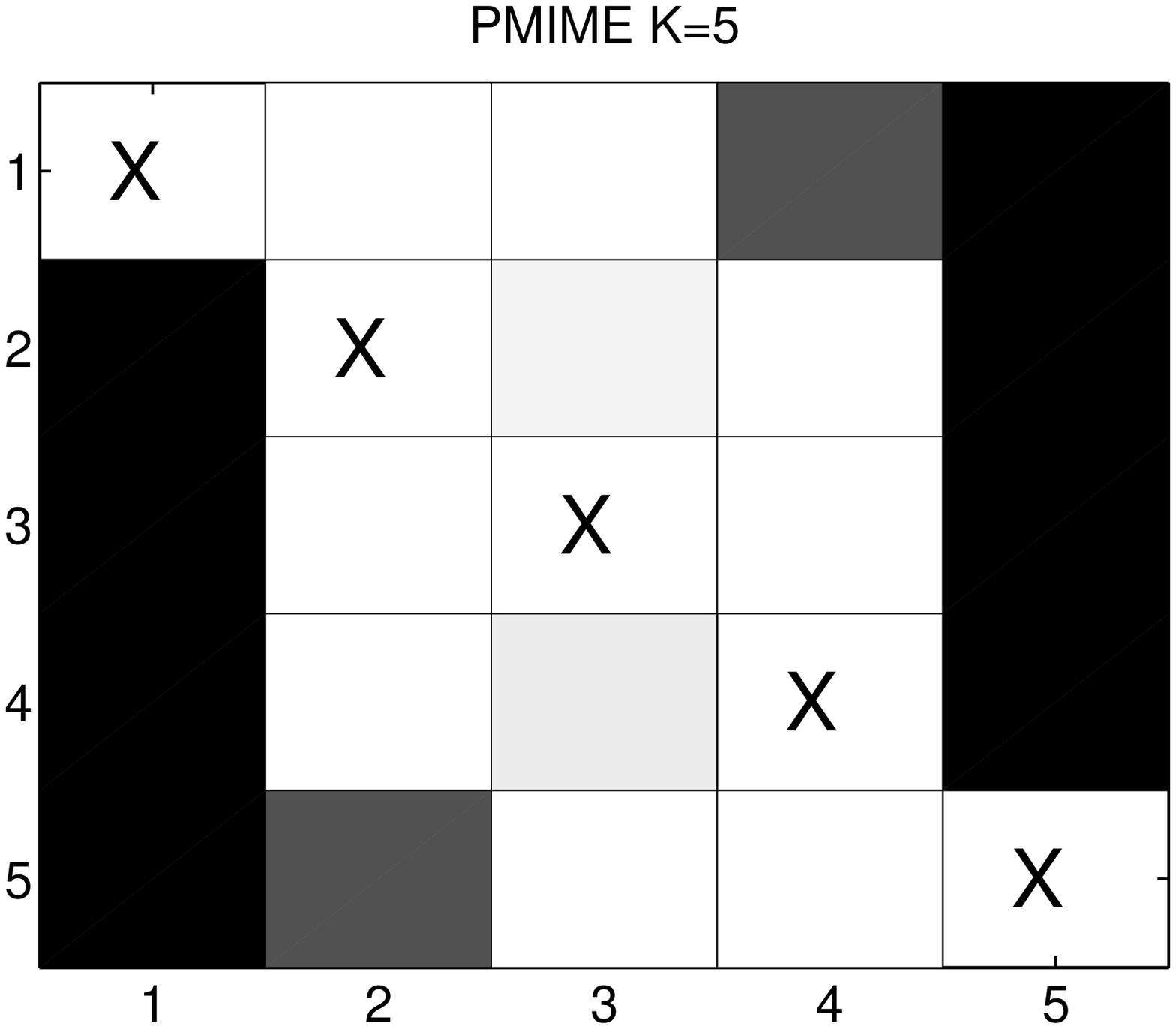}
\includegraphics[width=4.5cm]{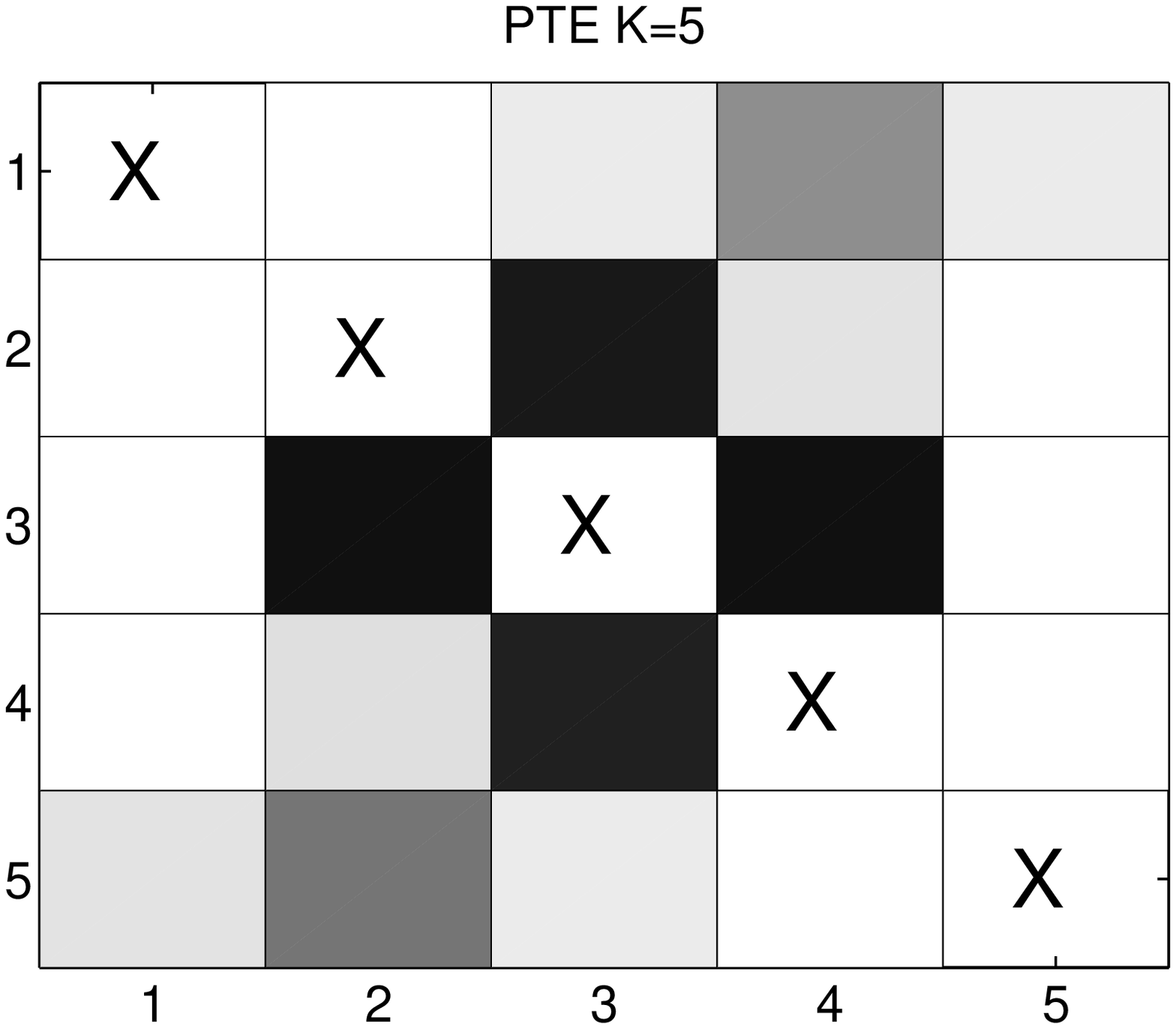}
\includegraphics[width=4.5cm]{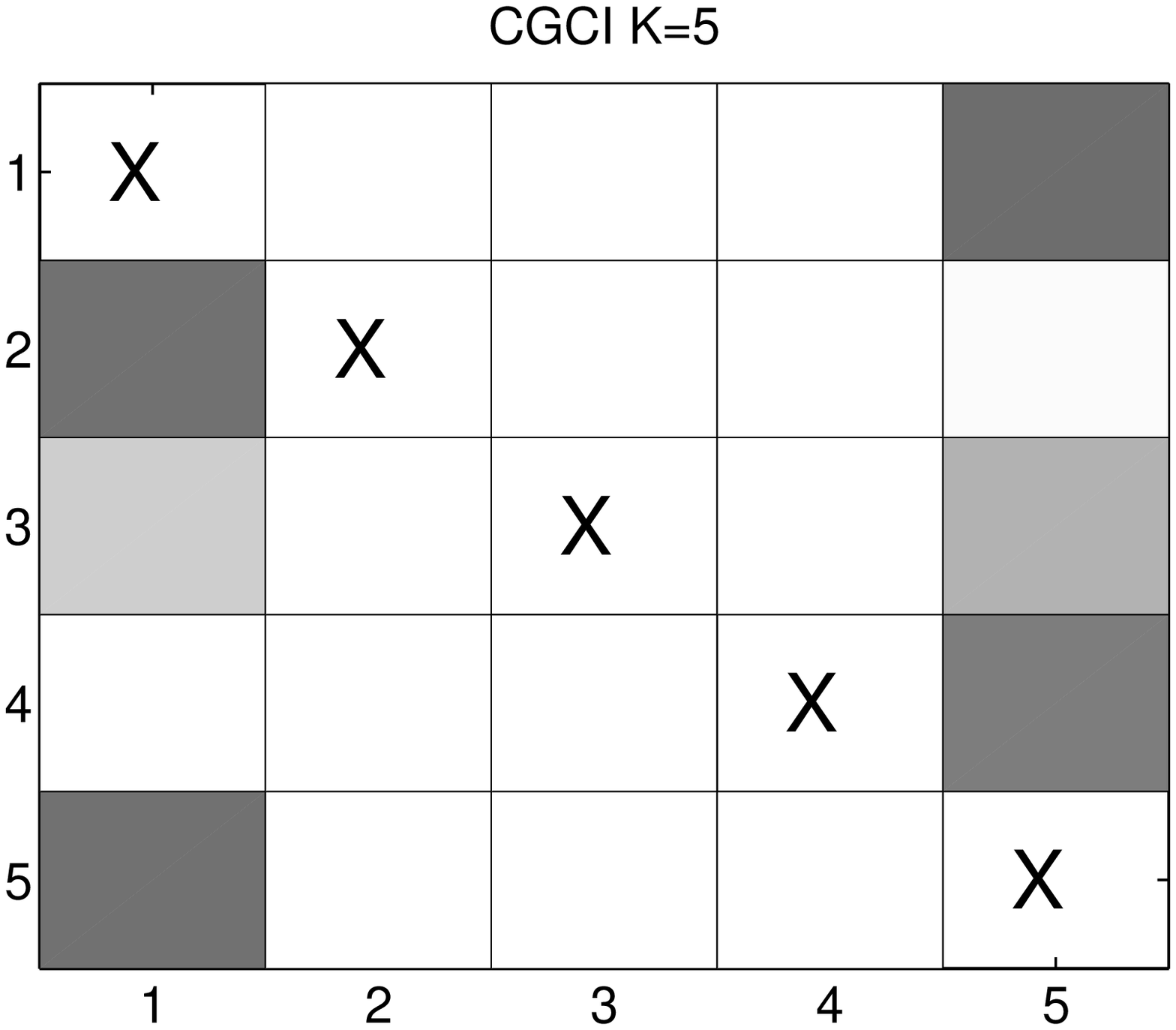}}}
\centerline{\hbox{\includegraphics[width=4.5cm]{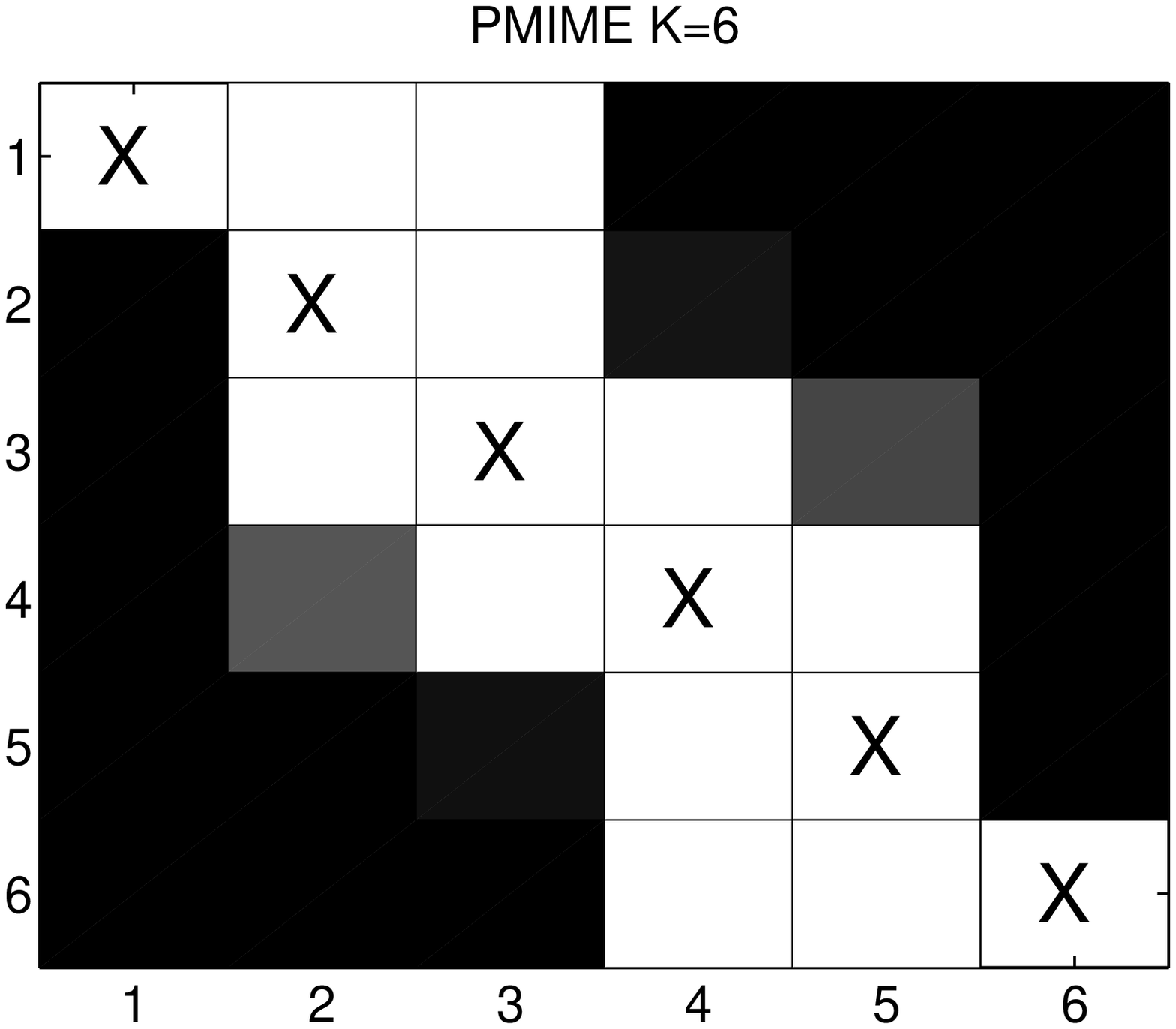}
\includegraphics[width=4.5cm]{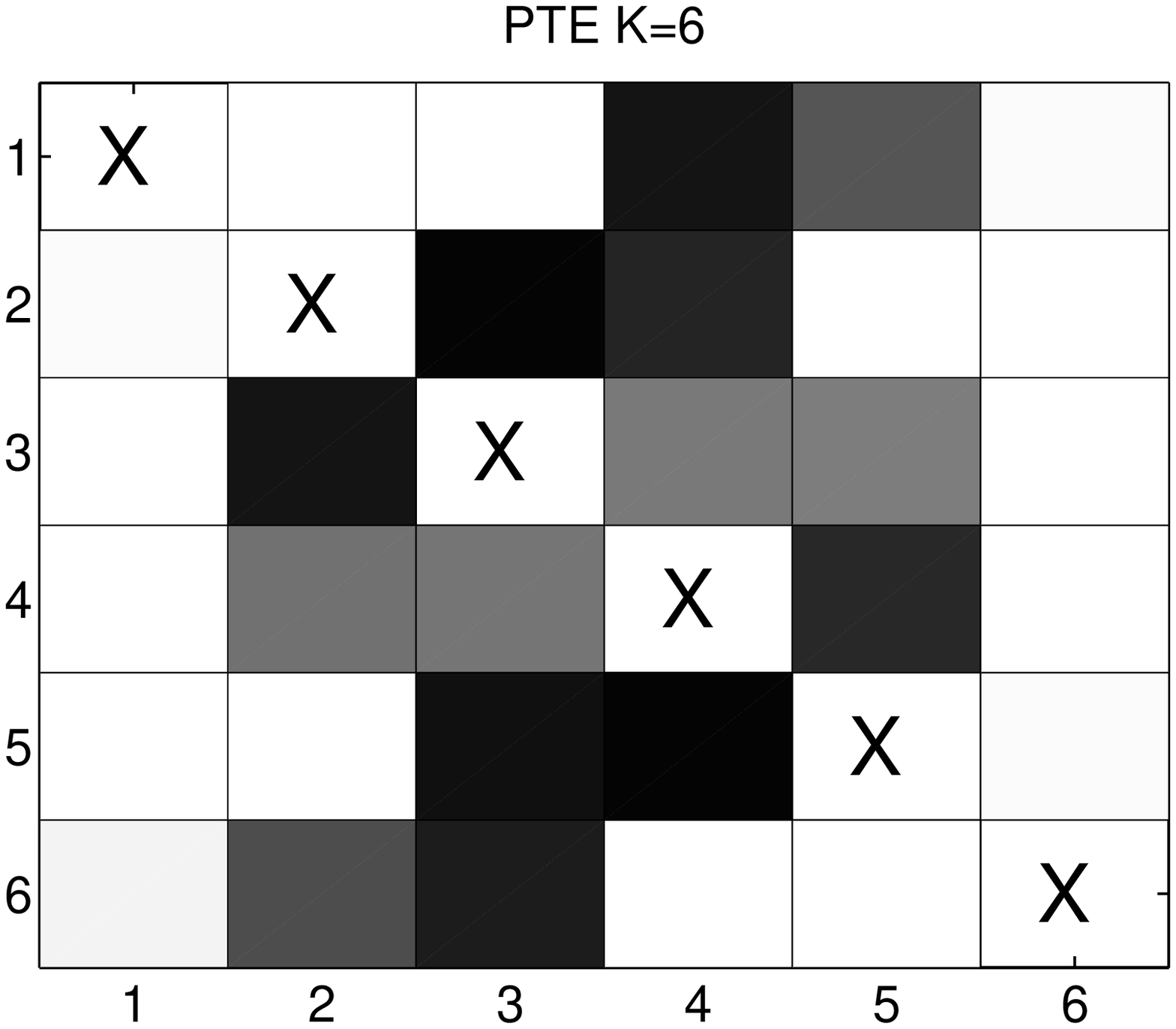}
\includegraphics[width=4.5cm]{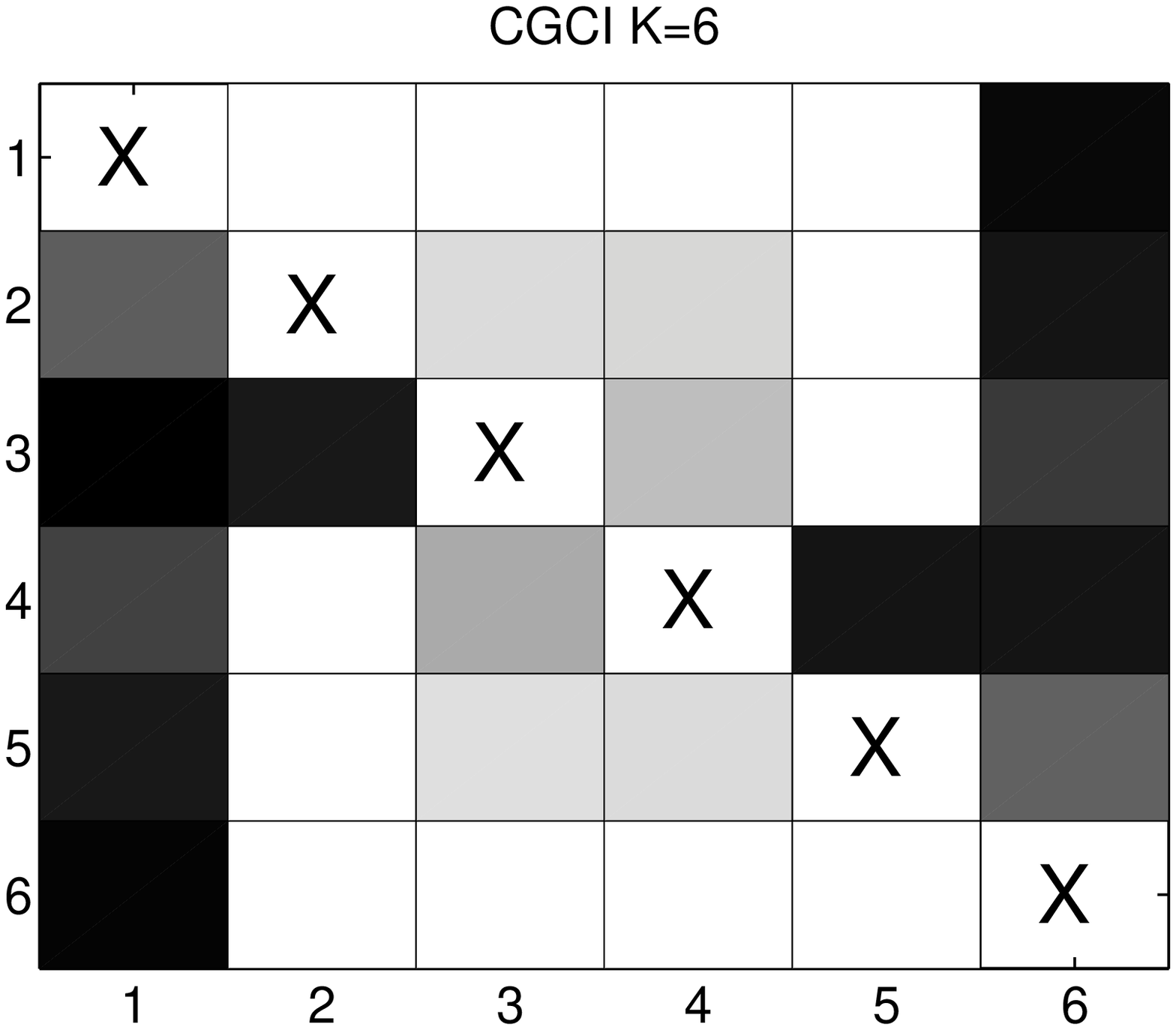}}}
\caption{Color maps for the statistical significance of the three measures, PMIME in first column, PTE in the second column and CGCI in the third column, for all possible couplings of the $K$ coupled Mackey-Glass subsystems, where $K=4$ in the first row, $K=5$ in the second row and $K=6$ in the third row. The time series are noise-free and have length $n=2048$, and the subsystems have all the same delay $\Delta=20$ and each subsystem $X_i$ drives $X_{i-1}$ ($X_i \rightarrow X_{i-1}$) and $X_{i+1}$ ($X_i \rightarrow X_{i+1}$) with the same coupling strength $C$ ($X_1$ only drives $X_2$ and $X_K$ only drives $X_{K-1}$). The statistical significance of PMIME is given by the number of positive values in the 100 realizations and for PTE and CGCI by the number of rejections of the null hypothesis of no coupling. The gray color scale is from 0 (black) to 100 (white).}
\label{fig:MGdiffKnoisefreen2048}       
\end{figure}

For larger $K$ the differences of PMIME from PTE and CGCI become clearer. In Figure~\ref{fig:MGdiffKnoisefreen2048}, the results are shown in the form of color maps for the statistical significance of the three measures for increasing number $K$ of weakly coupled Mackey-Glass subsystems.
The variables (subsystems) have all the same delay $\Delta=20$ and
each variable $X_i$ drives the variable next in the left ($X_i
\rightarrow X_{i-1}$) and in the right ($X_i \rightarrow X_{i+1}$)
with the same coupling strength $C$, where $X_1$ only drives $X_2$
and $X_K$ only drives $X_{K-1}$. The time series are noise-free
and have length $n=2048$. An interesting feature of PMIME is that
for any of $K=4,5,6$, there is no driving to the first variable
$X_1$ and the last variable $X_K$, which are designed not to
receive any causal effect from another variable, and the
corresponding PMIME values are zero for all realizations. This is
not preserved for the other two measures, with CGCI scoring close
to PMIME with regard to driving of $X_1$ and $X_K$ for $K=6$. With
further regard to specificity, PMIME gives positive values for the
indirect couplings, but this inadequacy of PMIME improves with $K$
and for $K=6$ PMIME is zero for the most of the indirect
couplings. The sensitivity of PMIME is the highest for all $K$ and
PMIME is positive at all realizations (almost all for $K=5$) for
all direct couplings. On the other hand, PTE and CGCI fail to
detect the coupling structure of the system for any $K$ with PTE
giving very low sensitivity and specificity.

We note here that the coupled Mackey-Glass system may involve
complicated scenaria of coupling structures that are difficult to
detect, and indeed PMIME gets also fooled and identifies spurious
couplings. We observed this especially when the subsystems are not
identical, setting different $\Delta_i$. For such situations, it
remains an open problem whether the Granger causality measures can
distinguish intrinsic dynamics from the inter-dependence
structure, e.g. see the discussion in \cite{Sugihara12}. It should
also be noted that the performance of PMIME could possibly be
improved if we would choose a maximum lag $L>15$, e.g. in
\cite{Vlachos10} $L=50$ was used giving good results for $K=2$,
but such large $L$ was not used here due to increased demand of
computation time. However, for the simulations with $K>3$ we
experienced that the computation of PTE with $m=5$ using 100
surrogates for the significance test is much more time demanding
than for PMIME.

\subsection{Real world example}

\begin{figure}[h!]
\centerline{\hbox{\includegraphics[width=5cm]{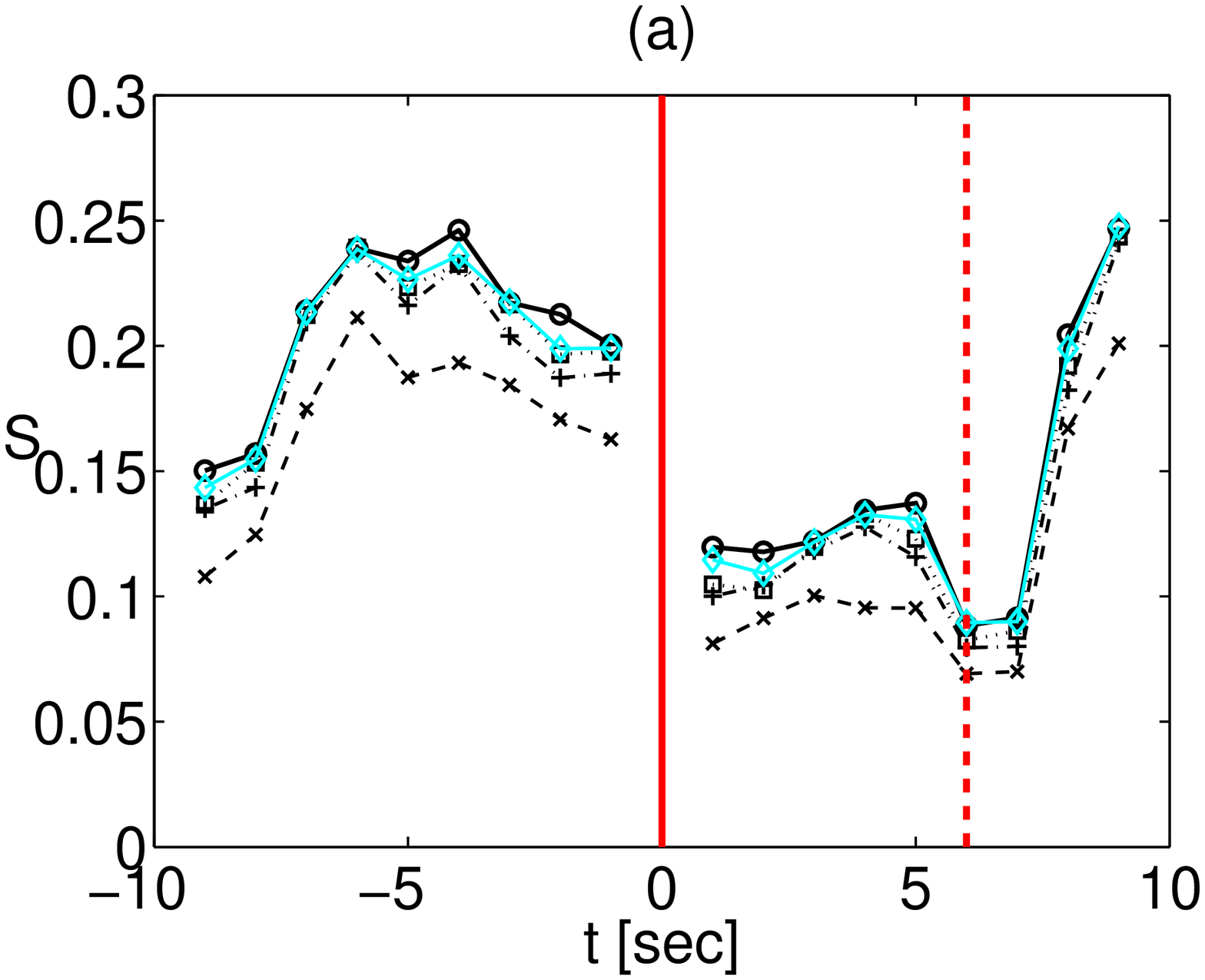}
\includegraphics[width=5cm]{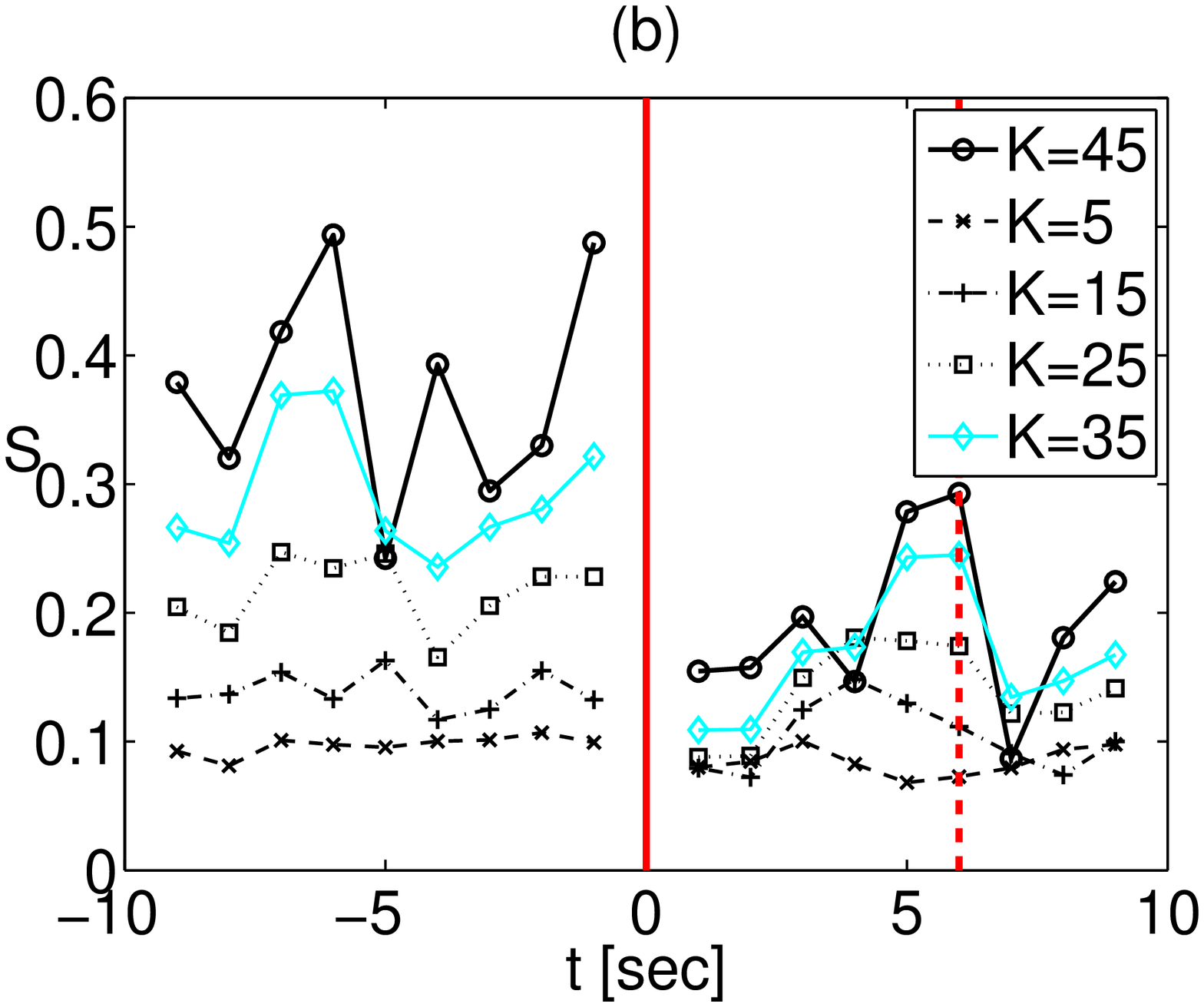}}}
\centerline{\includegraphics[width=5cm]{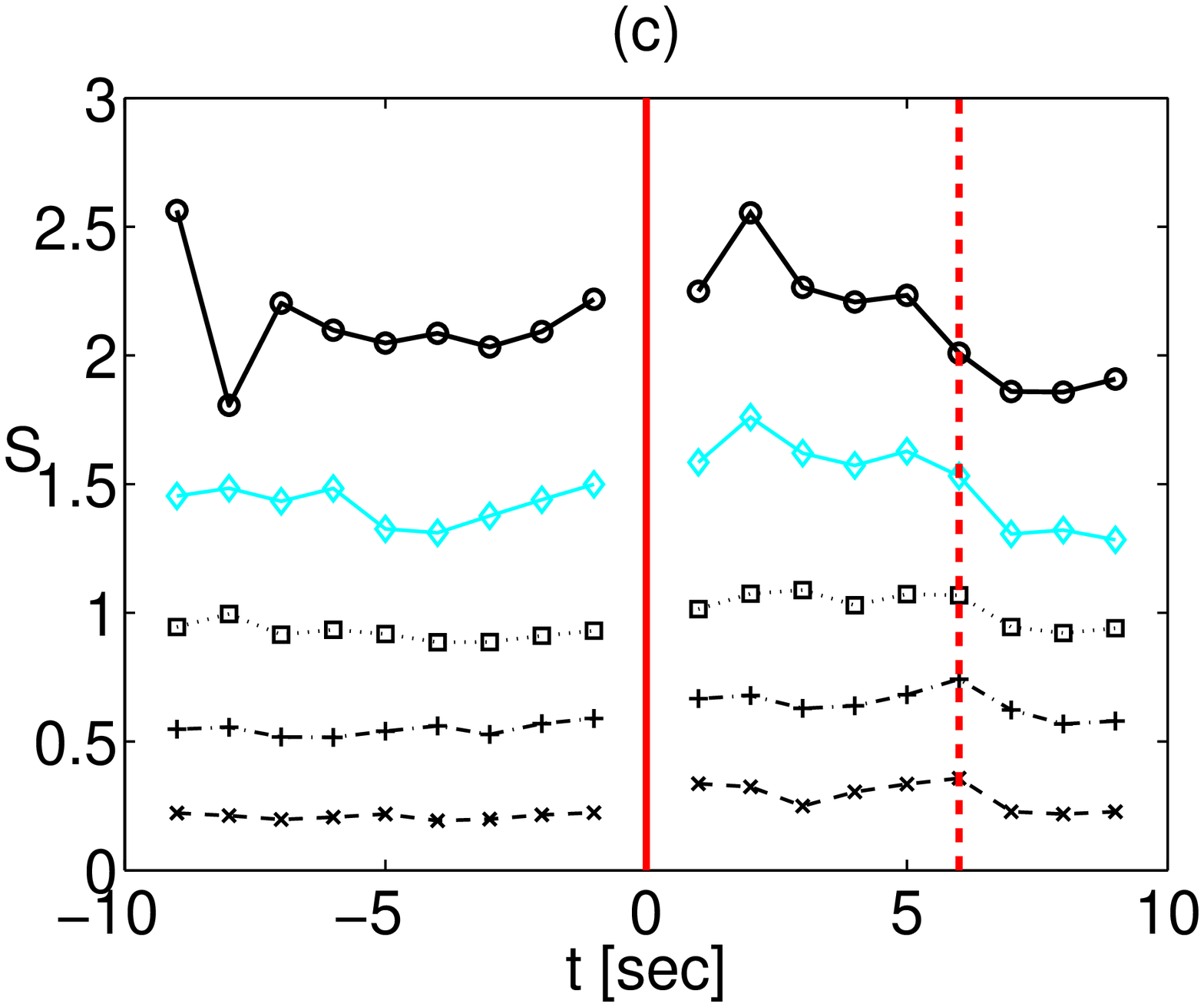}}
\caption{Mean average strength $S$ for PMIME in (a), PTE in (b) and CGCI in (c), on sliding windows along the period before and after the start of ED (marked with a vertical line at time zero) for the number of channels as shown in the legend. The vertical dashed line denotes the end of the ED.}
\label{fig:EEG}
\end{figure}

Finally, we demonstrate the robustness of PMIME and its
appropriateness in connectivity analysis and complex networks with
an example of a human scalp multi-channel electroencephalogram
(EEG) recording during an epileptic discharge (ED), i.e. an
electrographic seizure of short duration \footnote{The data were
provided by V.K. Kimiskidis at the Department of  Neurology III,
Medical School, Aristotle University of Thessaloniki.}. After
artifact rejection, filtering and re-referencing, a set of 45 EEG
signals were obtained and downsampled to 200 Hz frequency,
covering 10 sec before and 10 sec after the start of an ED of
duration 5.7 sec. We computed in each of the two periods PMIME,
PTE and CGCI ($m\! = \!5$), all for $T\! = \!1$ and for all
possible channel pairs on sliding windows of 2 sec with a step of
1 sec. To assess the strength of connectivity in the brain network
estimated by each causality measure at each 2 sec window, we
computed the average strength $S$ (mean of the measure values over
all channel pairs). In an attempt to test the robustness of the
causality measures to the network size, we repeated the same
analysis 12 times on randomly selected subsets of the set of 45
channels. The results are shown in Fig.~\ref{fig:EEG} for subset
sizes 5, 15, 25 and 35.
We observe that only PMIME gives a stable pattern of connectivity strength over the two periods for any subset (some deviation can be seen for subset size 5). Moreover, PMIME distinguishes readily the period before ED, during ED and after ED. Similar results were obtained using the average degree, i.e. the mean binary connections obtained by the significance test for PTE and CGCI for $\alpha\! = \!0.05$ and when PMIME is positive.

\section{Conclusions}
\label{sec:conclusions}

The presented measure PMIME addresses successfully the problem of
identifying direct causal effects in the presence of many
variables. Intensive simulations on discrete- and continuous-timed
coupled systems have confirmed this. While Taken's embedding
theorem advocates against the estimation of direct Granger
causality in nonlinear systems \cite{Sugihara12}, and a vector
with lagged components only from the response variable may be
representing equivalently the mixed embedding vector, in practice
PMIME pinpoints the set of the most and significantly contributing
lagged components, identifying thus the direct causal effects.

PMIME does not rely on embedding parameters, and the structure of
the mixed embedding vector allows for identification of the active
lags of the driving variable affecting the response. The latter is
currently an active research direction \cite{Runge12b,Wibral13},
but we did not take it up in this study. Our initial results for
detecting the true active lags in the bivariate analysis with MIME
were promising, and work on this with PMIME is in progress.

We have improved the termination criterion in the progressive
building of the mixed embedding vector, initially set for the
bivariate measure MIME, and instead of using a fixed threshold we
let the threshold be adjusted by the estimated bias using
randomized replicates.

We have showed that PMIME scores highest in sensitivity and
specificity as compared to PTE and CGCI, and moreover it does not
require computationally intensive randomization (surrogate)
significance test. While PMIME is much slower than PTE it is less
computational intensive if PTE has to be combined with
randomization test or when the number of observed variables gets
large. The example on an EEG record of epileptic discharge
demonstrates the usefulness of PMIME in analyzing multivariate
time series from real complex systems and constructing causal
networks.


%

\end{document}